%% file: rememberingGeorgeFinal.tex
\documentclass[a4paper,12pt]{article}

\usepackage{geometry}
\usepackage{times}
\usepackage{graphicx}
\usepackage{layout}
\usepackage{tikz}
\usepackage{setspace}
\usepackage{caption}
\usepackage{pdfpages}

\usepackage{graphicx}

\usepackage[utf8]{inputenc}

\usepackage{hyperref}
\hypersetup{
    linktocpage}

\begin{document} 

\pagenumbering{roman}





\thispagestyle{empty}

\hspace*{-0.27in}
\includegraphics[height=297mm,keepaspectratio]{
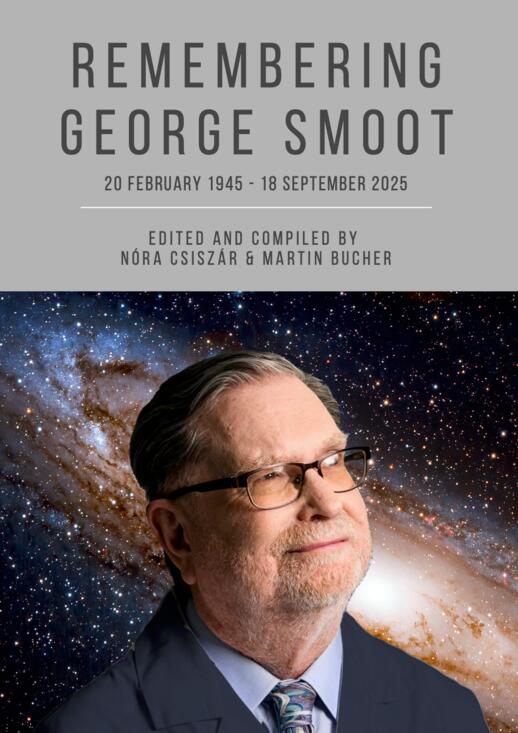}

\newpage

\thispagestyle{empty}

\phantom{Raul}

\newpage

\thispagestyle{empty}

\hspace*{-0.27in}
\includegraphics[height=297mm,keepaspectratio]{
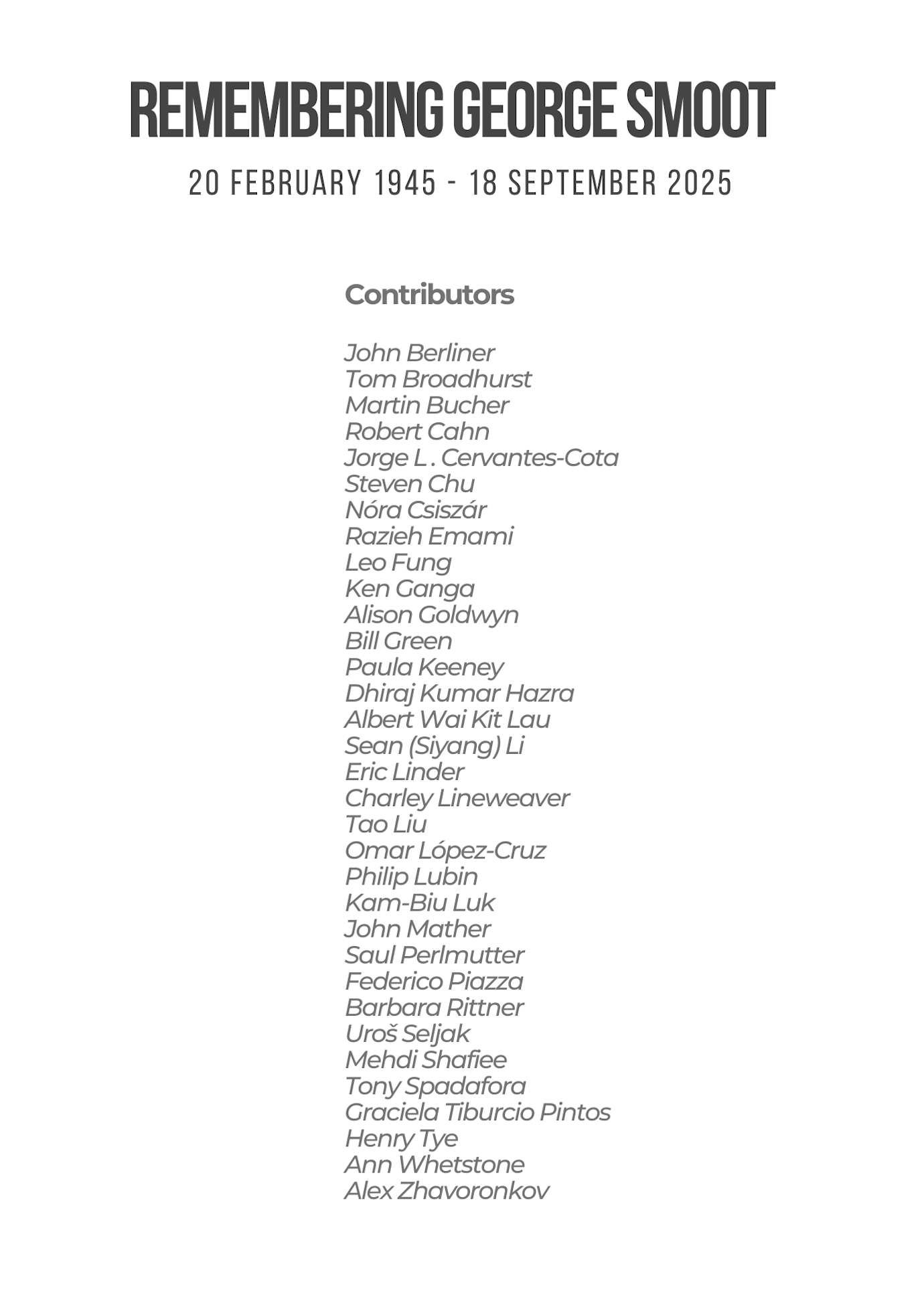}

\newpage

\newgeometry{
  left=15mm, width=180mm,right=15mm, 
height=257mm, top=20mm, bottom=20mm
}

\thispagestyle{empty}

\phantom{Raul}

\vskip 5in 

\noindent 
\phantom{ Rauuuuuuul}

\begin{quote}
\copyright 2026 N\'ora Csisz\'ar

\vskip 16pt 

\noindent 
This publication is in copyright. Subject to statutory exception,
no reproduction of any part may take
place without the written permission of the copyright holder.  

\vskip 0.75in 

\noindent
Atelier DMNB

\vskip 0.75in 

Front cover image credit: Jacopo Landi

Back cover image credit: NASA

\vskip 1cm

www.georgesmoot.com 

\end{quote}

\newpage

%
%

\newpage

\newgeometry{
  left=15mm, width=180mm,right=15mm, 
height=257mm, top=20mm, bottom=20mm
}

\renewcommand\baselinestretch{1.5}
\selectfont
\vskip 4in

%
%
%

\tableofcontents 

\clearpage 

\addcontentsline{toc}{section}{Preface}
\begin{center}
\Large 
Preface
\end{center}

\input{preface.tex}

\clearpage 

\newpage

\thispagestyle{empty}

\phantom{Raul}

\newpage 
\pagenumbering{arabic}

\section*{}
\addcontentsline{toc}{section}{Obituaries and News Articles}

\vskip 4in 
\begin{center}
\Huge
\textsc{Obituaries and News Articles}
\end{center}
\clearpage 

\subsection*{APC Announcement}
\addcontentsline{toc}{subsection}{\hskip 0.5in 
Laboratoire AstroParticules et Cosmologie}

\input{apc_statement.tex}

\clearpage 

\subsection*{LBNL Statement}
\addcontentsline{toc}{subsection}{\hskip 0.5in 
Lawrence Berkeley National Laboratory Announcement}

\input{witherell.tex}

\clearpage 

\subsection*{Le Monde Nécrologie}
\addcontentsline{toc}{subsection}{\hskip 0.5in 
Le Monde Obituary}

\input{leMonde.tex}

\clearpage 

\subsection*{}
\addcontentsline{toc}{subsection}{\hskip 0.5in 
UC Berkeley News Article}

\input{berkeley_announcement.tex}

\clearpage 

\subsection*{} 
\addcontentsline{toc}{subsection}{\hskip 0.5in 
New York Times Obituary}

\input{nyt.tex}

\newpage 

\thispagestyle{empty}

\phantom{Raul} 

\newpage


\section*{}
\addcontentsline{toc}{section}{Testimonials}

\vskip 4in 
\begin{center}
\Huge
\textsc{Testimonials}
\end{center}

\newpage 

\subsection*{John Berliner}
\addcontentsline{toc}{subsection}{\hskip 0.5in John Berliner}

\input%
{johnBerliner.tex}

\newpage 

\subsection*{Tom Broadhurst}
\addcontentsline{toc}{subsection}{\hskip 0.5in Tom Broadhurst}

\input%
{broadhurst.tex}

\newpage 

\subsection*{Martin Bucher}
\addcontentsline{toc}{subsection}{\hskip 0.5in Martin Bucher}

\input{martinBucher.tex}

\newpage 

\section*{Robert Cahn}
\addcontentsline{toc}{subsection}{\hskip 0.5in 
Robert Cahn}

\input{robertCahn.tex}

\newpage 

\section*{Jorge L.~Cervantes-Cota}
\addcontentsline{toc}{subsection}{\hskip 0.5in 
Jorge L.~Cervantes-Cota}

\input{jorge-cervantes-cota.tex}

\newpage 

\section*{Steven Chu}
\addcontentsline{toc}{subsection}{\hskip 0.5in 
Steven Chu}

\input{steve_c.tex}

\newpage 

\section*{Razieh Emami}
\addcontentsline{toc}{subsection}{\hskip 0.5in 
Razieh Emami}

\input{raz.tex}

\newpage 

\section*{Leo Fung}
\addcontentsline{toc}{subsection}{\hskip 0.5in 
Leo Fung}

\input{leo_fun.tex}

\newpage 

\section*{Ken Ganga}
\addcontentsline{toc}{subsection}{\hskip 0.5in 
Ken Ganga
}

\input{ken2.tex}

\newpage 

\section*{Alison Goldwyn}
\addcontentsline{toc}{subsection}{\hskip 0.5in 
Alison Goldwyn}

\input{alisonGoldwyn.tex}

\newpage 

\section*{Bill Green}
\addcontentsline{toc}{subsection}{\hskip 0.5in 
Bill Green}

\input{billGreen.tex}

\section*{Dhiraj Kumar Hazra}
\addcontentsline{toc}{subsection}{\hskip 0.5in 
Dhiraj Kumar Hazra}

\input{George-From-DhirajHazra.tex}

\newpage 

\section*{Paula Keeney}
\addcontentsline{toc}{subsection}{\hskip 0.5in 
Paula Keeney}

\input{paulaKeeney.tex}

\newpage 

\section*{Albert Wai Kit Lau}
\addcontentsline{toc}{subsection}{\hskip 0.5in 
Albert Wai Kit Lau}

\input{albert_wai_kit_lau.tex}

\newpage 

\section*{Sean (Siyang) Li}
\addcontentsline{toc}{subsection}{\hskip 0.5in 
Sean (Siyang) Li} 

\input{seanLi.tex}

\clearpage 
\newpage 

\section*{%
Eric Linder}

\addcontentsline{toc}{subsection}{\hskip 0.5in 
Eric Linder}

\input{%
linder.tex}

\newpage 

\section*{%
Charley Lineweaver}

\addcontentsline{toc}{subsection}{\hskip 0.5in 
Charley Lineweaver}

\input{%
lineweaverV3.tex}

\newpage 

\section*{Tao Liu}
\addcontentsline{toc}{subsection}{\hskip 0.5in 
Tao Liu}

\input{tao_liu.tex}

\newpage 

\section*{Omar López-Cruz}
\addcontentsline{toc}{subsection}{\hskip 0.5in
Omar López-Cruz}

\input{omar.tex}

\newpage 

\section*{Philip Lubin}
\addcontentsline{toc}{subsection}{\hskip 0.5in 
Philip Lubin}

\input{lubin3.tex}

\newpage 

\section*{Kam-Biu Luk}
\addcontentsline{toc}{subsection}{\hskip 0.5in 
Kam-Biu Luk}

\input{luk.tex}

\newpage 

\section*{John Mather}
\addcontentsline{toc}{subsection}{\hskip 0.5in 
John Mather}

\input{mather.tex}

\newpage 

\section*{%
Saul Perlmutter}
\addcontentsline{toc}{subsection}{\hskip 0.5in 
Saul Perlmutter}

\input{perlmutter.tex}

\newpage 

\section*{Federico Piazza}
\addcontentsline{toc}{subsection}{\hskip 0.5in 
Federico Piazza}

\input{federico_piazza_smoot_memory.tex}

\newpage 

\section*{Barbara Rittner}
\addcontentsline{toc}{subsection}{\hskip 0.5in
Barbara Rittner}

\input{rittner.tex}

\newpage 

\section*{Uro\v s Seljak}
\addcontentsline{toc}{subsection}{\hskip 0.5in 
Uro\v s Seljak}

\input{seljak.tex}

\newpage 

\section*{Mehdi Shafiee}
\addcontentsline{toc}{subsection}{\hskip 0.5in 
Mehdi Shafiee}

\input{mehdi.tex}

\newpage 

\section*{Tony Spadafora}
\addcontentsline{toc}{subsection}{\hskip 0.5in 
Tony Spadafora}

\input{spadafora.tex}

\newpage 

\section*{Graciela Tiburcio Pintos}
\addcontentsline{toc}{subsection}{\hskip 0.5in 
Graciela Tiburcio Pintos}

\input{pintos.tex}

\newpage 

\section*{Henry Tye}
\addcontentsline{toc}{subsection}{\hskip 0.5in 
Henry Tye}

\input{tye.tex}

\newpage 

\section*{Ann Whetstone\footnote{See also the related testimonial of Paula Keeney in this booklet.}}
\addcontentsline{toc}{subsection}{\hskip 0.5in 
Ann Whetstone}

\input{annWhetstone.tex}

\newpage 

\section*{Alex Zhavoronkov}
\addcontentsline{toc}{subsection}{\hskip 0.5in 
Alex Zhavoronkov}

\input{alex.tex}

\newpage

\section*{%
Epilogue: Nóra Csiszár}
\addcontentsline{toc}{subsection}{\hskip 0.5in 
Epilogue: Nóra Csiszár}
\input{epilogue.tex}

\newpage

\thispagestyle{empty}

\phantom{Raul} 

\newpage 

\newgeometry{
  left=0mm, width=210mm,right=0mm, 
height=297mm, top=0mm, bottom=0mm
}


\hspace*{-0.27in}
\includegraphics[height=300mm,keepaspectratio]
{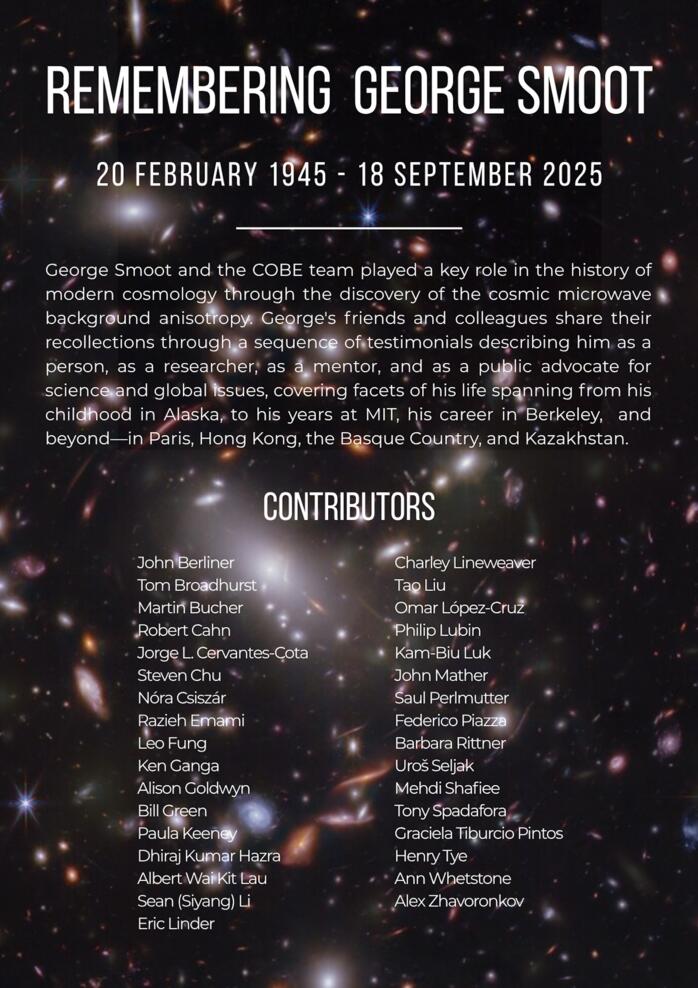}

\end{document}

%% file: preface.tex
\begin{quote}
We gather here thirty-three remembrances of George Smoot from his collaborators, fellow 
scientists, colleagues, and close friends, which have not previously been published along 
with a number of newspaper articles and official announcements reflecting on his life.
To many George Smoot is mainly known for his role in the NASA COBE mission, in which he 
together with his collaborators first discovered the anisotropy of the cosmic microwave 
background radiation, which may be described as the remnant of the Big Bang. Cosmology in 
the course of the 20th century has undergone a major revolution. What once was a domain of 
almost boundless speculation---or one might say,
the realm of theorists and philosophers---has transformed into 
a precise science with stringent observational constraints. It is no longer so easy to 
concoct a new theory of the origins of the universe that is consistent with what has now 
been established through observation.
The story of the rise of modern cosmology largely remains to be written, and it is our hope 
that these recollections may serve as raw material for future historians of science.
We would like to thank all the contributors for their thoughtful essays. We would also like to 
thank the New York Times and Le Monde for permission to reprint their obituaries of George 
Smoot. We thank Barbara Rittner and Doris Rojas for their help in proofreading the 
manuscript. 

\vskip 1cm

\noindent
N\'ora Csisz\'ar \hfill \break
Martin Bucher \hfill \break 
June 2026 

\vfill 

\end{quote}

%% file: apc_statement.tex
%

\noindent 
{\large 
In Memory -- George Fitzgerald Smoot, III
(September 25, 2025)
}

\begin{figure*}[h]
\begin{center}
\includegraphics[width=0.25\textwidth ]{%
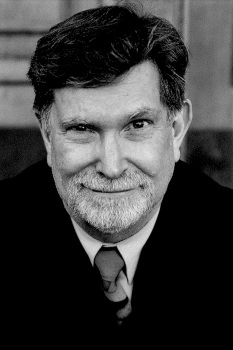}
\end{center}
\caption{[Credit: Peter Badge]}
\end{figure*}


The APC Laboratory mourns the loss of George Smoot, who passed away unexpectedly at his home in Paris. George Smoot was 
one of the pioneers of observations of the cosmic microwave background (CMB), which revolutionized our understanding of 
the cosmos and placed cosmology on a firm experimental footing.

Born in Yukon, Florida in 1945, Smoot was passionate about science and technology. As an undergraduate, he studied 
mathematics and physics at MIT, graduating with a dual major. He stayed on at MIT for a PhD in experimental elementary 
particle physics under Henry Frisch. His thesis was titled ``Charge exchange of $K^+$ on platinum at three GeV/c.'' After 
MIT, Smoot went to UC Berkeley as a postdoc in the group of Luis Alvarez. Alvarez, a participant in the Manhattan 
Project and 1968 Nobel Physics Laureate, had become interested in experiments relating to the cosmos. At Berkeley, Smoot 
initially worked on the HAPPE experiment, using balloons to study high-energy cosmic rays, partially escaping the shield 
of the Earth’s atmosphere. This work included searching for an antimatter cosmic ray component. In 1973 Smoot changed 
his focus to studying the CMB, working on a number of COBE precursors. Most notably, early work by Smoot and 
collaborators led to the definitive measurement of the CMB dipole in 1977 using an instrument aboard a U-2 spy plane. 
This work confirmed earlier measurements of the dipole at lower statistical significance. As part of the NASA Cosmic 
Background Explorer (COBE) team, Smoot led the Differential Microwave Radiometer (DMR) instrument, which announced the 
discovery of the CMB anisotropies in 1992. As a result of this work he was awarded numerous prizes and honors, 
including, together with John Mather, the 2006 Nobel Prize in Physics.

The 1992 DMR detection of the CMB anisotropies was a milestone, establishing the amplitude of the primordial 
cosmological perturbations, which ultimately grew into the large scale structure of the Universe we see today. The 
discovery of the CMB anisotropies started what might be described as the Golden Age of Cosmology. Before these 
measurements, cosmology had a reputation as a field where theoretical speculation abounded, little constrained by 
observations. Today cosmology is at the center stage of physics, due in no small part to this and subsequent 
measurements of the CMB. This discovery prompted many researchers to switch to cosmology. A host of experiments, from 
the ground, from stratospheric balloons, and from space have now followed the pioneering COBE measurements.

\begin{figure*}[h]
\begin{center}
\includegraphics[width=0.99\textwidth ]{%
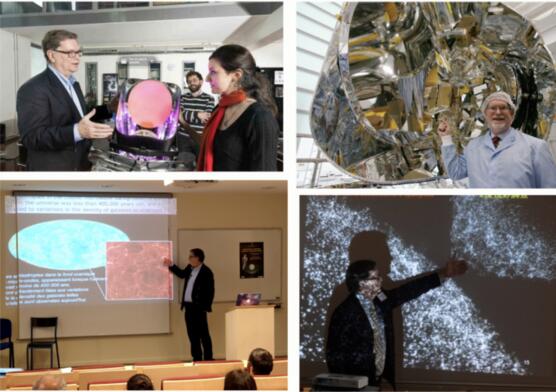}
\end{center}
\caption{\textbf{George at Laboratoire APC.}
[Credit: Courtesy of APC]}
\end{figure*}

In 2009, George Smoot joined Université Paris Cité (then Université Paris-Diderot VII)  as Professor of Physics. He was 
affiliated with the Laboratoire Astroparticule et Cosmologie (APC). Smoot played an instrumental role in the founding of 
the Paris Center for Cosmological Physics and the opening of the endowment fund ``Physics of the Universe,'' attracting 
top postdocs, establishing a conference center, and forging ties with broader society. He was deeply committed to the 
idea that cosmology research should reach beyond the walls of the laboratory to promote the public understanding of 
science and connect science to major societal issues. For this reason, he took a keen interest in educating the younger 
generation. He founded the ``Teaching the Universe'' program for secondary school teachers, and the MOOC ``Gravity!'' with 
Pierre Binétruy and George Smoot was an international success.  While at APC, Smoot initiated a programme to develop 
optical KIDs (Kinetic Inductance Detectors) and established new international partnerships. He mentored a large number of postdocs.  He was also the 
president of the Scientific Council of the LabEx UnivEarthS bringing together geosciences and particle astrophysics.

We will remember him as a larger than life character, with a broad range of interests beyond the discoveries for which 
he is best known. He travelled the world, took a keen interest in societal issues such as climate change, was proud to 
have appeared in cameos on the celebrated sitcom ``The Big Bang Theory,'' and was one of only two people to win the top 
prize on the game show ``Are You Smarter than a Fifth Grader?''

Our deep condolences are with his friends, colleagues, family and his partner Nóra Csiszár.


%% file: witherell.tex
\begin{figure*}[h]
\begin{center}
\includegraphics[width=0.75\textwidth ]{%
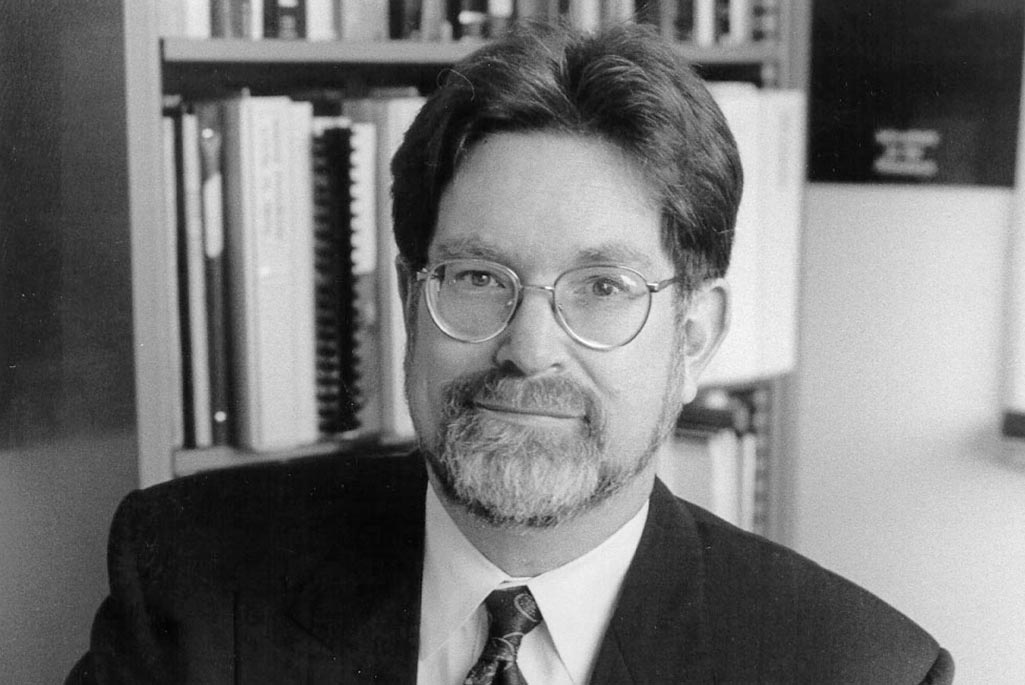}
\end{center}
\caption{[Photo credit: J. Bauer]}
\end{figure*}

{\bf 
\noindent 
Statement by Lawrence Berkeley National Laboratory Director Mike Witherell:\\
\rm (September 26, 2025)
}

\begin{quote}
I am saddened to share that George F. Smoot III, a pioneering astrophysicist and Nobel Laureate in Physics, has passed away at the age of 80. George joined UC Berkeley in 1971 and Berkeley Lab in 1974, where he spent a distinguished career uncovering the secrets of the universe.

At Berkeley Lab, George was the leader of a research team that produced detailed maps of the infant universe. They revealed a pattern of minuscule temperature variations in the cosmic microwave background (CMB), relic light from billions of years ago. Those early tiny fluctuations evolved into the galaxies we observe today. George was awarded the 2006 Nobel Prize in Physics, along with John Mather of the NASA Goddard Space Flight Center, for “their discovery of the blackbody form and anisotropy of the cosmic microwave background radiation.”

After graduating from MIT with degrees in math and physics, George turned his attention to cosmology – then considered a fringe field. He used an instrument called a Differential Microwave Radiometer (DMR), flown on a U-2 airplane, to find evidence of the Doppler effect as our galaxy moves through the universe relative to the CMB. He went on to lead the team that developed the DMRs that flew on NASA’s Cosmic Background Explorer (COBE) satellite and made the Nobel Prize-winning “baby picture” of the universe. The findings provided strong support for the Big Bang theory and established cosmology as a truly quantitative science.

George’s contributions to cosmology were recognized with many other awards, including NASA’s Exceptional Scientific Achievement Medal (1992), the Department of Energy’s E.O. Lawrence Award (1994), the Albert Einstein Medal (2003), and the Gruber Prize in Cosmology (2006). He was elected a member of the U.S. National Academy of Sciences, a fellow of the American Physical Society, and a fellow of the National Academy of Inventors for his development of instrumentation to study the CMB.

In addition to his groundbreaking research, George was committed to engaging with the public and helping the next generation of scientists. He wrote a popular book about cosmology and the CMB experiments entitled Wrinkles in Time. In 2007, he used a portion of his Nobel award to launch the Berkeley Center for Cosmological Physics at UC Berkeley, which was established as a premier cosmological research center to support postdoctoral students, graduate students, and faculty.

After retiring from Berkeley Lab in 2014, Smoot continued to engage actively in scientific research and education. In 2009, he joined Université Paris-Cité as an affiliate of the Laboratoire Astroparticule et Cosmologie, and as a professor of physics, he mentored postdocs and played an instrumental role in the founding of the Paris Center for Cosmological Physics. He also launched the long-running Physics In and Through Cosmology Workshop, an annual outreach event hosted by the Berkeley Lab Physics Division for Bay Area high school physics teachers and students, now in its nineteenth year.

Please join me in honoring the legacy of George Smoot.
\end{quote}



%% file: leMonde.tex
{

\fontencoding{T1}\selectfont

\addtolength{\baselineskip}{6pt}


{\Large \bf 
George Smoot, explorateur de la soupe cosmique primordiale et Prix Nobel de physique, est mort}
\normalsize Par Pierre Barthélémy (25 septembre 2025)\footnote{Reproduit avec l'aimable autorisation 
de M. Pierre Barthélémy et du Monde.}

\vskip 10pt 

{\large \it \noindent 
L’astrophysicien américain avait reçu la plus haute récompense de sa discipline en 2006 avec son compatriote John Mather. Il est décédé à Paris le 18 septembre, à 80 ans.
}

\begin{figure*}[h]
\begin{center}
\includegraphics[width=0.7\textwidth ]{%
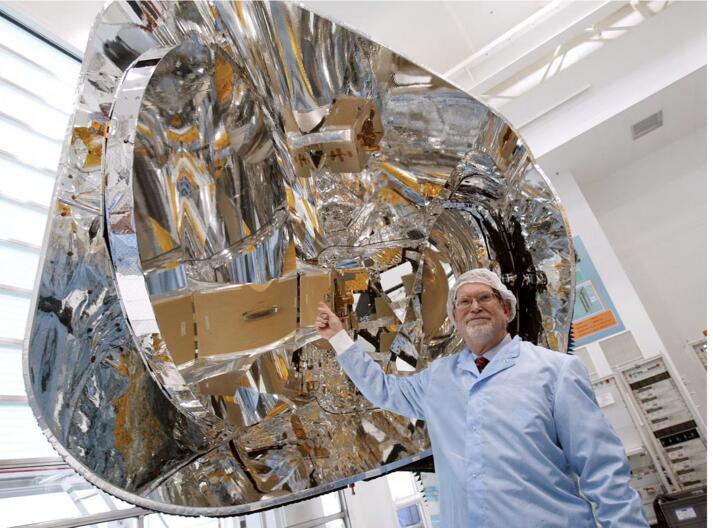}
\end{center}
\caption{%
George Smoot, devant le miroir du satellite Planck, dans les ateliers d’Alcatel Alenia Space, à Cannes (Alpes-Maritimes), 
en février 2007. \it ERIC ESTRADE/AFP}
\end{figure*}
\vskip 10pt

Il avait vu le jour le 20 février 1945 à Yukon (Etats-Unis), aujourd’hui une ville fantôme, et il a quitté ce monde huit 
décennies plus tard loin de sa Floride natale, dans une ville bien vivante, Paris, qu’il avait choisie il y a quinze ans 
pour la dernière partie de sa carrière. Le 18 septembre s’est éteint l’astrophysicien George Smoot, Prix Nobel de physique 
en 2006 pour avoir mis en évidence que, dans la soupe cosmique apparue après le Big Bang, il y avait des « grumeaux », 
essentiels pour expliquer l’Univers tel que nous le connaissons.

Quand, après avoir reçu cette récompense, George Smoot s’était retourné sur son passé, il se rappelait ce long voyage vers 
l’Alabama pendant son enfance, au cours duquel il s’était étonné que la Lune « suive » la voiture familiale. Avec 
pédagogie, ses parents – père hydrologue, mère enseignante – lui avaient expliqué que, la Lune étant loin et grosse, 
l’angle selon laquelle on la voyait ne changeait pas malgré les kilomètres parcourus, contrairement au reste du paysage. « 
Ce fut, disait-il, une révélation surprenante que le monde puisse être compris par une simple évaluation rationnelle. »

Le travail de son père emmène la famille, assez pauvre, en Alaska. Il y a des élans dans la cour de récréation, mais pas de 
télévision. Il regarde le ciel et étudie, explore la nature. Plus tard, quand les Smoot s’installent dans l’Ohio, le monde 
s’éveille à l’espace avec le lancement par l’Union soviétique, en 1957, du premier satellite artificiel, Spoutnik, et le 
bip-bip qu’il émet. L’adolescent fabrique des postes de radio. Ses parents renforcent son éducation scolaire, sa mère en 
lui donnant des cours supplémentaires d’histoire et de sciences, son père en l’obligeant à bûcher la trigonométrie et le 
calcul différentiel. Le jeune homme travaille pour payer ses études au prestigieux Massachusetts Institute of Technology 
(MIT), où il passe un doctorat en physique des particules en 1970.

L’aventure scientifique commence pour lui à Berkeley, en Californie, où, avec le Nobel Luis Walter Alvarez (1911-1988), il 
tente de détecter de l’antimatière avec des ballons-sondes. En vain. George Smoot se tourne alors, en 1973, vers l’étude du 
fond diffus cosmologique (FDC).

Découvert par hasard en 1964, le FDC, plus connu sous le nom de « rayonnement fossile », est, selon la théorie, le vestige 
du moment où l’Univers s’est allumé, 380 000 ans après le Big Bang. Mais cette soupe cosmique très chaude de particules 
s’est depuis fort refroidie, à moins de 3 degrés au-dessus du zéro absolu. Elle semble parfaitement homogène, et c’est bien 
embêtant pour les cosmologistes : pour expliquer la formation des galaxies, il faudrait y trouver des « grumeaux » – des 
fluctuations de température et de densité à partir desquelles la matière se serait ensuite accumulée sous l’effet de la 
gravité. 

\noindent
\textbf{Apparitions dans « The Big Bang Theory »}

Au milieu des années 1970, Smoot va donc, pour traquer ces variations, concevoir un des instruments du satellite COBE 
(Cosmic Background Explorer), qui, financé par la NASA, décolle en 1989. COBE sera une révolution. Très vite, les résultats 
montrent que le rayonnement fossile est en parfaite adéquation avec le Big Bang, ce pourquoi John Mather sera co-lauréat de 
ce Nobel 2006.

George Smoot met plus de temps à livrer ses conclusions – il faut dire que son instrument cherche des fluctuations infimes, 
de l’ordre de 30 millionièmes de degré – mais, le 23 avril 1992, l’Américain annonce, dans une conférence de presse devenue 
célèbre, avoir détecté ces « grumeaux » qui sont à la source des grandes structures de l’Univers. « Si vous êtes croyant, 
c’est comme voir Dieu », commente-t-il – phrase qui lui sera beaucoup reprochée.

Le prix Nobel arrive en 2006, et sa vie change. George Smoot voit désormais sa mission comme celle d’un « mentor ». C’est 
dans cette optique qu’en 2010 il atterrit en France, à l’université Paris-Diderot (aujourd’hui université Paris Cité), au 
laboratoire Astroparticules et cosmologie. Il s’investit dans le Paris Centre for Cosmological Physics, devenu Cosmos, 
sciences et sociétés (qu’il présidait encore au moment de son décès).

La transmission de la science est devenue son credo, auprès des étudiants, mais aussi du grand public, et il fait notamment 
deux apparitions dans la série The Big Bang Theory. Un goût du spectacle qui ne plaisait pas à tous, et George Smoot en 
avait conscience, déclarant au Guardian en 2014 : « De nombreux scientifiques pensent que vulgariser la science revient en 
quelque sorte à la dégrader, mais je pense que c’est absolument crucial.»

}

%% file: berkeley_announcement.tex
{\large \textbf{UC Berkeley News} (September 29, 2025)

\vskip 5pt

\noindent 
\textbf{Nobelist George Smoot, whose satellite experiments validated the Big Bang theory, dies at 80}}\\
By Robert Sanders

\vskip 7pt

\noindent 
{\large \it 
Smoot, a physicist at UC Berkeley and Berkeley Lab, shared the 2006 Nobel Prize in Physics for detecting 
minute temperature variations in the cosmic microwave background, a prediction of the Big Bang theory.
}
 
\begin{figure*}[h] 
\begin{center}
\includegraphics[width=0.99\textwidth ]{%
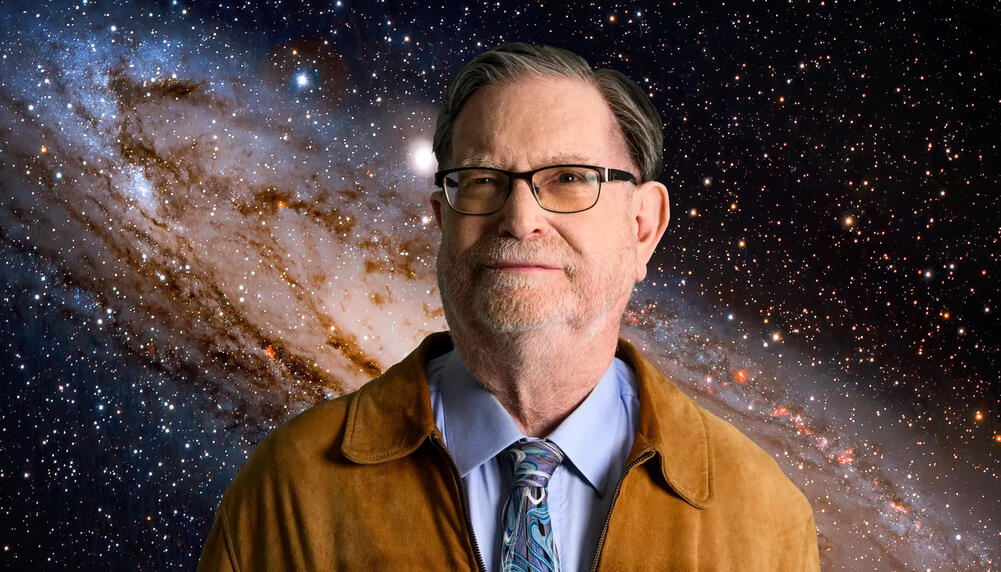}
\end{center}
\caption{George F. Smoot III, a UC Berkeley professor emeritus of physics, died in Paris on Sept. 18, 2025, at the 
age of 80. [Photo credit: Jacopo Landi]}
\end{figure*}

Physicist George Smoot told a packed press conference in 1992, ``If you’re religious, it’s like seeing God.''

He was referring to the cosmic microwave background, which he and colleague John Mather imaged with NASA’s 
Cosmic Background Explorer (COBE), marking the first detection of the minute temperature fluctuations in the 
radiation surrounding us. That detection was a confirmation of the Big Bang theory — the idea that the 
universe was born in a rapid cosmic expansion nearly 14 billion years ago — and earned him and Mather the 
2006 Nobel Prize in Physics.

Smoot, a professor emeritus of physics at the University of California, Berkeley, and an emeritus faculty 
senior scientist at Lawrence Berkeley National Laboratory, died on Sept. 18 in Paris of a heart attack. He 
was 80 and, since 2009, had been a physics professor at the Université Paris-Cité and an affiliate of the 
Laboratoire Astroparticule et Cosmologie (APC).

Detection of the CMB was a triumph of precision cosmology — the detailed measurement of the temperature of 
the universe that continues to reveal new details about the universe’s infancy and its evolution since.

According to an obituary posted by APC on Sept. 24, ``Today cosmology is at the center stage of physics, due 
in no small part to this [Smoot’s] and subsequent measurements of the CMB. This discovery prompted many 
researchers to switch to cosmology. A host of experiments, from the ground, from stratospheric balloons and 
from space have now followed the pioneering COBE measurements.'' 


\begin{figure*}[h] 
\begin{center}
\includegraphics[width=0.99\textwidth ]{%
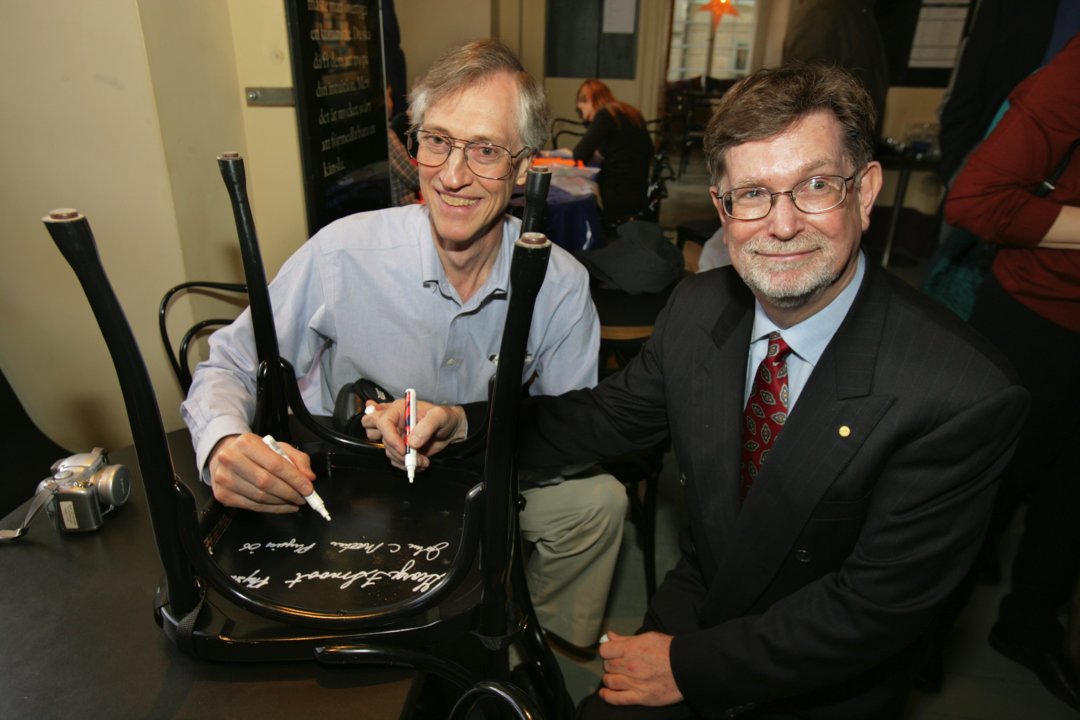}
\end{center}
\caption{%
George F. Smoot (right) and John C. Mather (left), like many Nobel Laureates before them, 
autographed a chair at Kafé Satir at the Nobel Museum in Stockholm on Dec. 6, 2006.
[Photo credit: Fredrik Persson
\copyright 2006 The Nobel Museum]
}%
\end{figure*}

Smoot and Mather, who earned his Ph.D. in physics from UC Berkeley in 1974, together led the building and 
launch of the COBE satellite in 1989 in a highly competitive race to detect the signature of the primordial 
explosion that birthed the universe. According to the reigning theory of the origin of the universe, the Big 
Bang fireball 13.7 billion years ago filled the universe with heat that has since cooled to a mere 2.7 
degrees above absolute zero. But the theory also predicted that the temperature should vary across the sky, 
though previous experiments had failed to detect any variation down to 1 part in 1,000

Mather’s instrument confirmed that the microwave background radiation matched perfectly the spectrum of 
colors that astronomers predicted if the universe formed in a fireball.

Smoot’s instruments went further, detecting fluctuations equivalent to 1 part in 100,\,000 in the 2.7 degree 
Kelvin microwave glow. These slight variations in temperature and density of the early universe grew over 
billions of years into the galaxies and clusters of galaxies we see today.

When Smoot announced the detection in 1992, the late Stephen Hawking called it the greatest scientific 
discovery of the century. 


\begin{figure*}[h]
\begin{center}
\includegraphics[width=0.99\textwidth ]{%
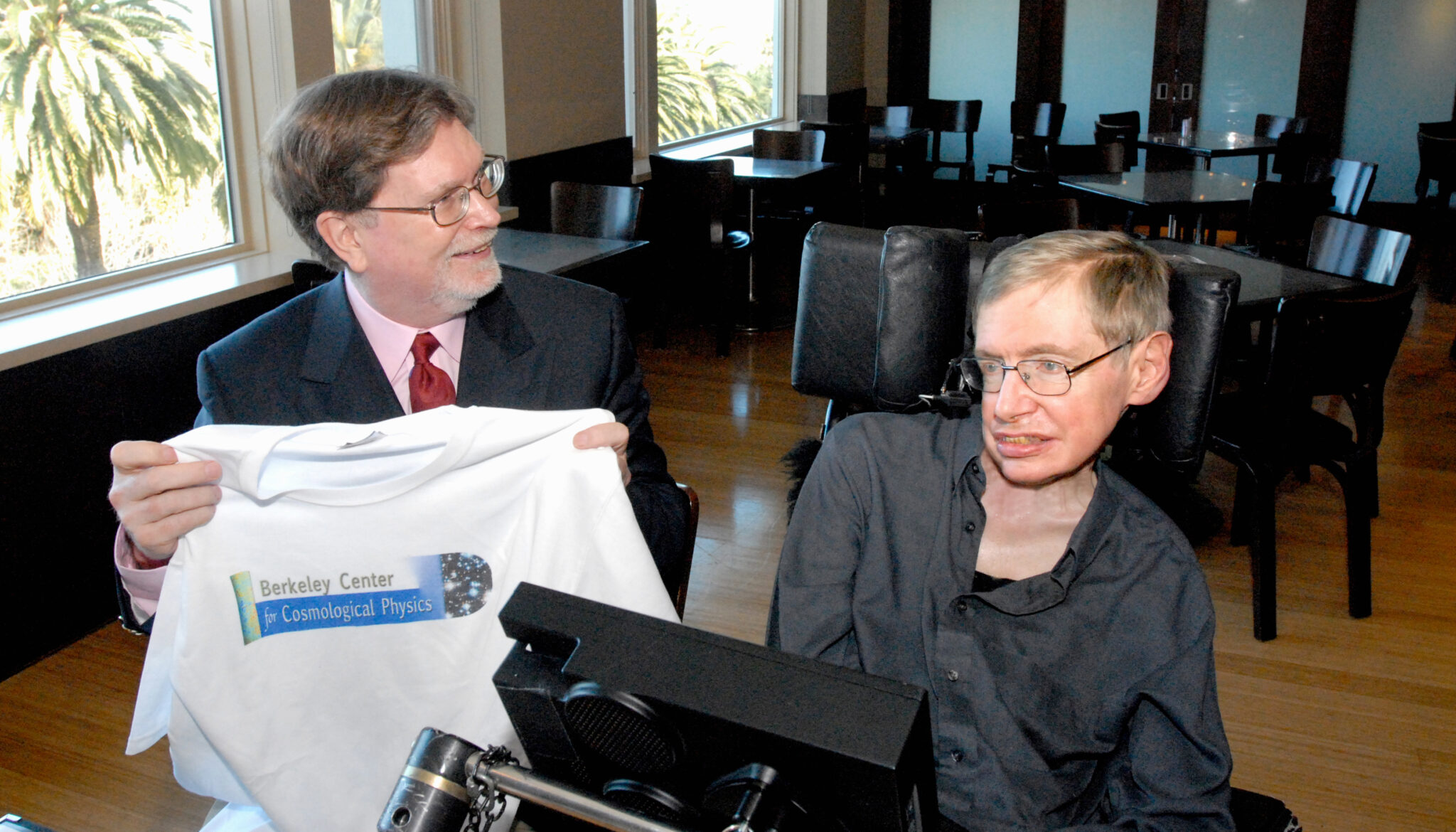}
\end{center}
\caption{%
Hawking called Smoot’s detection of the fluctuations in the 
microwave background the greatest scientific discovery of the 20th century. Smoot offered Hawking a T-shirt 
advertising the Berkeley Center for Cosmological Physics, which he directed.
[Photo credit: Peg Skorpinski for UC Berkeley] 
}%
\end{figure*}

``Those measurements really confirmed our picture of the Big Bang,'' Smoot said at the time. ``By studying the 
fluctuations in the microwave background, we found a tool that allowed us to explore the early universe, to 
see how it evolved and what it’s made of.''

Smoot, who had been at Berkeley Lab since 1971, was appointed to the UC Berkeley physics faculty in 1994. 
The Nobel Prize committee cited Smoot and Mather for ''the discovery of the blackbody form and anisotropy of 
the cosmic microwave background radiation.''

In 2007, Smoot used \$500,\,000 of his Nobel Prize winnings to help endow the Berkeley Center for Cosmological 
Physics.

As Smoot acknowledged on his website, the Nobel Prize ``brought a new dimension to his life.'' In addition to 
research and teaching, he began traveling the world as a speaker and commentator on science-related issues. 
He also appeared in cameos on the celebrated sitcom ``The Big Bang Theory,'' and, in 2008, was one of only two 
people to win the \$1 million prize on the game show ``Are you smarter than a fifth grader?''

``He was somebody who always enjoyed looking ahead to what was happening in the world and what was coming 
next,'' said colleague Saul Perlmutter, a UC Berkeley professor of physics and himself a Nobel Prize winner for 
discovering dark energy. ``He would come to your office and explain to you why you had to change something 
you were doing because of what was about to happen. Almost always, I had that feeling that he was right, 
but it was very hard to operationalize any of the advice!''

The APC obituary called Smoot a ``larger than life character'' who ``revolutionized our understanding of the 
cosmos and placed cosmology on a firm experimental footing.''

Most recently, he had focused on applying basic physics discoveries to improve peoples’ lives, particularly 
in the fields of air quality and medicine. According to his website, Smoot ``$\ldots $sees his role to inspire and 
encourage smart young people, to set them off on the path to discovery.''

“He did really enjoy the question of how do you reach a public and how do you spread the excitement about 
science,” Perlmutter said. 

George Fitzgerald Smoot III was born in Yukon, Florida, on Feb. 20, 1945. He graduated from high school in 
Ohio, but spent part of his childhood in Alaska with his father, a hydrologist for the U.S. Geological 
Survey. His mother was a science teacher and school principal. On Smoot’s website, he wrote that in Alaska 
he ``discovered a new way of life more directly linked with nature and the juxtaposition to modern technology 
and understanding of mankind.'' 


\begin{figure*}[h]
\begin{center}
\includegraphics[width=0.99\textwidth ]{%
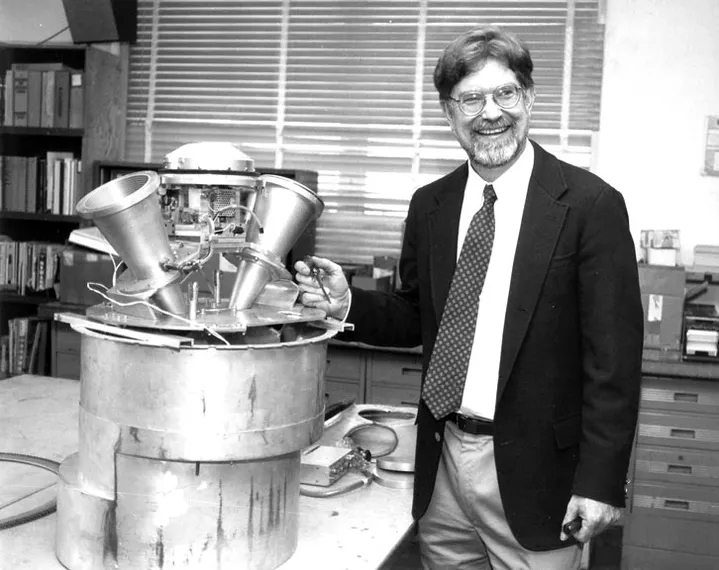}
\end{center}
\caption{%
Smoot with a model of the Differential Microwave Radiometer (DMR) that flew aboard the COBE satellite in 1989.
[Photo Credit: Lawrence Berkeley National Laboratory. 
Courtesy of AIP Emilio Segrè Visual Archives]
}%
\end{figure*}

He graduated from MIT with a dual major in math and physics and completed his PhD there in 1970 in the field 
of experimental elementary particle physics. He then moved to Berkeley Lab as a postdoc in the group of Luis 
Alvarez, a Nobel Prize winner and UC Berkeley professor of physics. Alvarez had become interested in 
experiments relating to cosmology, and Smoot followed in his footsteps, working on the High Altitude Particle 
Physics Experiment (HAPPE), a stratospheric weather balloon designed to detect antimatter in cosmic rays. 
They were unsuccessful.

In 1973, Smoot changed his focus to studying the CMB, which had only been detected in 1964. He worked with 
Richard Muller, another member of the Alvarez group and now a UC Berkeley professor emeritus of physics, to 
develop microwave detectors that they flew on balloons to reach heights above much of the atmosphere. Their 
team eventually developed an instrument, called a Differential Microwave Radiometer (DMR), to detect 
differences in the CMB temperature in spots 60 degrees apart on the sky, which they flew aboard a U-2 spy 
plane. It worked, finding an imbalance in the temperature of the sky that implied our galaxy is traveling 
about 1 million miles per hour through the universe.

At the time, Smoot said, cosmology was a fringe field of study. ``Back then, you could get all of us in the 
field into a single room. I remember the teasing from my particle physics colleagues that real physics is 
done at accelerators. Today, opinions have changed. We have begun to explore the early universe, the 
original accelerator. The fields of particle physics and cosmology have been joined.''

Particle physicists, who are used to measuring miniscule signals from the interactions of elementary 
particles, helped ramp up the precision of cosmological measurements, Perlmutter noted.

``George came from particle physics and along with several others brought a can-do style of experimental 
design to bear on cosmological questions, where they were working at more and more extreme conditions of 
size and energy,” Perlmutter said. ``The drive for greater and greater precision to actually see the 
anisotropies in the cosmic microwave background, in some ways, led the way in precision cosmology.''

In 1974, he submitted a proposal to NASA to build a satellite to get the DMR instrument above the atmosphere 
to search for, measure and map even smaller temperature fluctuations, joining many competitors. His proposal 
was combined with two others and he began collaborating with Mather at NASA’s Goddard Space Flight Center, 
who served as project director. Fifteen years later, the COBE satellite was launched. Smoot’s team at 
Berkeley Lab involved more than 40 people, while the COBE satellite project included an estimated 1000 
individuals. 

After Smoot announced the discovery of the CMB anisotropies in 1992, he continued to work on experiments to 
refine the measurements, including as a collaborator on a third generation CMB anisotropy observatory, the 
Planck satellite. These experiments have refined maps of the CMB to the point that they now mark the first 
notch in a ``cosmic ruler'' used to measure the history of an expanding universe driven by dark energy. 


\begin{figure*}[h]
\begin{center}
\includegraphics[width=0.99\textwidth ]{%
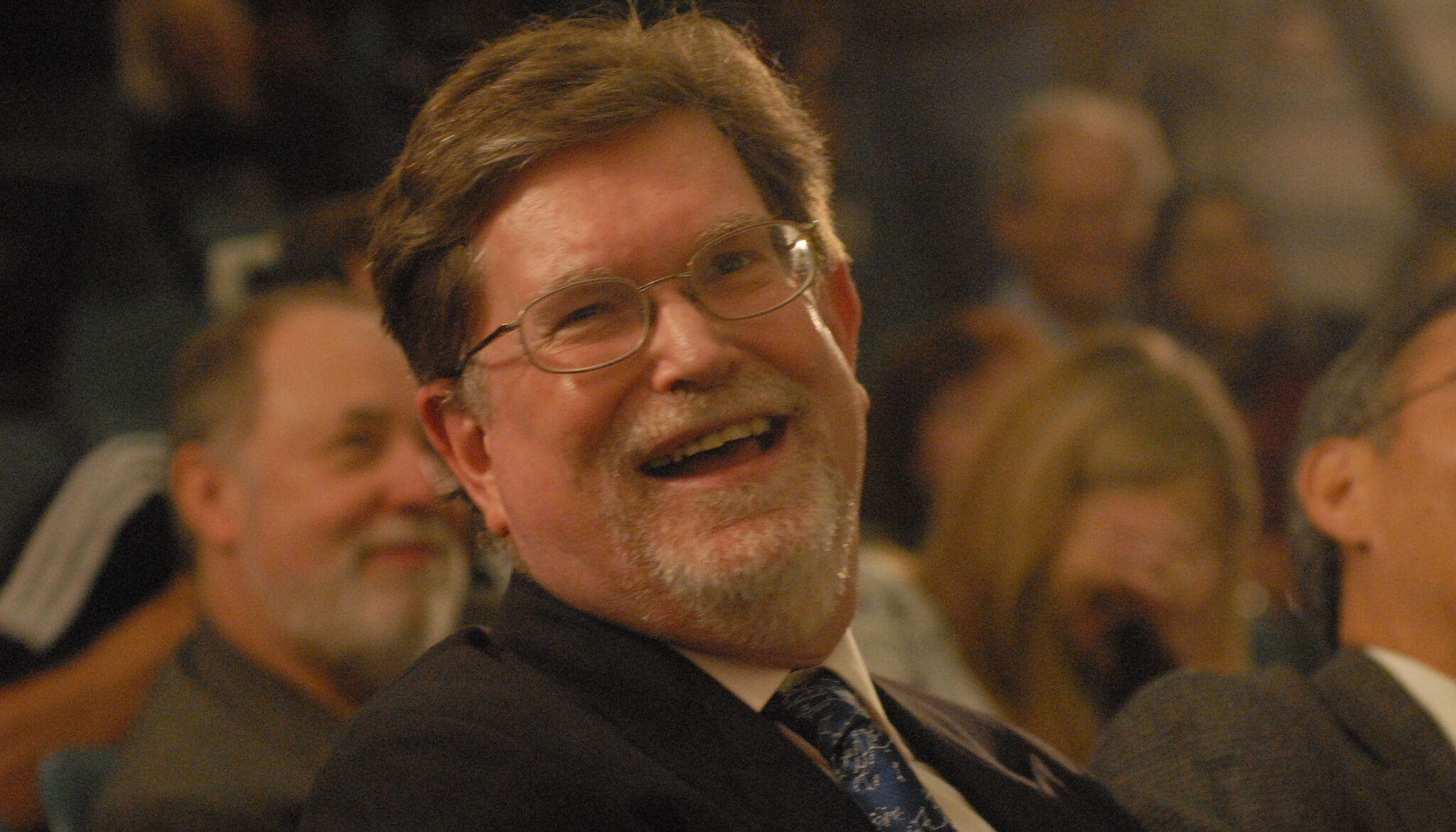}
\end{center}
\caption{%
George Smoot at the press conference following his selection as co-winner of the 2006 Nobel Prize in Physics.
[Photo credit: Peg Skorpinski for UC Berkeley]
}%
\end{figure*}

He also helped found various research institutes around the world, including in South Korea, Spain and 
France. At APC in Paris, Smoot played an instrumental role in the founding of the Paris Center for 
Cosmological Physics and the opening of the endowment fund “Physics of the Universe” to attract top 
postdocs, according to the center’s website.

He took a keen interest in educating the younger generation, founding the ``Teaching the Universe'' program 
for secondary school teachers. He also created an internationally popular MOOC (Massive Open Online Course) 
called ``Gravity!'' with Pierre Binétruy.

Smoot also collaborated with journalist Keay Davidson to write a general-audience book, ``Wrinkles in Time'' 
(1994), about the COBE team’s work.

Among his awards were the 2006 Gruber Prize, given jointly with Mather, the 2003 Einstein Medal of 
Switzerland’s Albert Einstein Society, the 1995 Lawrence Award from the U.S. Department of Energy, the 1993 
Kilby Award and the 1991 NASA Medal for Exceptional Scientific Achievement. He was a member of the National 
Academy of Sciences.

Smoot leaves behind a sister, Sharon Smoot Bowie, of New London, New Hampshire, two nieces, and his partner, 
Nóra Csiszár of Paris.

%% file: nyt.tex
\noindent
\textbf{New York Times, Oct. 20, 2025}

\vskip 10pt 

\noindent 
{\Large \bf
George F. Smoot, Who Showed How the Cosmos Began, Is Dead \mbox{at 80}}
\rm \\(By Katrina Miller)\footnote{Reproduced with permission from
The New York Times, Oct. 20, 2025
\copyright 2025 The New
York Times. All rights reserved. Used by permission and protected by the
Copyright Laws of the United States. The printing, copying,
redistribution, or retransmission of this Content without express written
permission is prohibited.}

\vskip 6pt 

{\it \Large \noindent 
He led a team of scientists who helped confirm that a Big Bang was the source of the universe. The discovery earned 
him a Nobel Prize.
}

\begin{center}
\includegraphics[width=0.95\textwidth ]%
{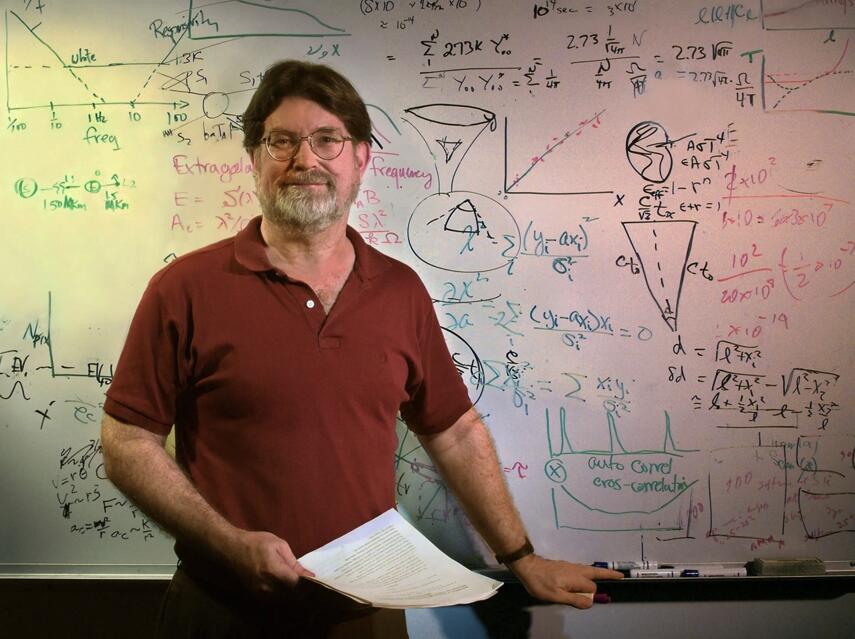}
\leftline{[Credit: Roy Kaltschmidt, LBL, Berkeley]}
\end{center}

\noindent
George F. Smoot, an American physicist and Nobel laureate who helped elucidate the story of cosmic creation, 
providing evidence of what he called the primordial seeds that grew into galaxies and galaxy clusters, died on Sept. 
18 at his home in Paris. He was 80.

His death, from cardiac arrest, was confirmed by his sister, Sharon Bowie.

Dr. Smoot was a research scientist at Lawrence Berkeley National Laboratory and the Space Sciences Laboratory at the 
University of California, Berkeley, when he led a team that constructed a picture of the infant universe using an 
instrument he developed in the 1970s.

The instrument was launched into space in 1989 aboard a NASA satellite, the Cosmic Background Explorer, or COBE, 
from which it detected tiny variations in the temperature of the light that was left over from what most scientists 
presumed was a Big Bang. The pattern of those temperature variations was a record of how unevenly cosmic matter was 
distributed billions of years ago — the seeds from which the current design of the universe, dense in some parts and 
empty in others, sprouted.

``If you’re religious, it’s like seeing God,'' Dr. Smoot said when he announced the COBE findings in 1992 at an 
American Physical Society conference, making front-page news around the world. (The account in The New York Times — 
its lead story in the paper of April 24 — appeared under the headline ``Scientists Report Profound Insight on How 
Time Began.'')

The theoretical physicist Stephen Hawking called it ``the greatest discovery of the century, if not of all time.''

Dr. Smoot’s research built on that of the physicists Arno A. Penzias and Robert W. Wilson, who in 1964 discovered 
that the universe was bathed in a sea of ancient light, known as the cosmic microwave background. It was evidence 
that the universe had a beginning and had exploded into existence with a Big Bang.

Before that discovery, cosmology had been a theoretical playground, full of imaginative ideas virtually untethered 
to data. Measurements taken with the COBE satellite provided data by which scientists could test those various 
theories about how the universe began, what it was made of and how it had evolved. It helped transform cosmology 
from a field largely based in speculation to a science grounded in precise measurement.

Dr. Smoot shared the 2006 Nobel Prize in Physics with John C. Mather, a cosmologist at the NASA Goddard Space Flight 
Center in Maryland, for groundbreaking discoveries made by the COBE team. Each was a team leader in the project.

\begin{figure*}[t]
\begin{center}
\includegraphics[width=0.5\textwidth ]{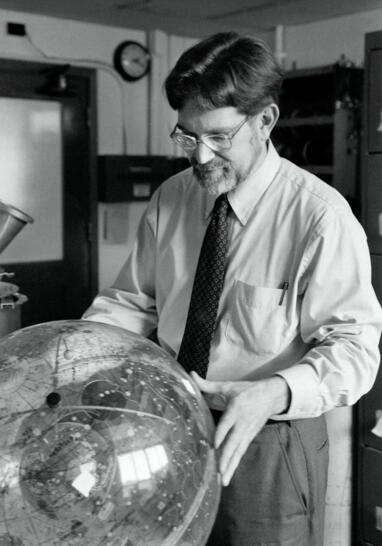}
\end{center}p
\caption{%
Dr. Smoot in 1992, the year he announced findings that helped transform 
cosmology from a field largely based in speculation to a science grounded in precise measurement. [Credit: Susan Spann]}
\end{figure*}

As a scientist, Dr. Smoot was revered for his brilliance but also resented by some colleagues, who felt he took 
undue credit for scientific results that others had achieved. Days before the physics conference announcement in 
1992, Berkeley Lab published a news release that many viewed as unfairly attributing the discovery of light 
temperature variations solely to Dr. Smoot and the lab, overlooking other team leaders and NASA.

In his 1993 book, ``Wrinkles in Time: Witness to the Birth of the Universe,'' written with Keay Davidson, Dr. Smoot 
documented the process leading up to the discovery. Some of his collaborators disagreed with his version of events 
and encouraged Dr. Mather to write his own account. He did, publishing it three years later as ``The Very First 
Light: The True Inside Story of the Scientific Journey Back to the Dawn of the Universe,'' which differed from Dr. 
Smoot’s version in some details.

Dr. Smoot ``was a lot of trouble,'' Dr. Mather allowed in an interview, but also ``ingenious and thoughtful and 
enthusiastic, as everybody knows.''

George Fitzgerald Smoot III was born on Feb. 20, 1945, in Yukon, Fla., to George Smoot II, who served as a fighter 
pilot during World War II, and Talicia (Crawford) Smoot, a science teacher and school principal.

The family moved to Alabama after the war. George’s father was a hydrologist with the United States Geological 
Survey, a job that also took the family to Alaska, Ohio and Virginia. By Dr. Smoot’s account, his mother instilled 
in him a love of science and education; from his father, who traveled the world measuring river flows, he learned to 
appreciate the value of invention and instrumentation.

Dr. Smoot attended the Massachusetts Institute of Technology, where he earned degrees in physics and mathematics in 
1966 and his Ph.D. in particle physics in 1970, before moving to Berkeley to study under Luis Alvarez, a particle 
physicist who had won the Nobel Prize in 1968.

There, Dr. Smoot pivoted to cosmology. With the physicist Richard Muller, he developed an instrument that could 
measure temperature differences in the cosmic microwave background; he put it to use by mounting it on a U-2 spy 
plane operated by NASA. The experiment led to one of the first measurements indicating that our galaxy, the Milky 
Way, is hurtling through space at more than a million miles an hour, suggesting that it is being pulled by the 
gravitational force of an even more gigantic mass.

In 1974, Dr. Smoot approached NASA proposing a mission to send the instrument into space. NASA combined the proposal 
with two others, establishing the team behind COBE, which carried three instruments into orbit.

Weeks after the launch, data analyzed by the team overseen by Dr. Mather, the Goddard cosmologist, solidified the 
link between the cosmic microwave background and the Big Bang. It also measured the temperature of that background: 
a chilly 2.7 kelvin (about minus 455 degrees Fahrenheit), just a smidgen above absolute zero.

Dr. Smoot’s team measured minute variations in that temperature across the universe, at the level of about 10 
millionths of a degree. He would describe this measurement in his book as similar to ``listening for a whisper during 
a noisy beach party while radios blare, waves crash, people yell, dogs bark and dune buggies roar.''

\begin{figure*}
\begin{center}
\includegraphics[width=0.85\textwidth ]{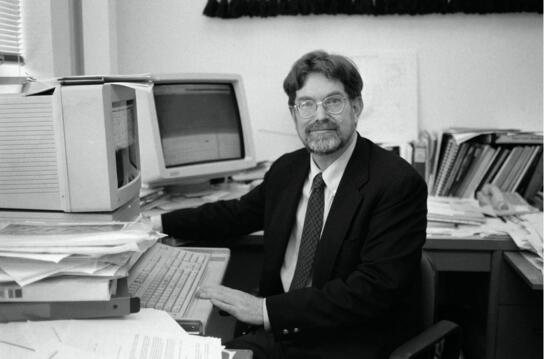}
\end{center}
\caption{%
Dr. Smoot in 1992. ``If you’re religious, it’s like seeing God,'' he said of the discoveries about the origins of the 
universe. 
[Credit: Susan Spann]}
\end{figure*}

The announcement of this discovery inspired countless creation analogies. ``It really is like finding the driving 
mechanism for the universe,'' Dr. Smoot told The Times in 1992. ``And isn’t that what God is?''

In addition to confirming the Big Bang picture of the universe, the COBE discoveries strengthened evidence for the 
existence of dark matter and the theory of cosmic inflation, which posits that the universe went through a period of 
rapid expansion shortly after its birth.

Two more space missions refined COBE’s measurements of the cosmic microwave background; one, the Planck observatory, 
launched by the European Space Agency in 2009, had been proposed by a team that included Dr. Smoot.

Dr. Smoot, who became a professor at Berkeley in 1994, donated a large share of his Nobel Prize award money (he and 
Dr. Mather split a gift of about \$1.37 million) to endow the Berkeley Center for Cosmological Physics, where he was 
the founding director. He helped establish cosmology institutes around the world, including in France and South 
Korea, and was elected to the National Academy of Sciences and the National Academy of Inventors. In 2009, he joined 
the faculty of Paris Cité University and became an affiliate of the Astroparticle and Cosmology Laboratory.

He married Maxine Bixby in 1969; they divorced in 1979. In addition to his sister, Dr. Smoot is survived by his 
partner, Nóra Csiszár.

Toward the end of his career, Dr. Smoot grew more involved in public outreach and science education. He helped start 
programs to teach high school teachers and students about cosmology, and taught an online course about gravity that 
attracted an audience of more than 87,\,000.

He also made appearances on the CBS sitcom ``The Big Bang Theory'' and, in 2009, competed on the Fox network game show 
``Are You Smarter Than a Fifth Grader?'' He went on the show, he said, to encourage his students to try new things; in 
doing so, he became the second person to win \$1 million.

During his appearance, Dr. Smoot noted that he had donated his Nobel winnings to create scholarships and fellowships 
for young scientists. Pointing to the students on the stage as the crowd cheered, he said, ``I’m hoping one of these 
guys gets one of them.''

\noindent 
\textit{Katrina Miller is a science reporter for The Times based in Chicago. She earned a Ph.D. in physics from the 
University of Chicago.}

%% file: martinBucher.tex

George came to Paris in 2009 as a part-time professor in our lab, I believe at 25\% time. Starting then, Ken Ganga and 
I had many interesting conversations and interactions with George, with whom we would typically lunch when he 
was at Laboratoire Astroparticule et Cosmologie (APC).

However my first and probably most influential encounters with George were indirect and date much further back. I 
think my story in some similar form is shared by many physicists of my generation. I did my first postdoc at the 
Institute for Advanced Study in Princeton when the COBE DMR 
(Differential Microwave Radiometers) detection was first announced. Although I knew some 
basic cosmology at that time, my focus was theoretical physics, and in particular exotic statistics in two 
dimensions, as for example in the Fractional Quantum Hall Effect (FQHE). It was anticipated that I would follow this path 
with Frank Wilczek and others, but I found the subject frustrating. Whatever I discovered seemed already to have 
been done, as I found out on many occasions after a few days in the library. Then came the COBE announcement. I did 
not then entirely understand its meaning, but knew this was something big, and within a year my research had 
switched almost entirely to theoretical cosmology, where I had much better luck than with the FQHE.

In the now 34 years following the COBE DMR detection of the 
Cosmic Microwave Background (CMB) anisotropy, the study of the CMB has 
progressed immensely. 
Many of the promises from the immediate post-COBE era, which at that time seemed far-fetched, have now 
been realized. George, John Mather, and the COBE team were pioneers in this endeavor.

I greatly enjoyed the discussions with George during his Paris years. George had an unrivaled mastery of experimental 
physics and electronics. He was the go-to person for any question in this area, and also remarkable in his 
understanding of theory and the breadth of his interests, which ranged well beyond physics.

George also had a great sense of humor, even when he was being serious. I remember one lunch when he was explaining 
to a younger colleague how he came to work on the CMB. He explained, as one might expect, how the CMB anisotropies 
were a clean observable, there being a simple, linear link between the theory and what is measured, except for some 
simple small nonlinear corrections---so far all quite standard. He then he gave an example of an area that in his 
view was not so: relativistic heavy-ion collisions, which he described as ``shit-shit scattering.'' I cannot to this 
day listen to someone talk about relativistic heavy-ion physics without being reminded of “shit-shit scattering.” 
Discussions with George were often colorful. He was not always right, but most of what he said, even when it seemed 
outrageous and exaggerated, had at least a grain of truth, often much more than one might at first suspect.

Another George story pertains to teaching introductory physics for UC Berkeley engineers. George recounted
a meeting with some irate parents, whose son did not pass his class. The meeting opened with a question
from George, ``What kind of car do you drive?'' I do not recall the precise answer, but it was something
like an Audi or a Toyota. George replied, ``Why don't you drive an American car?'' ``Because they are 
not as well made,'' to which George replied, ``We at UC Berkeley are trying to train engineers in 
such a way that American
cars will be of the same high quality as German or Japanese cars.'' This may not have been tactful, nor how
deans would desire professors to deal with complaining parents, who might give money to the 
university. But it was pure George uncensored, which endeared him to many around him but not necessarily
everybody. 

George often spoke of his postdoctoral mentor Luis Alvarez toward whom he had great admiration. Alvarez is perhaps 
most striking for the breadth of his contributions, from hadronic resonances, to cosmic ray physics (George’s first 
project at Berkeley), to the extinction of the dinosaurs. Recently in a Paris used bookstore I came across the 
collection of papers with commentary titled \textit{Discovering Alvarez: Selected 
Works with Commentary by his Students and Colleagues.} 
I was astounded with the range of original contributions. George too as a physicist was a generalist. 
Through his questions and commentary, George encouraged those around him to think broadly and about the most 
important questions of physics, of science, and also of society. This approach today is not so common. But George’s 
example helps this style of doing science live on.

%% file: robertCahn.tex

One day in 1992, George walked into my office where I sat early in my tenure as the Director of the Physics Division at 
Lawrence Berkeley Laboratory (LBL).  For many years George had been supported partially by LBL and partly through the Space 
Sciences Lab of UC Berkeley.  The Physics Division was funded by the Office of High Energy Physics of the U.S. Department 
of Energy (DoE).  
We could hardly claim that the cosmic microwave background (CMB) was ``high energy physics,"  but in those days there 
was a tolerant attitude at the DoE.  We took advantage of that tolerance to support George's work and a small effort in 
the search for supernovae.  As a hgh energy physicist, I knew next-to-nothing about the CMB and only a little more about the 
supernova work led by Rich Muller, Gerson Goldhaber, and their young colleague, Saul Perlmutter.

George declared, ``Bob, I have something to show you."  Indeed he did.  He presented me with the image showing the 
variation---at the level of parts per hundred thousand---of the CMB across the sky.  I understood nothing.  I asked, ``Is 
this going to be big?"  ``Yes," George assured me.  This was made especially apparent to me when 
\textit{People} magazine in 
December 1992 came out featuring ``The 25 Most Intriguing People of 1992,"  with George's full-page picture falling 
between those of Katie Couric and Madonna.

George's fame was a mixed blessing for me.  Naturally, George was soon offered outstanding faculty opportunities elsewhere 
and I wanted him to stay in Berkeley.  The recalcitrant Berkeley physics department failed at the time to generate the 
appropriate offer to George.  I tried to persuade George not to leave and whether my persuasion was effective or not, he 
did stay. Given another year to think about it, the faculty, perhaps fearing the embarrassment they would endure if George 
won a Nobel Prize, finally did the right thing and gave George a faculty position.

As I drove to the lab one early October day in 2006, I got a phone call from a friend, not a physicist, but a Cal football 
fan who always wanted a good parking spot on the campus for tailgating.  I had assured him previously that I wasn't going 
to win a Nobel Prize and thereby also win a permanent parking place. I did say I had friends who would win the prize.  
He called me that morning to say someone named George had won it.  When I got to the lab, I entered Building 50 and walked up 
the half-flight of stairs to the fifth floor, where I saw George huddled in a room with the Lab Director, Steve Chu, who 
had won a Nobel Prize himself in 1997.  George was already talking to Steve about what he could do with his prize money to 
support cosmology at Berkeley.  This led to the Berkeley Center for Cosmological Physics (BCCP).

This displayed George's generosity and also his dedication to the field of cosmology.  George's commitment to public 
education was displayed in the ``MOOC"  (Massively Open Online Course) he and Pierre Bin\'etruy, our mutual friend, created 
in Paris, France, and which attracted tens of thousands of participants.

George was an iconoclast.  Not everyone found this endearing, but I was always delighted when George wandered into my 
office, in the years after his great discovery. He always had his own thoughtful view of events, a view often not widely 
shared.  In later years I saw him as often in Paris as I did in Berkeley.

Winning the Nobel Prize affects different laureates in different ways.  Some seize the opportunity to switch fields 
entirely, as did Donald Glaser, inventor of the bubble chamber.  Some drop out of science altogether. George's enthusiasm 
for research never wavered; it expanded in topics and most dramatically in geographical location.

Having George as a colleague and friend over the years was one of the great pleasures of my years at LBL.


%% file: jorge-cervantes-cota.tex
\phantom{George F. Smoot: His science and journey in Mexico}

\begin{figure}[h] 
\begin{center}
\includegraphics[width=0.7\textwidth]{%
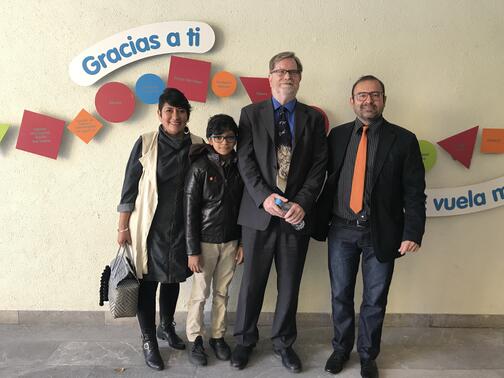}
\end{center}
\caption{%
\textbf{%
George with Jorge L.~Cervantes Cota his wife Olinca and son Sebasti\'an.}
[Photo credit: N\'ora Csisz\'ar]
}
\end{figure}

\begin{figure}
\begin{center}
\includegraphics[width=0.7\textwidth]{%
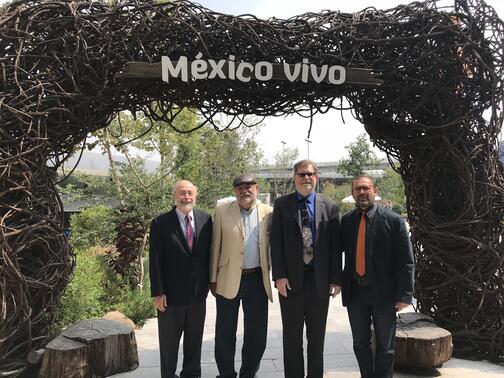}
\end{center}
\caption{%
\textbf{%
At the Papalote Museum, M\'exico (October 2017).}
(Left to right):
Michael Barnett, Salvador Galindo, George Smoot, and Jorge Cervantes-Cota.
[Photo credit: N\'ora Csisz\'ar]
}
\end{figure}

In the 1990s, George was a well-known figure. As a doctoral student, I met him at one of the cosmology schools organized by 
Norma Sánchez of the Paris Observatory under the political and financial umbrella of Antonino Zichichi, who recently also 
passed away. The lectures were held in Erice, La Città della Scienza (The City of Science), a small town in the western 
part of Sicily, the land of Cyclopes. Norma invited renowned professors and George was a frequent guest. Students could 
receive full scholarships, an option I always chose during my visits to that School. There I saw him alone once and seized 
the opportunity to speak with him. It wasn't easy given my limited English and the depth and complexity of his 
personality. He
was not only a physicist but also a highly cultured man. And, unsurprisingly, he was always surrounded by people. Although on that 
occasion we were only able to have brief conversations about physics, it set a precedent for me.

Years later, we organized the Fifth School of the Division of Gravitation and Mathematical Physics of the Mexican Physical 
Society. We wanted to invite leading researchers, so we thought of George among others. It was 2002 and George kindly 
agreed to come to Playa del Carmen near Canc\'un. It was our first School to combine such a pleasant location with 
top-notch facilities. During the event, George spoke to us about the fluctuations of the 
Cosmic Microwave Background (CMB) and how they vary in space, 
not only at large angles (7°) as he had discovered in 1992, but also at angles much smaller than one degree. It was a very 
motivating plenary session. We would later learn that he had been revealing to us the fingerprints of the Universe at 
scales below the horizon. It was something new for everyone, and we had that scoop at the Fifth School! The idea was to 
understand that when the Universe was about 380,000 years old, the transition from an opaque to a transparent medium 
occurred, and photons were released at what is called the surface of last scattering. That light reached Earth 
13.8 billion 
years 
later and was measured with angular precision by the WMAP probe. These details would then also be imprinted on the 
gravitational perturbations that would form cosmic structures.

At that same event, we also sought to include Mexicans as plenary speakers. I discovered that Axel de la Macorra from the 
Institute of Physics of the National Autonomous University of Mexico (UNAM), whom I didn't know, had recently written an 
article in Physical Review Letters, which led us to include him in the program. Axel and George met there, and it was some 
years later that the relationship between the two of them solidified.

Around 2006, some Mexican colleagues from UNAM, the Center for Research and Advanced Studies (CINVESTAV), and myself from 
the National Institute of Nuclear Research (ININ) used to meet to discuss different aspects of theoretical physics.  
Among other topics, we treated themes such as black hole physics and scalar hair, scalar field dark matter, and some 
cosmological results. We began to see the importance to collaborate in a more formal way. It was very interesting, although 
complicated by having people from many specialties discussing diverse subjects. During these discussions, the idea arose to 
form a larger group to systematically investigate these aspects. Meanwhile, Axel wanted to establish a new institute with 
similar ideas, but more focused on cosmology. Axel contacted me, and I contacted Tonatiuh Matos and Darío Nuñez. Thus it 
was easy to converge on the formation of this institute. It was in the fall of 2006.

Around that time, George was awarded the Nobel Prize in Physics along with his colleague John Mather. I emailed George, but 
he didn't reply directly. Days later, he posted on his website that he appreciated the countless congratulations on his 
award but couldn't respond to them personally. In November, I wrote to him again and let him know of our intention to 
create our institute. I didn't know that George was about to found his own in Berkeley. I told him that ours would be 
virtual, without offices, but with the intention of promoting cosmological science in Mexico. He quickly suggested that we 
become partner institutes and work together. He even proposed the name ``Mexican Center for Cosmological Physics," but the 
four founders had already decided on a name.

The Advanced Institute of Cosmology/Instituto Avanzado de Cosmolog\'\i a (IAC) was inaugurated in June 2007 with a keynote address by George at El Colegio 
Nacional, the most prestigious scientific institution in Mexico. The venue was full of colleagues, students, and the general public. It was 
a great event, and after the talk, George was surrounded by impatient people wanting his signature or just to shake hands 
with him. The IAC was the first virtual institute in Mexico and was part of the international scene where new cosmology 
institutes were being established, such as the aforementioned Berkeley Center for Cosmological Physics (BCCP) founded in 
2007 by George, the Center for Excellence (ECU, in 2008), which brought together the universities and Max Planck 
Institutes in southern Germany and the Centre for Cosmology at the University of Cambridge, England (2008). The rationale 
behind these institutes was the growing convergence of scientists from diverse fields of knowledge, such as quantum field 
theory, elementary particles, gravitation, astronomy, and astrophysics. Researchers in these fields worked at different 
institutions for historical reasons. Meanwhile, cosmology was experiencing a period of rapid progress, both because of 
the interesting theoretical problems it addressed and because of the massive acquisition of precise data from new 
telescopes and satellites. The IAC maintained a close relationship with the BCCP and ECU, organizing Schools of Cosmology 
and facilitating exchanges for students, postdoctoral fellows, and researchers. Some colleagues and postdoctoral fellows 
from Mexico visited the BCCP for research stays or scientific events, forging academic and even personal connections 
between our communities.

The BCCP and the IAC were always friends, just as George wanted. During one of George's visits to Mexico, George, Axel, and I 
during lunch came up with the idea to organize a cosmology school in Mexico. We founded the “Cosmology for the Next 
Generation” schools, informally known as “Cosmology on the Beach.” These schools not only fostered the training of new 
generations of scientists in cutting-edge topics worldwide, with international experts teaching courses and leading plenary 
sessions, but also put Mexico on the international map. Our schools inspired others, such as those in South Africa 
(Cosmology on a Safari), Venice (Cosmology on a Cruise), Switzerland (Cosmology in the Alps), and elsewhere. Furthermore, 
the BCCP and IAC jointly organized multiple academic events, public talks, and the production and presentation of various 
films about cosmology for the general public, such as \textit{The Phantom of the Universe} in the digital dome at the Papalote Museum 
(which ran for two years) and \textit{The Mayan Universe} in the Luis Enrique Erro Dome.

The relationship with George, his drive, and his actions allowed the IAC to play a decisive role in Mexico's incorporation 
into the international Dark Energy Spectroscopic Instrument (DESI) project, currently the world's most important effort 
measuring in large scale structure (LSS) observational cosmology. Its objective is to map tens of millions of galaxies and quasars to understand 
the evolution of the Universe as well as the nature of dark matter and dark energy. Dozens of Mexican students and 
researchers participate in the DESI project. George left an 
indeleble
mark on 
Mexican cosmology science by helping to promote, 
expand, and professionalize it with high standards.

George, on the other hand, generated an emotional closeness with the Mexican community that we will never forget, and that 
we 
carry on 
in our hearts.


%% file: steve_c.tex
\phantom{A tribute to George Smoot (Dec 1, 2025)}

\begin{figure}[h] 
\begin{center}
\includegraphics[width=0.7\textwidth]{%
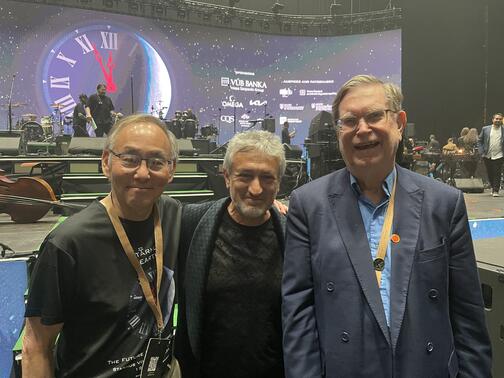}
\end{center}
\caption{%
\textbf{%
At Seventh Starmus Festival, Bratislava, Slovakia ( 15 May 2024).}
(Left to right): Steven Chu, Garik Israelian, and George Smoot.
[Photo credit: N\'ora Csisz\'ar]
}
\end{figure}

My path with George began while I was a graduate student and postdoc at Berkeley from 1970 to 1978. During that time, I 
began to attend the Luis Walter Alvarez’s Thursday afternoon “physics salons” that anybody could attend. George came to 
Berkeley as a postdoc to work with Luis in 1970. Another contemporary, colleague and collaborator of George, Rich Muller 
earned his PhD as a student of Luis in 1969. The style of those physics salons was open brainstorming where half-baked 
ideas could become fully baked or thrown in the trash. I suspect that both George and Rich picked up some of the Luis’ 
style of challenging ideas, the importance of developing radically new scientific instruments that could lead to monumental 
discoveries.

George and Rich used a converted U-2 spy plane to measure the lowest-order anisotropy of the Cosmic Microwave Background 
(CMB). This experiment was the first step in George’s lifelong quest to study fluctuations in the CMB. He built a team that 
constructed increasingly 
sensitive differential microwave amplifiers that led to the Differential Microwave Radiometers (DMR) part of the 
COBE satellite. The instrument with its tour de force resolution of $\Delta T/T \approx  10^{-5}$ led to his 2006 Nobel Prize. Stephen Hawking 
described this measurement as ``the greatest scientific discovery of the century.''

Clearly the COBE mission was a team effort, and there were some rough patches when George’s enthusiasm led to his 
“pre-announcement” of the COBE results in 1992. 
Also, his now famous declaration, “If you’re religious, it’s like seeing the face 
of God” seemed excessive to some of his more staid colleagues. While most scientists are better at hiding their egos than 
George, in the long run his theatrical flair helped popularize experimental cosmology. He also generously donated 
\$500,000 USD, 
his share of the 
Nobel Prize
money, to endow the Berkeley Center for Cosmological Physics.

As his interests broadened into climate change, biomedical research, and scientific outreach, our overlapping interests 
grew. On a more personal note, I always enjoyed talking with George. He was very smart, had a wry sense of humor, and 
enjoyed making wild speculations that led us to some friendly intellectual jousting. I will remember how I enjoyed most of 
his quirks, but above all else, 
his love and fascination of science shined most brightly. George was a real character, and I will 
miss him as a colleague and as a friend.


%% file: raz.tex

I first met Professor George Smoot at the Institute for Advanced Study (IAS) at
the Hong Kong University of Science and Technology (HKUST) in 2016, not yet aware that he was a 
Nobel Prize winner. At the time, I was also relatively new to IAS. To me he was 
simply a kind and respectful visitor in the office next door. As an early-career postdoctoral 
researcher eager to broaden my horizons in astrophysics, I was encouraged by others to speak with 
him. That introduction marked the beginning of a collaboration and mentorship that would profoundly 
shape my career.

George was remarkably generous with his time and insights. Despite frequent travel and a part-time 
presence at the institute, his door was always open. He was consistently approachable, cheerful, 
and deeply engaged. He read my many drafts with care and patience, and he listened thoughtfully to 
my ideas, no matter how preliminary. His feedback was always constructive, and his encouragement 
unwavering. It was through this environment of openness and intellectual curiosity that we began 
working together.

Our collaboration led to several publications, including work co-authored with early-career 
students and postdoctoral researchers, two of which appeared in \textit{Nature Astronomy}. 
George was 
especially supportive of my research directions, including projects on primordial black holes, 
neutrino clustering, and wave dark matter. His willingness to engage with new ideas and to support 
emerging researchers was a defining aspect of his scientific leadership.

Beyond our joint research, George played a pivotal role in my professional development. His strong 
letter of recommendation was instrumental in my transition to the United States, where I joined the 
Center for Astrophysics (CfA). That opportunity proved transformative, allowing me to further 
develop my research in astrophysics. More importantly, working with George taught me how to think 
as a scientist with clarity, rigor, and integrity. His calm demeanor, intellectual sharpness, and 
deep commitment to mentorship provided an ideal foundation for my growth during a critical stage in 
my career.

I was deeply saddened to learn of his passing. The scientific community has lost not only an 
extraordinary mind, but also a generous mentor and compassionate colleague. His influence endures 
through the many students and collaborators he supported, and through the ideas he helped shape.

I remain profoundly grateful for the time and privilege of working with George. His legacy will 
continue to guide and inspire those of us who were fortunate enough to learn from him.

%% file: leo_fun.tex
To many, Professor George Smoot was a Nobel Laureate who studied the mysterious origin of our Universe and all the other 
fancy stories. To most students studying astrophysics, he was the textbook figure who mapped out the remnants of the Big Bang for 
the first time in human history. To his own students, he was just a curious scientist who loved humans as much as the 
Universe, society, and everything in daily life.

His scientific methodology was distinctive, resembling the classical tradition of early 20th-century scholars more than 
that of his contemporary peers. At the outset of my doctoral studies, he challenged me to apply cosmological reasoning to 
biological systems by estimating cancer risks relative to age. ``If you can do cosmology,'' he told me, ``you should be 
capable of applying that same logical reasoning to daily life.'' This led to a semester of rigorous mentorship focused on 
`Fermi estimation.' Standing in front of a blackboard, he devoted a full hour each day to these exercises with me, 
covering topics from the cosmic to the sociological. This training became a cornerstone of my professional development. He 
remained a firm believer that while others might rely on simulations or digital databases for precision, the most 
profound scientific insights are often found through the simplicity of a back-of-the-envelope calculation.

Throughout his career, he observed several technological shifts that profoundly informed his view of the relationship 
between science and technology. He frequently noted that during his doctoral studies in the 1970s, the constraints of 
physical publishing meant that papers were brief and focused strictly on core theoretical concepts. Following the digital 
revolution, he noted a substantial paradigm shift: the research ecosystem moved toward exhaustive documentation and a 
heightened emphasis on reproducibility.

In his later years, he became particularly interested in the intersection of research and generative AI. Long before the 
maturation of tools like ChatGPT, he foresaw that AI would soon match the productivity of researchers in data-heavy 
disciplines. Consequently, he advised his students to prioritize creative inquiry over routine data processing---a warning 
that seemed premature in 2021 but has since proven remarkably prescient. His intellectual range was vast, allowing him to 
apply the same piercing logic to global technological trends as he did to the complexities of physics.

For a scientist of such high standing, he remained exceptionally accessible. He was a familiar sight on the
HKUST (Hong Kong University of Science and Technology) campus, often seen 
in a pair of sandals, taking his meals in the crowded canteen with students. Although he possessed a sophisticated palate and 
could offer discerning critiques of fine cuisine, he chose to overlook the canteen's mediocre offerings in favor of the 
student company. He took great pleasure in these informal interactions and was visibly energized whenever students 
approached him to share their thoughts.

Throughout our time together, he constantly emphasized the importance of bringing science to the public. He understood 
that as we dive deeper into the universe's most challenging questions, the cost of testing our theories has skyrocketed. 
Fifty years ago it was still possible to conduct groundbreaking experiments in small laboratories on a modest budget; 
however, as his own work on the COBE experiment pushed cosmology into the era of ``precision cosmology,'' the complexity and 
cost of research scaled up by orders of magnitude.

Foreseeing this shift, he became a tireless advocate for public engagement, knowing that science needs the world’s 
understanding and support to thrive. Rather than sticking solely to academic seminars, he stepped into the spotlight of 
the entertainment industry, making appearances on TV shows like \textit{The Big Bang Theory.} He felt a profound commitment to 
using his reputation as a Nobel Laureate as a tool for the advancement of science. While progress is always a 
collaborative effort, his unique standing made him the perfect ambassador for the wonders of the natural world.

My time as George’s student provided an education that extended far beyond the campus. Following my graduation, he 
continued to provide invaluable mentorship regarding my scientific career and personal development. I fondly recall our 
many discussions, whether in the quiet of his office or amidst the bustle of a public train, where we explored the 
complexities of science and the excitement of discovery. Though some of his broader humanitarian ambitions remained in 
progress at the time of his passing, his profound impact on his students remains an undisputed part of his legacy. I am 
honored to pass his wisdom to the next generation of researchers, embracing his dual mission: to comprehend the origins of 
our universe and to improve the lives of those within it.

%% file: ken2.tex
I first met George in late 1992 or early 1993, I think, right after the COBE 
Differential Microwave Radiometers (DMR) Cosmic Microwave Background (CMB) anisotropy discovery announcement. 
Since CMB anisotropy maps just look like noise, it's hard to be sure you've really measured what you think you've 
measured. So, while people were excited about the DMR announcement, there was a need for some other experiment to see the 
same thing.

I was a graduate student and had just started analysing data from the Far-Infrared Survey (or FIRS), a balloon-borne 
experiment that had mapped the CMB over some of the same sky as had DMR. Lyman Page and I went to NASA/Goddard to discuss 
a comparison of the maps from the two experiments with the COBE Science Team (which included Stephan Meyers and Ed Cheng, 
who were members of both FIRS and of COBE). When I think back on this meeting now, I'm sort of amazed. Two future Nobel 
Prize-winning results, from FIRAS and from DMR, had recently come out of work from this team. The  
Diffuse Infrared Background Experiment (DIRBE) 
data was being 
discussed, and it bears mentioning that DIRBE data is used far more today than that from either FIRAS or DMR. Rai 
Weiss, who would win the Nobel Prize for later work on gravitational waves, was also a part of the team. At the time, 
however, I didn't really understand much of what was going on around me. Honestly, what I remember most about this meeting 
was that George, ever the early technology adopter, had a laptop computer! I had heard about laptops by that point, but I 
had never actually seen one.

We all agreed to work together, and with Charley Lineweaver, Gary Hinshaw, Chuck Bennett, and the rest of the team, we 
showed that DMR and FIRS were indeed seeing the same ``structure'' on the sky. After our initial meeting, all of our work 
was done via email and telephone, and once we published, I didn't have much more to do with George during the rest of my 
thesis work. And as the 1990s progressed and ended, we still didn't interact much, except for occasional encounters at 
meetings for preparations for the Planck satellite.

Then Planck whisked me away to France, so while George's work obviously cast a long shadow over our work, I didn't see 
much of him for a while. But fifteen years ago or so, George appeared in Paris, with an appointment at the 
Laboratoire Astroparticles et Cosmologie (APC), 
where I work and where he helped found the Paris Centre for Cosmological Physics.  He had a 
few such appointments around the globe, with contracts that had him moving between them and spending a few months at each 
every year. So we would see George a couple of months a year, every year. It was here that I finally started to get to 
know him and became fond of him.

Most of our interactions centred around lunch. Whenever he was in town and free, a few of us would gather at the faculty 
canteen or a nearby Lebanese lunch place. The routine often included taking pictures of whatever we were eating and 
sending them in particular to N\'ora to emphasize how well George was eating and taking care of himself (unless he was cheating). I think 
one of the reasons he liked France was that it was easier to eat healthy food, in addition to the healthcare and the 
walkability of Paris. At these lunch gatherings, our conversations often ended up with him talking about his ideas for 
technology for improving health, be it our own or the world's. He knew he had already made memorable contributions to 
physics, but I suspect that he still wanted to make even larger contributions to society in general.

George often had something on his mind, and would pretty much pursue this, and only this, during conversations. A 
selection of topics, off the top of my head, might be: How should one invest? (He seemed to like individual stocks and 
real estate, though he may have soured on real estate towards the end.) Where should we live if we believe that the world 
order would collapse soon? (New Zealand---and yes, he was at least a bit serious.) What's wrong with politics? (We never did 
solve this one.) How to efficiently mix and taste a selection of 1000 wine bottles if you believe one of them might be 
poisoned?

Once his attention was fixated, it was very hard to tug him away from whatever it was. We lost people from our little 
group after a few too many extended conversations about the American stock market (I may have been complicit in this). And 
we lost graduate students when they realized that George was likely to ask them ``Fermi questions,'' and not let it go until 
they got to the answer.

George never did really learn much French, but he was very much an ``internationalist.'' I haven't met many other Americans 
so willing, eager even, to actually live in other places. Lots of us Americans like to visit places. Some of us even end 
up living abroad. But George seemed to thrive in multiple, vastly different places for the long-term. Berkeley, Mexico 
and France, okay. But also Hong Kong and South Korea. And various commitments to China, Russia, Kazakhstan, and I'm sure others 
I'm not aware of. This is not normal for an American. On the other hand, he did usually plan to spend his time in Paris 
around the summer, when all the normal people were away on vacation, so that he could get work done. Not a typical 
American, but perhaps a very typical, workaholic scientist.

%% file: alisonGoldwyn.tex

\phantom{falseStart}

\begin{figure}[h] 
\begin{center}
\includegraphics[width=0.7\textwidth]{%
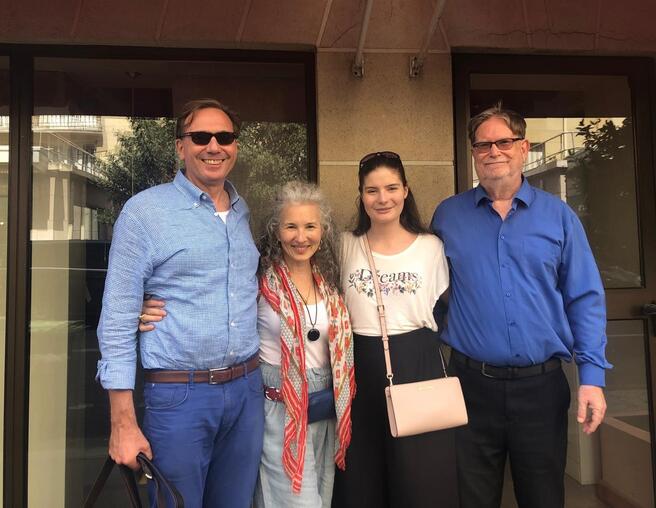}
\end{center}
\caption{%
\textbf{%
(Left to right):
Rainer Kabelitz, Alison Goldwyn, N\'ora Csisz\'ar, and George Smoot
(Paris 2019).}
}
\end{figure}

\noindent 
George was brilliant. Cosmic. And a riot. A veritable ``standup cosmic'' whom I met through dear N\'ora. 

His combination of eclectic information punctuated with humour often gushed out of his mind so fast that his 
sentences would trail off into the universe before completing. Trying to keep pace usually occurred not over 
some quiet home-cooked meal, but in bustling Paris bistro lunches in the 7th arrondissement, crammed into 
tiny-noisy-smokey tables where I would strain to interpret the missing sentence particle against the backdrop 
of fast-French chatter.

At one particularly upbeat meal, I tried to eat my chicken dish, understand his missing sentence, and not 
guffaw as he explained the climate crisis leading to the end of the world. N\'ora had a look of dread on her 
face as her appetite quickly faded. I was torn between assuaging her that we still had time left to finish 
our lunches before the destruction versus gobbling down my dish before the waiter came to remove it.

The one way to quiet George’s active mind was with talk of\ldots aliens. This was such a far out subject that he 
would go silent, drifting off into a faint smile as if his antennae were hearing the universe’s response. From 
where he presumably is now, he might have some newfound opinions :-).

Among other hilarities of note: his surprise birthday cake created from the stuff of chocolate fantasy, at a 
verrrry posh Parisian bakery. Knowing how much of a chocolate connoisseur he was, it was humorous to learn 
that he took a taste and didn’t like it. Too chocolatey. And days later proceeded to finish the entirety.

And of course the N\'ora photo shoots\ldots 
He loved capturing her aura---her beautiful ``N’aura,'' be it at hi-profile Nobel events she got him speaking 
engagements at, or her flying over the beautiful blue Cabo San Lucas sea---in a parachute---or her angelic 
smile while crouched nee crunched into a ball below the kitchen sink to fix a pipe.

They loved each other so … and I loved knowing this truly lively and loving duo. George will not be missed, 
because his vivacious energy permeates so many places, memories, and synchronicities. Including chocolate.

\begin{figure}[h] 
\begin{center}
\includegraphics[width=0.7\textwidth]{%
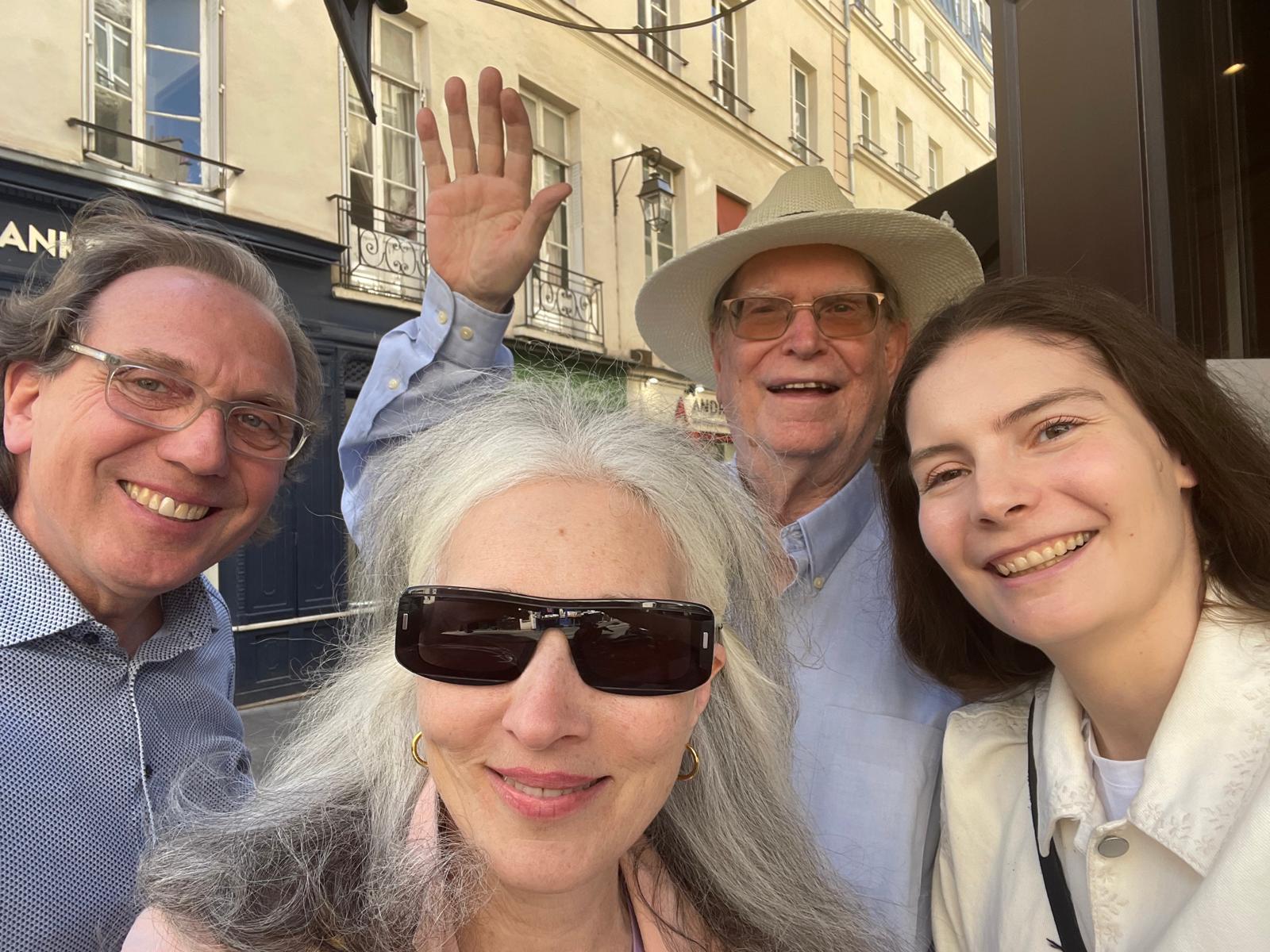}
\end{center}
\caption{%
\textbf{%
Rainer Kabelitz, Alison Goldwyn, Nora Csiszar, George Smoot in Paris.} (2024)
}
\end{figure}

%% file: billGreen.tex

\begin{figure}
\begin{center}
\includegraphics[width=0.7\textwidth]{%
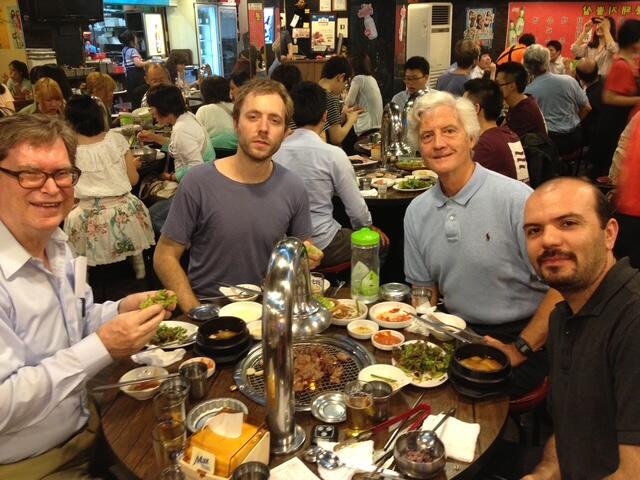}
\end{center}
\caption{%
\textbf{%
(Left to right):
George Smoot, Jonathan Regier, Bill Green, and Alireza Hojjati (2013).} 
[Photo credit: Bill Green]
}
\end{figure}

\begin{figure}
\begin{center}
\includegraphics[width=0.7\textwidth]{%
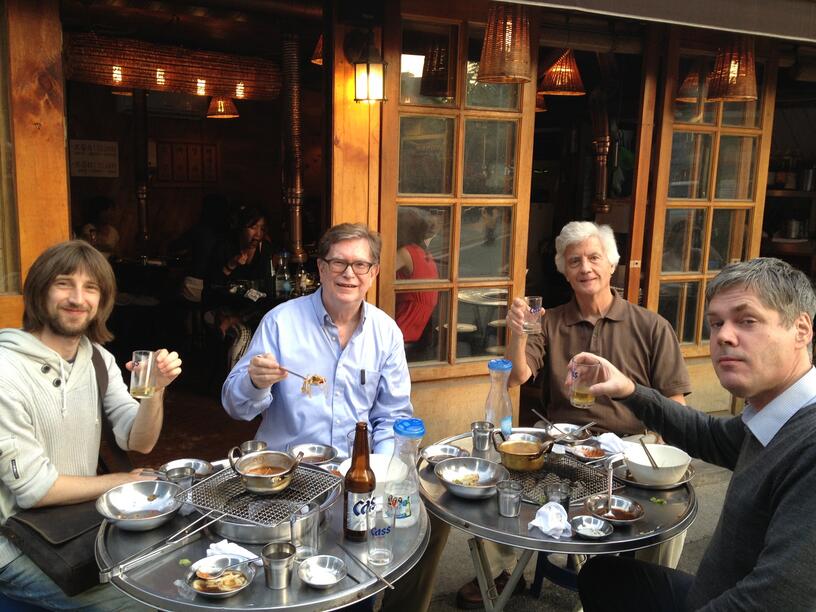}
\end{center}
\caption{%
\textbf{%
(Left to right):
Stephen Appleby, 
George Smoot, 
Bill Green, and Maurice van Putten (South Korea, 2013).}
[Photo credit: Bill Green]
}
\end{figure}

George and I became acquainted in the early 1990s while attending several International Schools on Cosmology and 
Gravitation at Erice, Sicily.  In those days, George was reminding everyone that more and more computational capacity 
(Moore’s Law---no pun intended) would be needed to handle the projected huge increase in the amount of cosmological data that 
would be coming in.  He foresaw from his work on the COBE satellite (for which he later shared a Nobel Prize) that it was 
only the beginning.

When I called George in 1997 to ask if I could be a sabbatical visitor in his group at Lawrence Berkeley Laboratory (LBL), 
his response was, ``Yes, but you will need to work when you get here.'' So my wife Diane and I drove from Florida to 
California in early 1998.  I showed up at LBL, got a parking pass, and entered George’s office in Building 50. After brief 
preliminaries, he said he would like for me to direct three Berkeley undergrads in their task of building and installing a 
small microwave radiometer on the Building 50 roof.  Off to the races with a fun project!  George and I had lunch at the 
LBL cafeteria most days. He always seemed to have a good appetite, resulting in part from the fact, I assumed, that he 
walked “up the hill” to work each day from his home about two miles away.  Conversations ranged widely.  Our daughter Amber 
(now McHugh) visited us there that summer where July 4 fireworks could be seen across the San Francisco Bay from LBL’s 
high elevation.

During our stay at LBL, I also had the good fortune to periodically fly down to Mt.~Wilson with Charley Townes to assist in 
measuring the spectra of red giant stars with his Infrared Spatial Interferometer (ISI).  It was quite fitting that the ISI 
at Mt. Wilson used a laser heterodyne system to amplify the incoming star radiation signals since Charley invented the 
maser that triggered the laser revolution.  In addition to staying up all night, I had the pleasure of assisting Charley 
and David Hale in renewing the aluminum coating on the large mirrors of the ISI, and of making runs down the mountain for 
supplies such as liquid nitrogen from Caltech needed to cool the detectors.  Like George, Charley was one-of-a-kind: 
brilliant, generous, down-to-earth, and fun to work with.  

It was remarkable that I had the great fortune to be at LBL 
while the understanding of the universe took two quantum jumps.  The first was that the expansion of the universe is not 
slowing down so Dark Energy must exist.  Three offices down the hall from mine was Gerson Goldhaber, who was analyzing 
supernovae data as part of the LBL supernova group.  I remember walking by each day and seeing the bearded Gerson at his 
computer wearing his trademark suspenders and often a red shirt.  Once I said, “Hi Gerson, is business booming today?” I 
think his reply was, ``You bet!'' Of course, at the time the LBL folks were observing the accelerated expansion of the 
universe that led to a 2011 Nobel Prize shared by Saul Perlmutter, Adam Reiss, and Brian Schmidt.  The second jump in 
knowledge came from the MAXIMA Cosmic Microwave Background (CMB) measuring balloon project lead by PI Paul Richards in 
which George collaborated and other suborbital measurements, which began to resolve the second and third peaks of the CMB 
multipole spectrum enough to show that the existence of Dark Matter was also necessary.  Of course, as 
resolution 
later improved with the WMAP and Planck satellite data, any doubt was removed.  Overall, we had a lot of fun and the 
student radiometer project, intended to measure the CMB polarization (in theory), was built.  To my knowledge it never 
succeeded but it was a great learning experience for the students.

In the following years, I flew out to LBL several times to help George, Eric Linder, and Tony Spadafora in their efforts to 
organize the Berkeley Center for Cosmological Physics, and also attended several of George and Eric’s ``Cosmology on the 
Beach'' conferences at various resort locations in Mexico. George visited us and Florida State University in Tallahassee a 
couple of times, the last one of which was in 2006 a few months before he and John Mather were awarded a Nobel Prize for 
their CMB work.  By then, George had become a very good friend of ours who frequently kept in touch.

George was much more than a scientist.  About 17 years ago we were on a Skype call with George from our son Dr.~Harris 
Green then home in North Carolina.  Harris’ three-year-old son Daniel had just been diagnosed with Acute Childhood Leukemia 
and was sitting on my lap feeling tired and sad.  George talked to him to try to cheer him up.  Amazingly, after several 
years of treatment and frequent follow-ups, Daniel is now a 6’2” tall young man with a very strong academic record and 
a promising future.  Thank you, George!

The last time Diane and I saw George was at a casual dinner with him and his partner N\'ora Csisz\'ar 
on May 7, 2024, the last 
day of a week-long stay in Paris.  They had generously facilitated this trip of a lifetime for our granddaughters Anna 
Palacio and Caroline Green by encouraging us to stay at their 
apartment.
After dinner the four of us said goodbye to George and N\'ora, returned to the rental, 
packed up, and waited on the sidewalk for a cab to take us to a hotel near Charles de Gaulle Airport.  As the cab was 
arriving, I noticed that we had only three suitcases. We immediately realized that Anna had left hers locked in the upstairs 
apartment where we had also left the key.  We called George with the bad news and left for the hotel.  To our total 
surprise, N\'ora went to the apartment, collected the suitcase, and had a taxi take her to our hotel miles away, where she 
delivered it to us.  Thank you, N\'ora!

As Diane says, George was a kind of Renaissance man.  On the one hand, he could understand the mysteries of the universe, 
setting up programs around the world with Eric Linder and Bruce Grossan’s help. (Sadly Bruce recently passed.)  On the other 
hand, he kept up with the political situation in America, the Olympics, Taylor Swift concerts that he and N\'ora attended, 
and many other current and historical events.

Last year when Skype went out of business, we could no longer directly communicate with George and I soon became concerned.  
I guess it was an omen.  Now George has left us.  But his memory is strong.  Most people think of him as a worldclass 
physicist.  We think of him as a true friend who will be missed and never forgotten.

%% file: George-From-DhirajHazra.tex
%
%
%

\phantom{singini}

\vskip -7mm 

\begin{figure}[h]
\begin{center}
\includegraphics[width=0.6\textwidth]{%
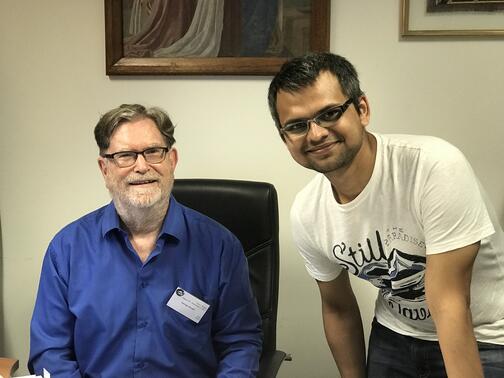}
\end{center}
\vskip -15pt
\caption{%
\textbf{%
George with Dhiraj Kumar Hazra at APC, Paris.}
[Photo credit: N\'ora Csisz\'ar]
}
\end{figure}

\vskip -10pt

When I started to write about George, I struggled a lot with the first line. I did not know where to begin. After a lot of 
thought, I decided to begin by mentioning this struggle. Yes, I typed and erased the first line several times. I thought of 
starting with George as my mentor, or George as my collaborator, or George as my friend, or George as a scientist. I think it is 
best if I simply share a few glimpses of my interactions with George.

I first met George during my visit to APC, Paris, in 2011. He attended my seminar and asked me a couple of questions. I did not 
get a chance to discuss things with him in detail during that visit, but I distinctly remember admiring the glass-door 
PCCP office rooms and George’s office opposite them. A couple of years later, I met him again during a conference organized 
by George, Arman Shafieloo, Eric Linder, and Uros 
Seljak
to celebrate Alexei Starobinsky@65. During that conference, George, Arman, and I began 
discussing a project in his office. Our collaboration started that day.

At that conference, we had a pleasant boat tour in Seoul, South Korea. 
Some of us went to the edge of the boat and leaned against the 
handrails, perhaps a bit more than was allowed, and the captain blew the horn. George immediately started laughing and said, 
“You guys are theoretical physicists, unaware of practicalities. That is why the captain reminded you.” George had a way of 
turning even a coincidence into a simple, funny story. During that same conference, he also showed me the clean room for 
detector development. He always loved showing people his labs. When I joined the Paris Center for Cosmological Physics (PCCP) 
as his postdoc, he showed me his mKIDs 
lab, and later he gave me a tour of the Berkeley lab at 
the Berkeley Center for Cosmological Physics (BCCP). It was always impressive to see how excited he remained about 
physics.

\begin{figure}[h] 
\begin{center}
\includegraphics[width=0.6\textwidth]{%
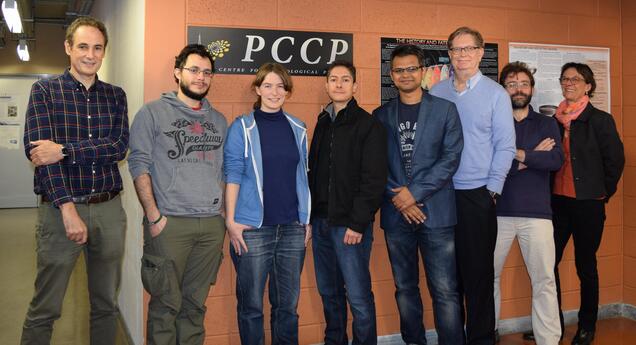}
\end{center}
\caption{%
\textbf{%
PCCP Group Photo.}
Left to Right: Pierre Bin\'etruy,
Mauro Pieroni, Valerie Domcke, Ivan Debono, Dhiraj Kumar Hazra,
George F. Smoot III, Andrea Tartari, and  Marie Verleure.
[Photo credit: Dhiraj Kumar Hazra]}
\end{figure}

George loved to spread physics rumors, and most of them eventually turned out to be true, always with remarkable detail. He told 
me about the possible BICEP B-mode detection announcement, and he also told me about the LIGO detection announcement. He 
explained to me in detail the differences between the engineering mock runs and the real events.

After the BICEP2 detection, George sent an email to discuss the possibility of building models that could reduce the tension 
between Planck and BICEP2. George, Arman, and I started working on the model, and Alexei joined soon after. From the first 
results to the first draft and then to submission, George was always one email or one Skype call away. We exchanged countless 
emails over the course of that month. Every result that improved the fit to the data made him excited, and he would immediately 
start thinking about the next step in the model building. Around that time, I realized that to be George Smoot, and to remain 
George Smoot, one had to be that enthusiastic about physics. All our collaborations were like that. In every paper I co-authored 
with George, and in every problem we were still discussing when he left us, he was just as enthusiastic and just as encouraging 
as he had been during my first paper with him. I promise to write about the details of our work-related discussions with George 
in another article.

At another meeting in Daejeon, South 
Korea, George brought up the subject of my next postdoc position. He started writing letters for 
me and also mentioned the possibility of an opening at PCCP, Paris. Well, I joined PCCP in 2015. Throughout the application 
process, George helped me shape my applications, wrote letters for me, and introduced me to his colleagues. He pushed me toward 
fundamental physics and, at the same time, discussed telescopes and observations in great detail.

When we connected through Skype, he started sending jokes regularly. He used to forward jokes and trivia all the time.

After I joined Paris, we used to go to a Lebanese place next to APC for shawarma. George suggested that I try the mixed-meat 
plate. While eating, he explained that this 10-euro plate, compared with the 6-euro shawarma (which was a more postdoc-level 
option), included several meats and vegetables, and that one could really enjoy it with the bread. The wrapped one, he said, 
simply could not hold that much. I had that plate once and really liked it, but from the next lunch onward I went back to the 
standard shawarma due to cosmological constraints.

On another day, while we were going for lunch, we saw some shattered glass on the road. The pieces were in a perfect cube shape. 
George asked, “Do you know what these are?” “Perhaps some window has fallen off a car,” I replied. “Then why are they 
cube-shaped?” he asked again, before explaining the safety reasons behind it. Yes, he loved bringing up and explaining trivia. 
On all the days we went to lunch with George, there was always some discussion of such trivia, or of world politics, or of the 
stock market.

He was always delighted by his first appearance on The Big Bang Theory. He said he still received \$45 USD every time it aired
in the US. 
In his PCCP office he had a framed photo 
of his meeting with Sheldon Cooper, 
played by Jim Parsons. 
In one episode, Sheldon wanted authorship of a paper with 
George to be in alphabetical order: Cooper and Smoot. George refused. When we wrote the paper on reionization in 2017, I told George 
that I had finally achieved a Hazra and Smoot in alphabetical order. He was even looking forward to the episode about Sheldon and 
Amy’s Nobel win. I once told him to check with the directors about the possibility that Young Sheldon had been inspired by his 
discoveries. He said the show could not make him look that young, but I could tell that he was really excited by the idea.

After I left PCCP, I never met George again in person. In 2023, when I visited APC, we had hopes of meeting. I was in Paris 
with my wife and daughter, and George kept us informed of the May political protests. He pinged me on Skype to check whether we had managed to 
reach our Airbnb safely from the airport. Unfortunately, I could not connect with him during our stay in Paris, and then I heard 
back from him again about a month later. I thought I would visit him the next time I was in Paris. That day did not come.

I never thought that George would leave us so soon. He was always full of life. He was very conscious about his health, and even 
today it is difficult for me to accept that George is no more. George and I discussed physics, teaching, world news, Nobel news, 
and many other things, and almost every month we communicated through email or Skype. My memory of George does not come only 
from our in-person interactions. It also lives in the thousands of emails and Skype chats we exchanged. A George email or a 
Skype ping about physics never stopped inspiring me. He was working on a certain inflation model, and he had sent me several 
emails about the modeling and a few projects we were planning next.

George’s passing has created a void, I believe, in the institutions he founded, among the students and postdocs he supervised, 
and in the lives of his loved ones. I am truly fortunate to have had him as my mentor, and lucky to have called him my friend. I 
miss him dearly, and I will continue to miss him.


%% file: paulaKeeney.tex

\phantom{nora}

\vskip -1cm

\begin{figure}[h] 
\begin{center}
\includegraphics[width=0.6\textwidth]{%
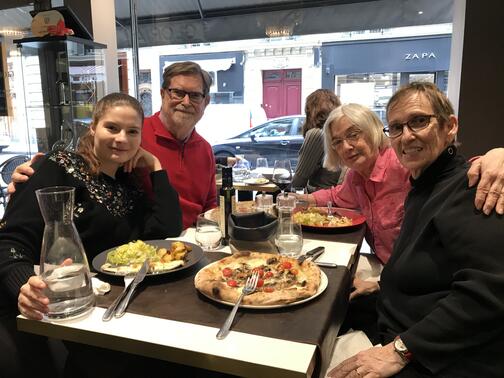}
\end{center}
\caption{%
\textbf{%
(Left to right):
N\'ora Csisz\'ar, George Smoot, Ann Whetstone, and Paula Keeney (Paris, France, 2019).}
}
\end{figure}

\vskip -5mm 

George entered my life 50 years ago--—obviously through Ann Whetstone. 
[See also the testimonial of Ann Whetstone in this booklet.]
At first he was just a voice on 
the telephone.
Of course, I’d long known of him. From the time we met, Ann regaled me with stories about 
their Alaska childhood.

It wasn’t until about five years later in the early 80s—when Ann and I had given up 
university teaching and moved to New York City---that I actually met the man who would become 
a cherished friend.
Shortly after George arrived in New York City for a meeting, I ended up in the hospital.
George called Ann in hopes of setting up a 
spur-of-the-moment, if long overdue, dinner.
Ann told him I was in the hospital.
When told of my hospital stay, he agreed ``bad timing,'' and talk moved on to ``next time'' when he 
visited the city.
But the next morning I awoke to find a bearded stranger sitting next to my bed. Then I 
was confronted by a big smile and a handshake, ``Hi, Paula, I’m George Smoot.'' He was no 
longer just a voice on the telephone.
It was just the beginning of a friendship I will always treasure.

Over the years, Ann’s friendship with George was a given---and I’d become part of it. There 
were his occasional East Coast visits, and our own trips west, and always those 
transcontinental telephone calls.
As his career flourished, we’d hear of George’s work—his discoveries, his awards. But our 
``connection'' was more personal, and stemmed from the one he and Ann had made long ago.
In later years, when he spent much of his time in Paris, we were lucky enough to share that 
life too---with visits there turning from an occasional week or two into months.
When our own Paris stays overlapped and we spent more time with George and N\'ora, we
were two couples 
dining in a restaurant or in their apartment, and the talk of science always seemed to take a 
backseat to tales of a lifelong friendship, starting with those Alaska years. N\'ora and I 
were happy to sit back and listen.

%% file: albert_wai_kit_lau.tex

\phantom{raul}

\vskip -1cm 

\begin{figure}[h] 
\begin{center}
\includegraphics[width=0.6\textwidth]{%
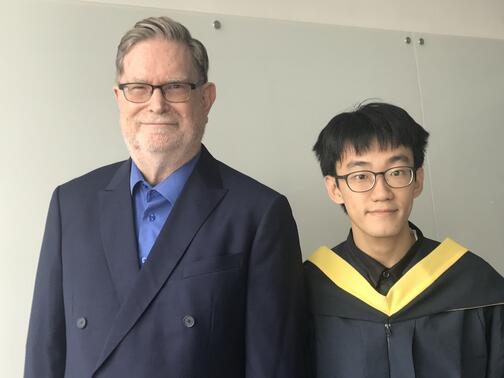}
\end{center}
\caption{%
\textbf{%
George Smoot and Albert Wai Kit Lau at IAS HKUST.}
[Photo credit: N\'ora Csisz\'ar]
}
\end{figure}

\begin{figure}
\begin{center}
\includegraphics[width=0.7\textwidth]{%
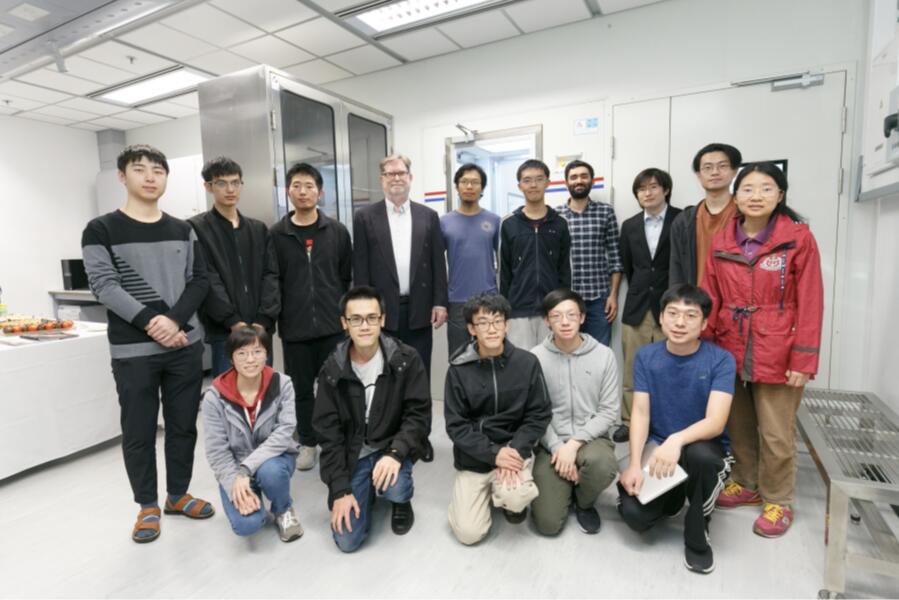}
\end{center}
\caption{%
\textbf{%
George Smoot in the lab with students at HKUST.}
}
\end{figure}


To the world, Professor George Smoot was a giant in cosmology and the Nobel Prize winner who helped figure out how our 
universe began. For those of us lucky enough to be his students, he was a mentor, a friend, and a wise elder. He was a 
brilliant scientist, but more importantly, he was someone who genuinely cared about us and fundamentally changed how we see 
the world.

I still remember how we first met. I was just a third-year undergraduate student 
at HKUST (Hong Kong University of Science and Technology) 
when I heard from other faculty members 
that Professor Smoot was coming to our university and was planning to take on students. I decided to be bold and sent him 
an email. To my surprise, he replied and told me to come to his office at the IAS for an interview the very next day.

That day turned out to be pouring rain. When I walked into his office, the very first thing he asked me was if I was wet. 
He then smiled and joked that if a student didn't carry an umbrella in such terrible weather, they probably didn't trust 
the weather report and might be ``too silly for research." After some friendly chatting, he agreed to take me on as an 
undergraduate research student. He gave me a project on the topic of instrument development for Ultra-Fast Astronomy (UFA). 
I had no idea at the time, but that project would become the main theme of my PhD and my future career.

George had an amazing ability to see the future of science long before anyone else did. When he introduced me to transient 
detection and optical intensity interferometry back then, almost no one in the community was looking in that direction. It 
was a lonely field. But George saw the hidden potential and strongly encouraged me to jump in. Today, as transient 
detection becomes a hugely popular and rapidly expanding field, it is incredible to look back and see that his early 
instincts were right.

During my PhD, we worked to bring those early ideas to life. I helped set up our new Quantum Optics for Astrophysics and 
Cosmology (QOAC) Lab and focused on my main project: the Single Photon Imager for Nanosecond Astrophysics (SPINA). Building 
an instrument from scratch during the COVID-19 pandemic was incredibly difficult, and we often had to meet through screens. 
Through it all George gave me a massive amount of freedom. He trusted me to try different directions, even when some 
turned out to be mistakes that cost us time and resources. Rather than stepping in to take over, he introduced me to his 
collaborators and friends, helping me gather the knowledge I needed to finally tackle the problems myself. He trusted us 
completely, but whenever we were truly stuck, his advice was always sharp and spot-on. He gave us the space to learn 
firsthand that breaking new ground means pushing our equipment to its absolute limits.

When we did talk, George loved keeping us on our toes. He had a brilliant, playful way of challenging our minds by dropping 
``Fermi questions" on us. These were fun but tough estimation problems designed to test our logical thinking. For example, 
he once asked us out of nowhere: ``How much of the air a human breathes in a lifetime was also breathed by one dinosaur?" 
He wasn't just being funny; he was actively training our intuition. To answer it, we had to quickly estimate the volume of 
the Earth's atmosphere, the size of a single human breath, how many breaths we take in a lifetime, and the breath of the 
dinosaur that shared the same air. He taught us how to break a seemingly impossible large and complex question into simple, 
manageable pieces. He was teaching us how to make quick estimates, cut through the noise, and spot the real physics before 
we wasted time on a bad idea.

He cared deeply about our growth outside the lab too. A couple of years ago, he brought me to the Lindau Nobel Laureate 
meeting. That trip was a milestone for me. Thanks to his guidance and introductions, I got to sit down and chat with some 
of the best minds in science. He used that experience to show me what it really means to be a scientist and how to figure 
out which frontier questions are truly worth dedicating a career to.

Despite his enormous achievements and global recognition, George never took himself too seriously. He was incredibly 
grounded and loved good food. Whenever we could finally meet in person, our lunch meetings were always a highlight of the 
week. Whenever he traveled around the world, he always made sure to bring back chocolates and candies for us. He even 
had a wonderful mischievous side. If there was another conference happening in our department building, he would joke with 
a twinkle in his eye that we should all sneak in just to grab some of the free food.

I will never forget the day we were chatting, and he casually mentioned he was flying back to the USA for a few days. We 
asked what academic meeting he was going to. He just smiled and said he was flying back to film a scene for 
the finale of the TV show \textit{The Big Bang Theory}.

George taught me how to understand the world. As I continue my career today, working on new telescope arrays and transient 
studies, his lessons remain the absolute foundation of everything I do. My ultimate tribute to him will be passing his 
legacy forward. I hope to guide my future students with the same warmth, exactness, and profound scientific curiosity that 
George so generously gave to all of us.
 

\begin{figure*}[h]
\begin{center}
\includegraphics[width=0.6\textwidth]{%
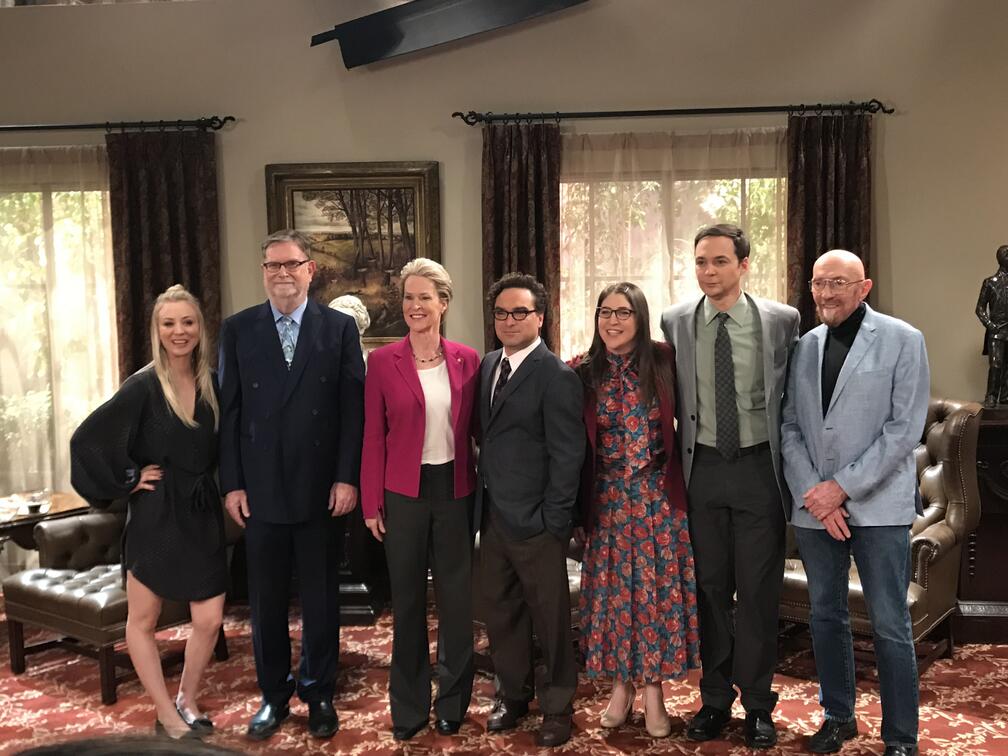}
\end{center}
\caption{%
\textbf{%
On set of US sitcom `The Big Bang Theory' Season 12, Episode 18, The Laureate Accumulation.} (February 2019) (left to right): 
Kaley Cuoco, George Smoot, Frances Arnold, Johnny Galecki, Mayim Bialik, Jim Parsons, Kip Thorne. 
[Photo credit: N\'ora Csisz\'ar]
}
\end{figure*}

%% file: seanLi.tex

%
%
%
%

\phantom{singini}

\begin{figure}[h]
\begin{center}
\includegraphics[width=0.7\textwidth]{%
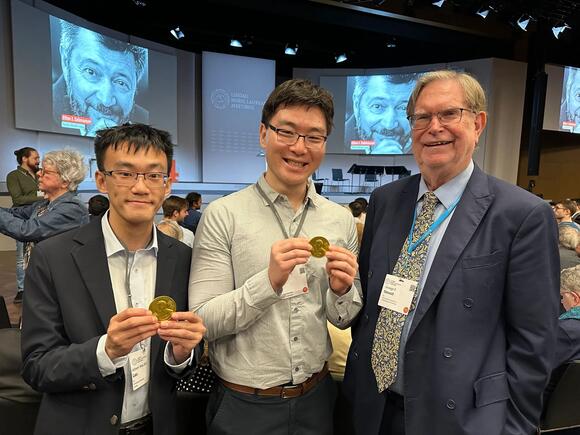}
\end{center}
\caption{%
\textbf{%
With Albert Wai Kit Lau and Sean (Siyang) Li at the 73rd Lindau Nobel Laureate Meeting (June 30, 2024).}
[Photo credit: N\'ora Csisz\'ar]
}
\end{figure}


I first met George when I was a freshman at the University of California, Berkeley. I went into undergrad with a 
fascination for cosmology and philosophy, and George's work with the cosmic microwave background deeply intrigued 
me. I decided to cold-email him to see if he was interested in mentoring a student.

To my complete surprise, he responded. He gave me a few physics problems to solve and offered to meet me on 
campus. Not wanting to inconvenience such a prominent figure, I insisted that I could meet him at his office at 
the Lawrence Berkeley National Lab. So I made the trek up the hill.

I vividly remember the physics problems he had given me to solve prior to that meeting. One of them asked me to 
model a Gaussian communication beam sweeping from Earth to a Mars colony, calculating the signal an observer would 
see as the beam swept over them from up to a thousand light-years away. Despite my best efforts, I failed to solve 
a single one. Yet instead of turning me away, George still took me on. At his stage in his career, he had 
absolutely no obligation to take a chance on an unproven freshman, especially one who had stumbled on his initial 
test. Yet he did so with open arms, ready to nurture my academic path.

I began by compiling parts lists for what later became the Ultra-Fast Astronomy (UFA) project he led. I contacted 
manufacturers like Hamamatsu to test Silicon Photomultipliers (SiPMs) that could be used to search for 
sub-millisecond transients. While part of this was to search for one of George's favorite signals---signs of 
extraterrestrial life, the broader concept was even more profound. By opening up a new phase space, in this case 
temporally, we could discover entirely new astrophysical phenomena.


\begin{figure}[h]
\begin{center}
\includegraphics[width=0.55\textwidth]{%
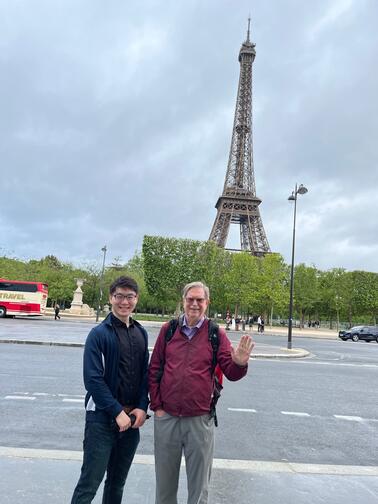}
\end{center}
\caption{%
[Photo credit: Sean Li]
}
\end{figure}

This early research experience set the tone for his mentorship style. George granted me an incredible amount of 
independence, but he was always there whenever I had questions or needed help. He nurtured my drive to design and 
push projects forward on my own, while expertly providing the gentle guidance necessary to put me back on track 
whenever I strayed.

Working with George also led me to visit him in Paris at the PCCP (Paris Center for Cosmological Physics). 
I recall that shortly after arriving, George 
warmly welcomed me. He proudly showed off his Fitbit and took me along the path he used to walk to the 
department---an hour-long journey!---taking the time to show me the Notre Dame Cathedral and other local 
landmarks. Later, I traveled to the Hong Kong University of Science and Technology and Nazarbayev University to 
continue working with George to characterize the SiPMs (Silicon Photomultipliers). There I set up dark boxes and ran extensive calibration 
tests, and George cheered me on as I published my findings in the SPIE proceedings and presented our research.

Even as I moved into the final stages of my doctoral work, George remained a fixture in my life. I had the 
incredible opportunity to meet him again at the Lindau Nobel Laureate Meeting in Germany before graduating with my 
PhD. Seeing him in that element, surrounded by the world's greatest minds, was awe-inspiring, but what stuck with 
me most was how beautifully human he remained. Here he was, a Nobel laureate, yet we spent our time walking to get 
ice cream together with his partner, N\'{o}ra, and dancing at the party on the boat trip to and from Mainau 
Island. He later attended my PhD thesis defense virtually, a continued testament to his dedication.

Just last summer 2025, I visited George in Paris while I was in Europe for various conferences. I will never forget the 
fascinating conversations we had during those two days, which ranged from the future of AI in science and society 
to space elevators and digital twins. I will also never forget how on the second day I met up with him at his 
apartment near the Eiffel Tower and we walked over to have brunch at McDonald's. We ordered Egg McMuffins, and I 
insisted on paying for both of our meals as a small token of my gratitude. Of course, their credit card machine 
happened to break and I ended up about two euros short in cash. George graciously laughed and pitched in. 
Afterward we strolled through an outdoor market under a bridge, where he excitedly introduced me to his favorite 
cheeses, including one unforgettable truffle cheese.

George was, and will always remain, a central figure in my life. He provided an incredibly impactful, unique 
mentorship that profoundly broadened my horizons. It is not an understatement to say I wouldn't be where I am 
today without his guidance. His memory, and his impact, will forever remain a guiding force in my life and my 
career.

\begin{figure}[h]
\begin{center}
\includegraphics[width=0.60\textwidth]{%
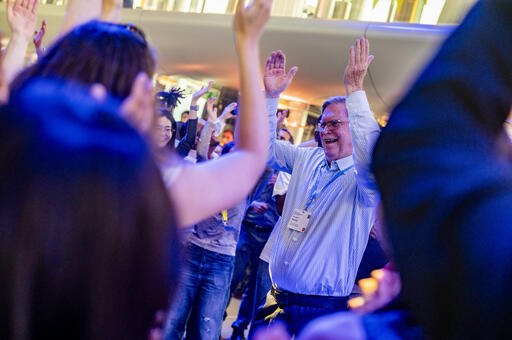}
\end{center}
\caption{%
\textbf{%
George dancing on boat en route to Mainau Island.} (2024) 
[Photo Credit: Lindau Nobel Laureate Meetings]
}
\end{figure}

%% file: tao_liu.tex

We celebrate the life of Prof.~George Smoot, a remarkable individual who touched the lives of so many. It 
is an honor to share a few words in memory of my dear friend, collaborator, and colleague.

As a Nobel laureate, Prof.~George Smoot is renowned for his groundbreaking work in discovering the anisotropy of the cosmic 
microwave background. As his colleague at 
Hong Kong University of Science and Technology 
(HKUST) since 2016, I had the privilege of working alongside him on the 
astronomical detection of dark matter. Together we proposed in 2019 utilizing linearly polarized pulsar light to detect one of the 
major dark matter candidates, ultralight axion-like dark matter. This work directly inspired the subsequent 
proposal to establish ``pulsar polarization arrays'' as an innovative astronomical tool for studying fundamental physics in 
2021, which has garnered significant attention and continues to have a growing impact. George's passion for 
science inspired everyone around him.

Beyond our fruitful collaborations, George was a strong advocate for junior researchers. In 2020 I had the 
opportunity to lead a team grant application focused on researching the nature of dark matter using astronomical tools. He 
wholeheartedly offered his support, sharing his experiences and warmth, and we ultimately succeeded in obtaining the grant. 
He had a remarkable ability to make everyone feel valued and appreciated.

As we remember George, let us honor his legacy by continuing to pursue our passions and support one 
another, just as he did. Although George may no longer be with us physically, his spirit will always remain in 
our hearts and in our work. Thank you for allowing me to share my thoughts and memories. Let us cherish the time we had 
with George and carry forward his light in our lives.


%% file: omar.tex
\phantom{Omar López-Cruz}

\begin{figure}[h]
\begin{center}
\includegraphics[width=0.5\textwidth]{%
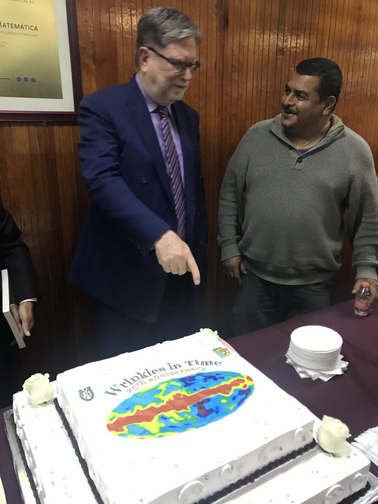}
\end{center}
\vskip -0.15in
\caption{%
\textbf{%
With Omar L\'opez-Cruz in Guadalajara, Mexico.} (2018)
[Photo credit: N\'ora Csisz\'ar]
}
\end{figure}

\begin{figure}
\begin{center}
\includegraphics[width=0.7\textwidth]{%
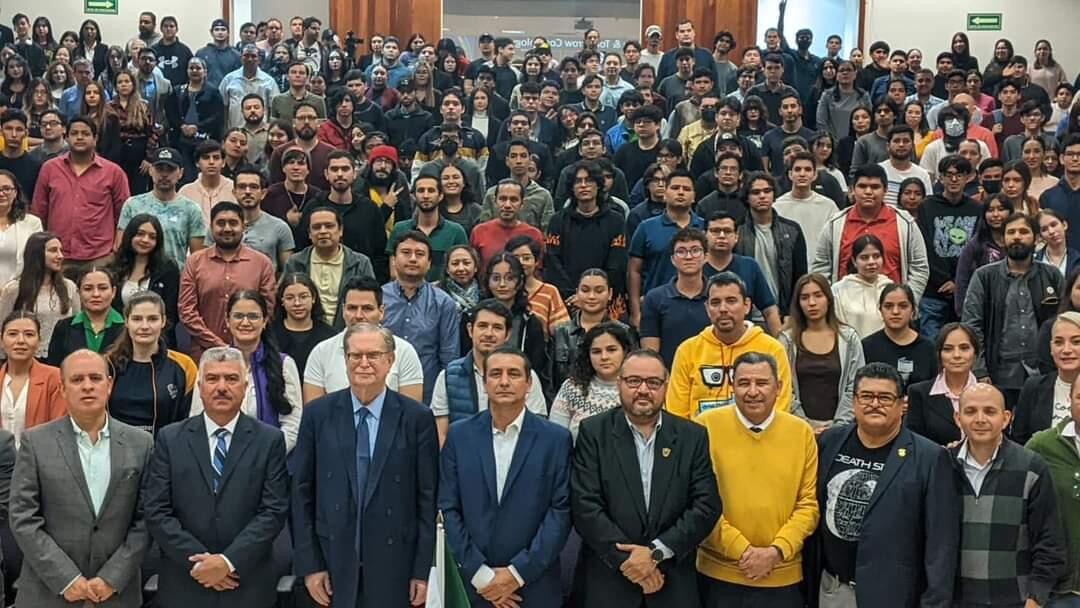}
\end{center}
\vskip -0.15in
\caption{%
\textbf{%
At Universidad Autónoma de Sinaloa (UAS) (December 12, 2023).}
[Photo credit: UAS]
}
\end{figure}

Nobel laureate George Fitzgerald Smoot III was born on February 20, 1945. He is best known for 
directing the Differential Microwave Radiometers (DMR) experiment on board the Cosmic Background 
Explorer (COBE) satellite and for leading the team that discovered primordial temperature 
fluctuations in the cosmic microwave background (CMB) in 1991. John Mather was the principal 
investigator of COBE’s Far Infrared Absolute Spectrophotometer (FIRAS), which measured the peak of 
the CMB spectrum. For these breakthroughs, Smoot and Mather were awarded the 2006 Physics Nobel Prize.
Smoot, Mather, and their colleagues from COBE ushered in a new epoch called high precision 
cosmology.

George was an intelligent and generous person with a very jovial and attractive personality, 
who after receiving the Nobel Prize, decided to make cosmology popular worldwide. As part of his 
efforts, he lectured widely and made television appearances, including two on the comedy show 
\textit{The Big Bang Theory}. George was also very popular in México, as he often visited us 
for scientific 
engagements, book presentations, lectures, and conferences. Other sporadic visits included spending 
holidays in Cabo San Lucas, Baja California Sur, México.

With Axel de la Macorra (IF--UNAM), Jorge Cervantes (ININ), Tonatiuh Matos (Cinvestav), and Darío 
Núñez (ICN--UNAM), George helped create the Instituto Avanzado de Cosmología (IAC) in 2007 in 
Mexico. In 2009 the IAC began a highly acclaimed series of winter schools known as Cosmology on 
the Beach (COTB): Essentials of Cosmology for the Next Generation. Nine COTB schools were 
organized between 2009 and 2024. The first COTB school was organized at Cabo San Lucas, his 
favorite.

In January 2014, COTB returned to the Cabo San Lucas area. The organizing committee asked me to 
make arrangements for George’s public lectures. I suggested the nearby city of La Paz in Baja 
California Sur. On January 23, George spoke about ``The New Map of the Cosmos'' before a packed 
Teatro de la Ciudad. In 2018 I decided to publish a Spanish translation of his book 
\textit{Wrinkles in Time/Arrugas en el tiempo}, 
which recounts George’s story about his discovery of the temperature anisotropies in the CMB. 
I searched for a publisher and funds to secure the book’s publication.


On December 4, 2018, George returned to La Paz to give a lecture and sign his book. We delivered 
500 books to the public for free on that occasion as part of a science dissemination program that 
we named La Paz, Puerto de la Ciencia. Before reaching La Paz, George stopped at the Feria 
Internacional del Libro de Guadalajara (FIL) and the Escuela Superior de Física y Matemáticas 
(ESFM--IPN). He again visited Mexico in December 2023 and we went to the Universidad Autónoma de 
Sinaloa (UAS), where he gave a brilliant lecture as well as a Q\&A session with the students. 
He was invited to throw the first pitch at the Tomateros baseball/beisball game in their stadium at Culiac\'an!

Gracias George!

\newpage

\begin{figure*}[h]
\begin{center}
\includegraphics[width=0.65\textwidth]{%
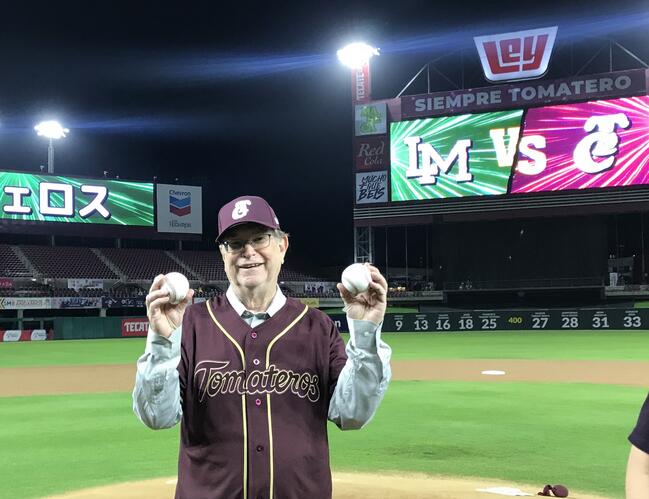}
\vskip 0.15in 
\includegraphics[width=0.65\textwidth]{%
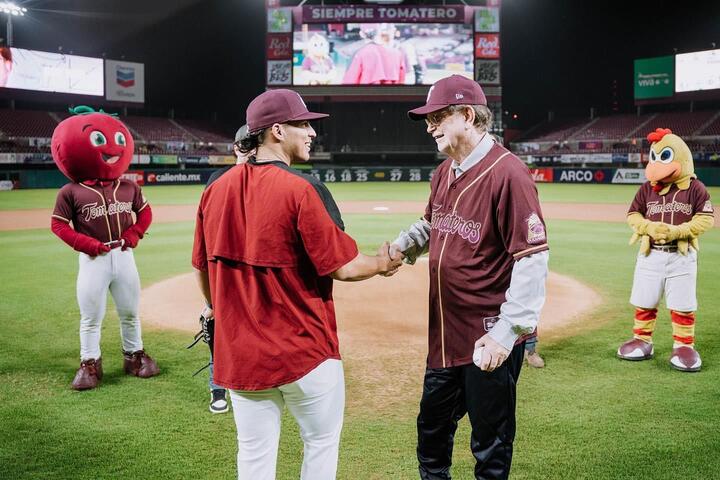}
\end{center}
\caption{%
\textbf{George supporting the Tomateros de Culiac\'an Beisball team in Mexico (2023).}
[Photo credit: N\'ora Csisz\'ar]
}
\end{figure*}


\newpage 

\begin{figure*}[t]
\begin{center}
\includegraphics[width=0.69\textwidth]{%
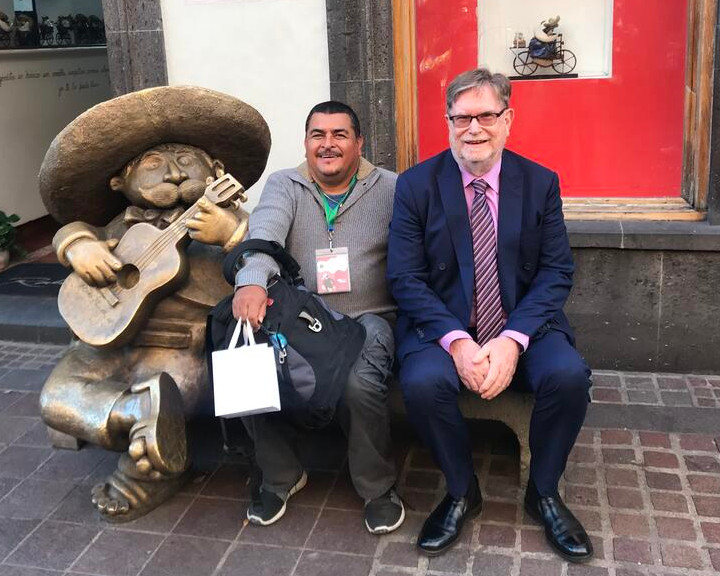}
\end{center}
\caption{%
\textbf{George with Omar.}
[Photo credit: N\'ora Csisz\'ar]
}
\end{figure*}

\begin{figure*}
\begin{center}
\includegraphics[width=0.69\textwidth]{%
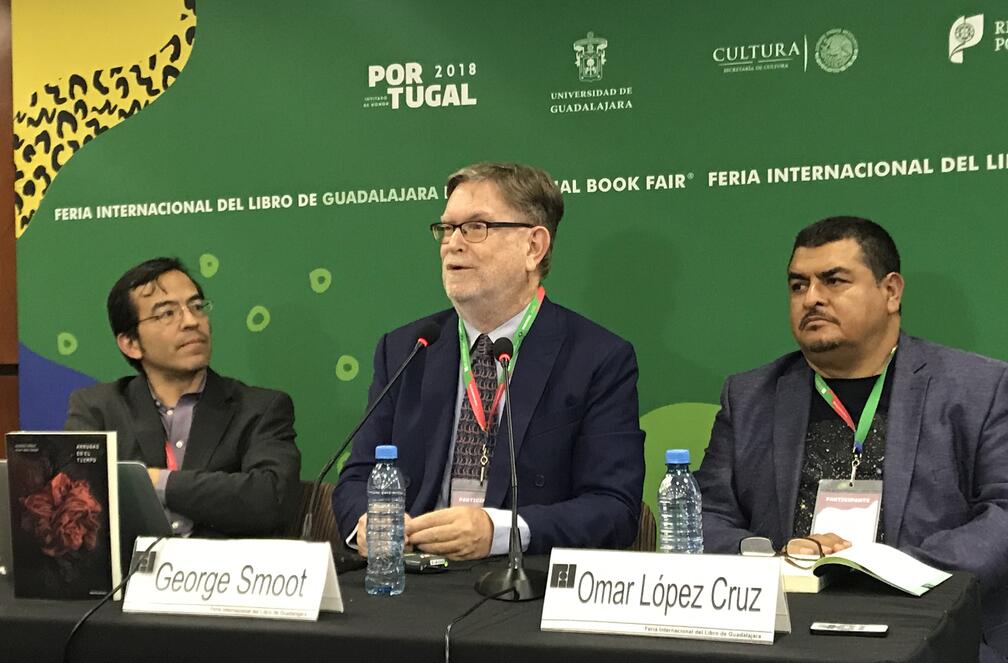}
\end{center}
\caption{%
\textbf{Tom\'as Granados Salinas, George Smoot, Omar L\'opez-Cruz at the Feria Internacional del Libro de Guadalajara.} (2018)
[Photo credit: N\'ora Csisz\'ar]}
\end{figure*}

\clearpage 
\newpage

%% file: lubin3.tex
I first met George in early 1974 when I was a junior physics and mathematics major at UC Berkeley. I was 
invited to come to Lawrence Berkeley Laboratory, now Lawrence Berkeley National Laboratory (LBNL), to do 
research with what was called Group A. George was a research scientist and a member of this group. George was a 
good fit given the dynamics of Group A. The Group A environment was unlike any I had seen before and has never 
been replicated since. In today’s language, ``it was not PC = Politically Correct, but it was IP = Incredibly 
Productive.'' Group A was largely split into accelerator particle physics and non-accelerator physics parts, 
though these were not absolute boundaries, with a healthy intellectual mix between them. George was in the 
non-accelerator portion of the group when I met him. The members of the group then were the nominal head Luis 
Alvarez (hence the internal affectionate name: group A), Richard Muller, Frank Crawford, Andy Buffington, and 
Charles Orth. All were researchers and not faculty with the exceptions of Luis Alvarez and Frank Crawford. The 
group was truly a gem in terms of creativity, breadth of capability, and some serious humor. Largely due to 
Luis’ extensive Department of Defense (DoD) contacts, the use of a high altitude U-2 aircraft was proposed for 
the observations. This program, in which I was involved both as an undergrad and later on as a graduate student,
was extremely successful.
I worked closely with George from 1974–2004 on multiple projects, 
some where George was PI such as the COBE Differential Microwave Radiometers (DMR) and the search for 
polarization of the CMB (my PhD thesis), and some where I was PI such as the TMSS (3mm Sky Survey) and the 
subhorizon scale ACME (Advanced Cosmic Microwave Explorer) balloon missions from 1988–1995.

(1976–1977) I initially started working back in Group A on a balloon-borne magnetic spectrometer to measure the 
galactic cosmic ray confinement time scale by measuring the ${}^{10}Be$ isotope using a very large superconducting 
magnetic spectrometer. This payload was very large and heavy for a balloon payload of about 3 metric tons. This 
experiment was a follow-up to the HAPPE (High Altitude Particle Physics Experiment) balloon program that Group 
A had previously embarked on. 

(1977–1979) George suggested working on searching for polarization in the CMB as well as working on the Aether 
Drift\footnote{%
 At the time, the CMB dipole, due largely (if not almost entirely) to the relative velocity between the observer and
the CMB rest frame, was sometimes called the ``aether drift,'' or ``new aether drift,'' even though according
to our modern understanding this effect does not involve any ``aether.'' Rather the matter-radiation (or more
generally, stress-energy) content of the universe explicitly breaks Lorentz invariance.} %
experiment. This was a wonderful collaboration. George gave me a Princeton thesis from the Wilkinson 
group on a recent polarization search and suggested I read it and look for a more sensitive method. Having worked 
on the Aether Drift experiment and having looked into using cryogenic receivers including a 90 GHz maser, I was 
already fairly well-versed in the technological aspects and fairly quickly had a CMB polarimeter put together 
that we operated on the roof of Building 50 at LBL. It worked extremely well and would constitute my thesis 
work. George and I worked closely on this. George also graciously invited me to join the nascent COBE 
collaboration in 1977. For this I am forever grateful. George and I were producing results within twelve months 
of starting the program and began publishing.

In 1978 George and I also took both the U-2 and the CMB polarimeter to Lima, Peru to get southern hemisphere 
sky coverage data. Both instruments worked flawlessly. I finished my PhD in March 1980 and put most of my 
energy into TMSS and COBE, and began pondering follow-up missions.

(1981–1983) I presented the idea of changing the DMR from room temperature to the cryogenic detectors I was 
flying on the TMSS payload. George and the entire COBE team was very supportive, and with the first 
demonstrated flight data from TMSS, it was clear that this would yield a much more sensitive DMR sky map. 
George and I flew TMSS and the maps were already clearly excellent. I arranged to have a southern hemisphere 
flight of TMSS in Brazil with the idea of having a flight to get full southern sky coverage and thus combined 
with full northern sky coverage would yield a full sky map of the CMB. George and I departed for Brazil in 
February 1983. The flight went perfectly until the end, when the payload was to separate from the balloon with 
an explosive charge and then the payload was to descend on a parachute. But the explosive charge failed to 
separate the payload and parachute from the balloon. Thus began an agonizing twelve hours of trying everything 
to force separation and landing. Nothing worked. It looked like the balloon would carry the payload into the 
ocean, lost forever. I left Brazil quite dejected, as I had worked years on this and now it was gone. The lesson 
was never make just one instrument if you are going to fly it on a balloon. In fact the payload came down in 
the jungles of Brazil. This is another ``adventures in cosmology'' story. Remarkably, the onboard data tape, 
having sat in the jungles for more than two years, was still usable. The data were downlinked via telemetry, 
but the flight tape, which had superior information, was on the balloon payload. I was able to make a map of 
about 60\% of the southern sky but the payload was gone and it was getting too close to the expected launch of 
COBE in 1986.

COBE was ``going to fly'' in 1986, so I decided to focus on searching for subhorizon fluctuations (subdegree) in 
the CMB as well as to focus on making sure the cryogenic detectors worked for the COBE DMR (superhorizon scale). 
These were some of the most difficult years at LBL but the outcome was the production of the ACME payload, which 
was designed to both fly on a balloon and could be used on the ground, particularly at the South Pole. The 
detectors I used were superconducting junctions at 3mm wavelength (from Tony Stark at Bell Labs), which 
exhibited exceptional sensitivity (about five times above the quantum limit) when operated at 4K and below. To 
achieve subdegree (subhorizon) performance, we used a 1 meter mirror provided by Robert Wilson at Bell Labs.

(1986) At 11:39 EST on January 28, 1986 the Challenger shuttle (STS-51-L) exploded. It was the 10th flight for 
the Challenger shuttle and the 25th flight for the shuttle program. This was a tragedy that while always a 
possibility never seemed likely, particularly given the many successful launches and recoveries of the shuttle 
program. All seven crew members---Francis R. Scobee, Michael J. Smith, Ronald McNair, Ellison Onizuka, Judith 
Resnik, Gregory Jarvis, and Christa McAuliffe---died. It was catastrophic for the families of these astronauts 
and for the entire NASA community as well as for the nation. The shuttle program was shut down for 32 months to 
understand and rectify whatever caused the failure. For the COBE team, there was an incredible sadness arising from 
the 
tragedy. We also knew that we would not fly on the shuttle in 1986 as planned. We gradually realized as NASA 
dealt with the implications that we would have to go on a dedicated rocket launch, not on the shuttle.

Late in in 1989, COBE was launched on a Delta rocket. Most of the COBE team members were there including George 
and my entire family. It was an absolutely beautiful launch and we were all relieved when COBE properly 
achieved orbit. It took time for the instruments to come to equilibrium. With our ACME superconducting 
detectors we were able to get higher overall sensitivity at the South Pole than a year of DMR on orbit. It 
would take almost 2.5 years after launch before we were able to see the superhorizon ($>7$ degrees) CMB 
signature.

Looking back this was an incredible period of exploration on many different levels. Working both with George, 
and George working with me as well as the independence of both of us, led to many discoveries and reflected the 
reality of a life in the sciences in its many forms. I wish I could go back and share with George all that 
unfolded for both of us. I greatly miss the wonderful days spent at LBL, particularly the early years when the 
streams of life ran so quickly and the possibilities seemed endless. But life was always to be finite for all 
of us on this journey. If I could speak once more with George, it would be to say that ``life is much more than 
scientific discovery. It is about all the beautiful possibilities that life presents to us. We should both use 
them wisely so we will remember fondly the days of our youth.'' I will and do miss George. I am thankful to 
those that were close and so good to him in his life.

\clearpage
\newpage 

\begin{figure}
\begin{center}
\includegraphics[width=0.7\textwidth]{%
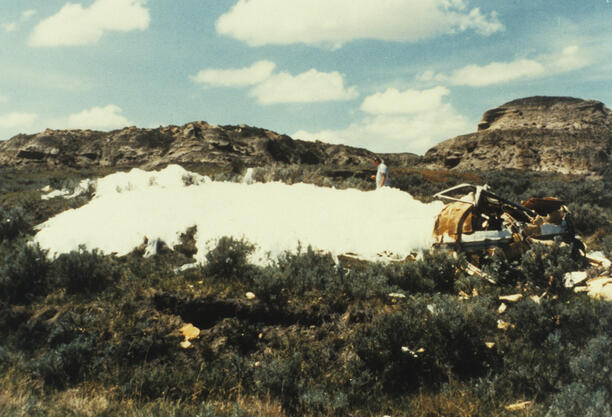}
\end{center}
\caption{%
\textbf{Badlands of North Dakota.}
Crash site remains of the LBL superconducting 
cosmic ray spectrometer designed to
measure the ${}^{10}Be$ isotope fraction in order to determine the
cosmic ray Galactic confinement time.
[Courtesy of Philip Lubin]
}
\end{figure}

\begin{figure}
\begin{center}
\includegraphics[width=0.7\textwidth]{%
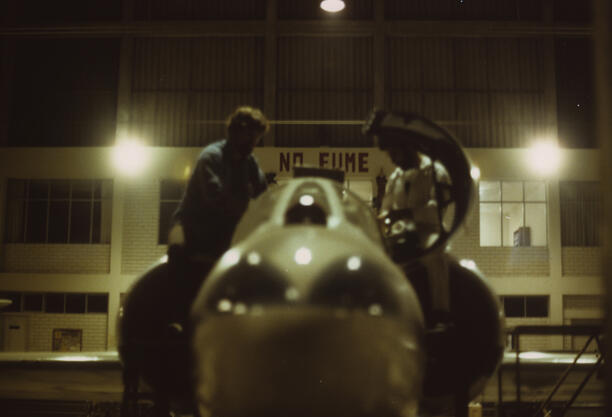}
\end{center}
\caption{%
\textbf{1978 Lima, Peru.}
Aircraft hangar with U-2 aircraft
for southern hemisphere measurements for further
characterizing the first-order dipole with
the ``aether drift'' experiment. George Smoot (left) 
and Philip Lubin (right). 
[Courtesy of Philip Lubin]
}
\end{figure}

\begin{figure}
\begin{center}
\includegraphics[width=0.7\textwidth]{%
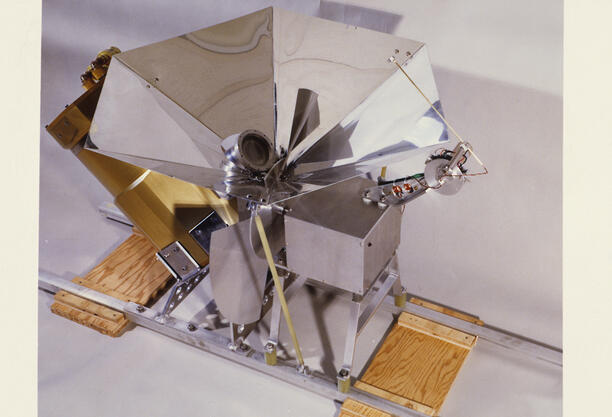}
\end{center}
\caption{
\textbf{1980 TMSS liquid helium cooled detector and radiometer.}
It was this cooled
detector design that I proposed we adopt for the COBE DMR instead 
of the room temperature detectors baselined. This was accepted by
the COBE SWG and implemented allowing COBE to detect the superhorizon
scale fluctuations. Had we not done this, COBE would not have been able to
make the detection. 
[Courtesy of Philip Lubin]
}
\end{figure}

\begin{figure}
\begin{center}
\includegraphics[width=0.7\textwidth]{%
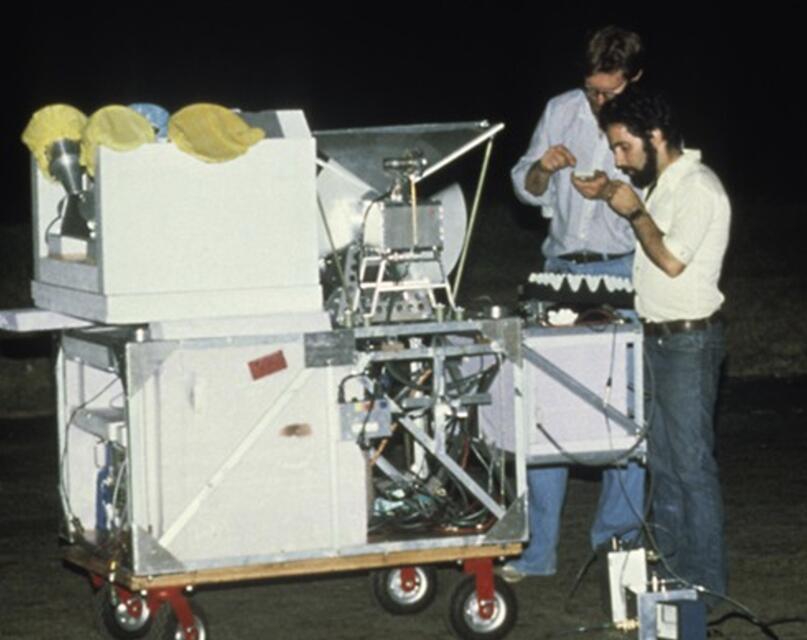}
\end{center}
\caption{%
\textbf{1981, Palestine, Texas.}
Testing the TMSS payload on the Princeton group gondola prior to launch a few days later.
George Smoot (background), Philip Lubin (foreground). 
[Courtesy of Philip Lubin]
}
\end{figure}

\begin{figure}
\begin{center}
\includegraphics[width=0.7\textwidth]{%
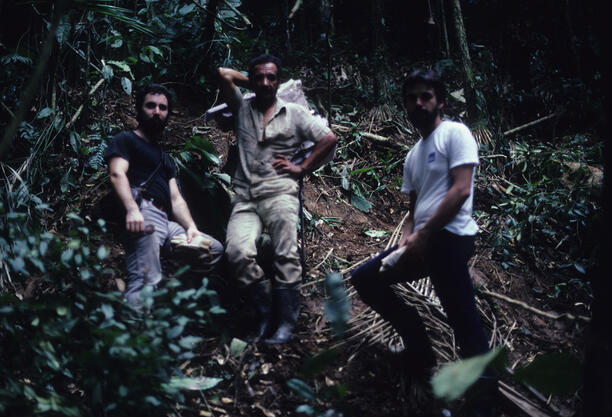}
\end{center}
\caption{%
\textbf{Recovery of the TMSS in the Brazilian jungle.}
[Courtesy of Philip Lubin]
}
\end{figure}

\begin{figure}
\begin{center}
\includegraphics[width=0.7\textwidth]{%
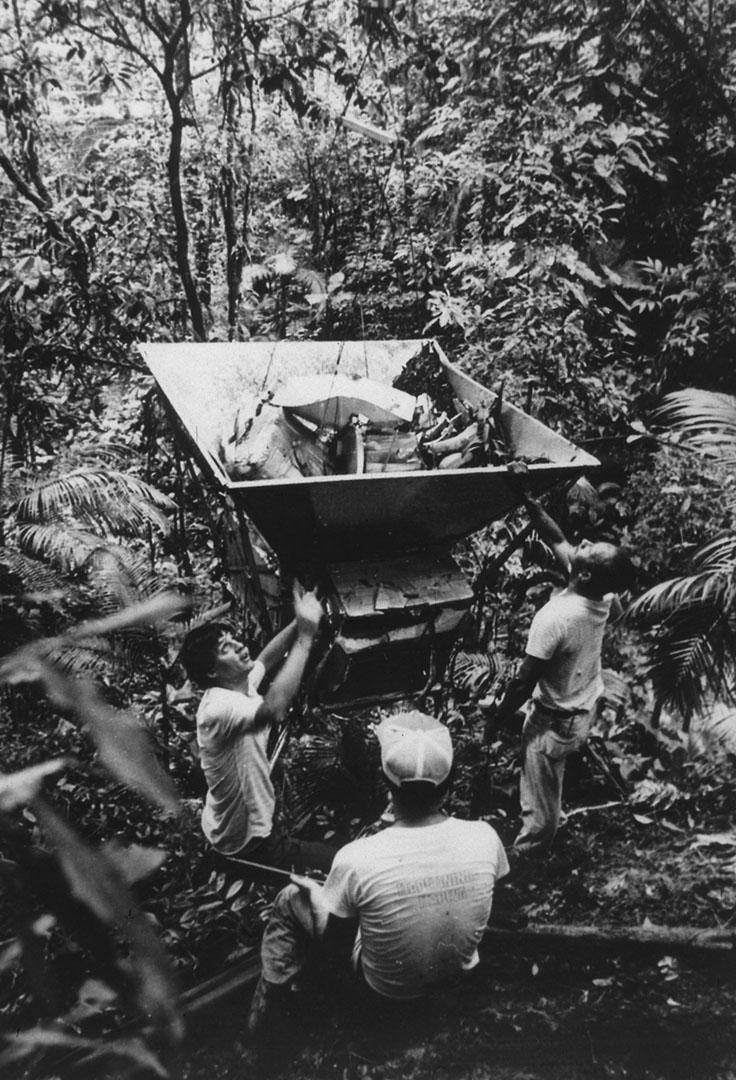}
\end{center}
\caption{%
\textbf{1985, Tapirai, Brazil.}
The TMSS payload ``lost in the jungle.''
All data onboard after three years in the jungle was still intact.
[Courtesy of Philip Lubin]
}
\end{figure}

%

\begin{figure}
\begin{center}
\includegraphics[width=0.7\textwidth]{%
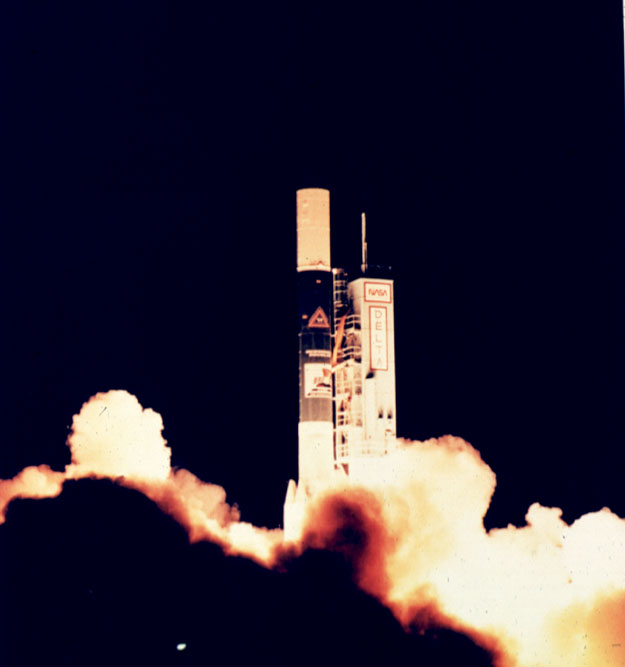}
\end{center}
\caption{%
\textbf{COBE launch on November 18, 1989 from
Vandenberg Air Force Base.}
[Courtesy of Philip Lubin]
}
\end{figure}

\begin{figure}
\begin{center}
\includegraphics[width=0.7\textwidth]{%
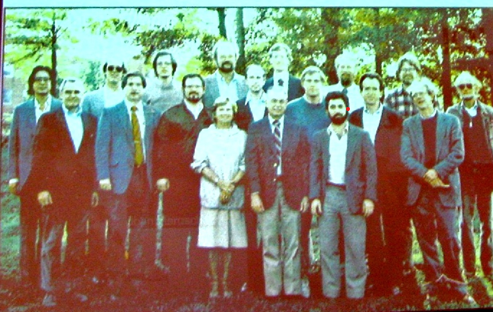}
\end{center}
\caption{%
\textbf{1990, COBE SWG at Goddard Space Flight Center (Greenbelt, Maryland).}
Back row:
George Smoot (upper right); front row: Rai Weiss (center), Philip Lubin (right).
[Courtesy of Philip Lubin]
}
\end{figure}

\begin{figure}
\begin{center}
\includegraphics[width=0.7\textwidth]{%
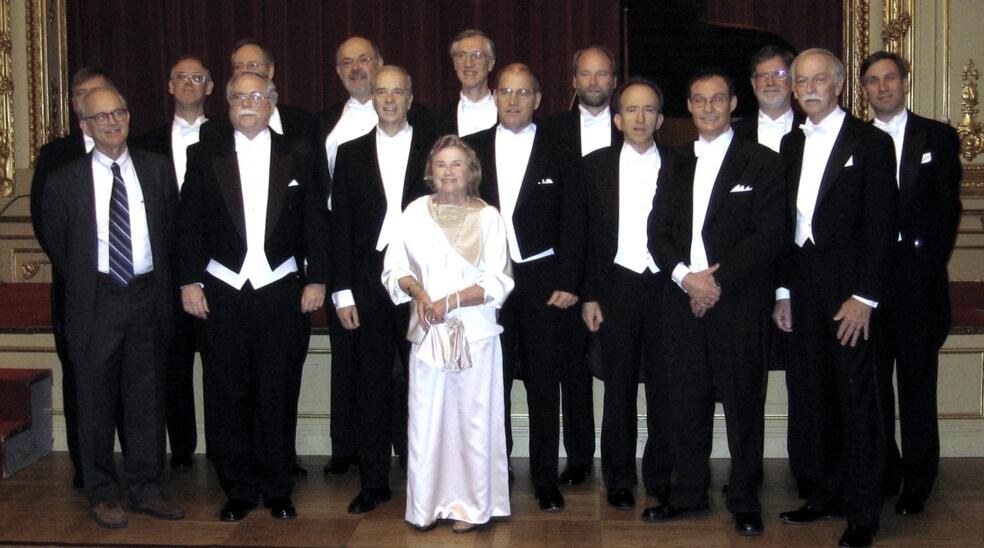}
\end{center}
\caption{%
\textbf{2006 Stockholm.}
COBE Science Working Group (George third from right).
[Courtesy of Philip Lubin]
}
\end{figure}

\begin{figure}
\begin{center}
\includegraphics[width=0.7\textwidth]{%
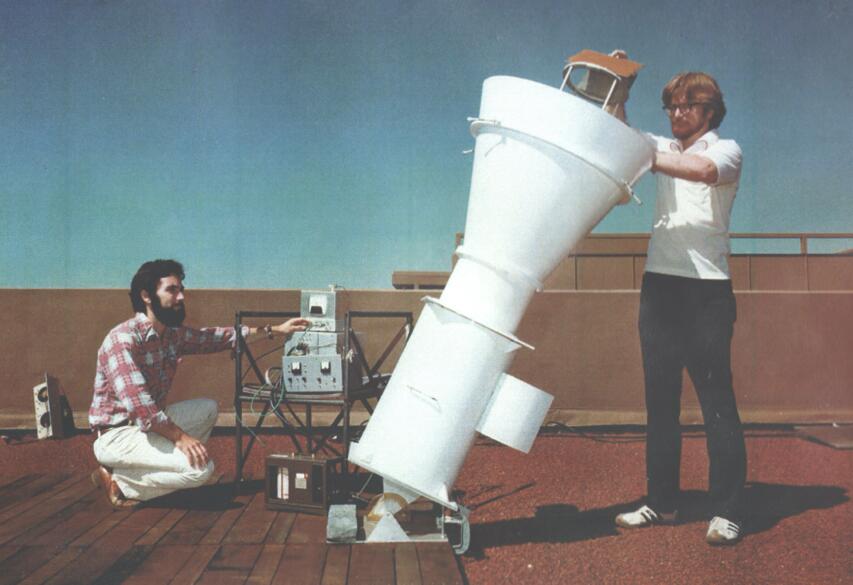}
\end{center}
\caption{%
\textbf{CMB Polarimeter.}
Testing on roof
of Building 50, LBL (George right, Phil left).
[Courtesy of Philip Lubin]
}
\end{figure}


\clearpage
\newpage

%% file: luk.tex
\phantom{singini}

\begin{figure}[h] 
\begin{center}
\includegraphics[width=0.9\textwidth]{%
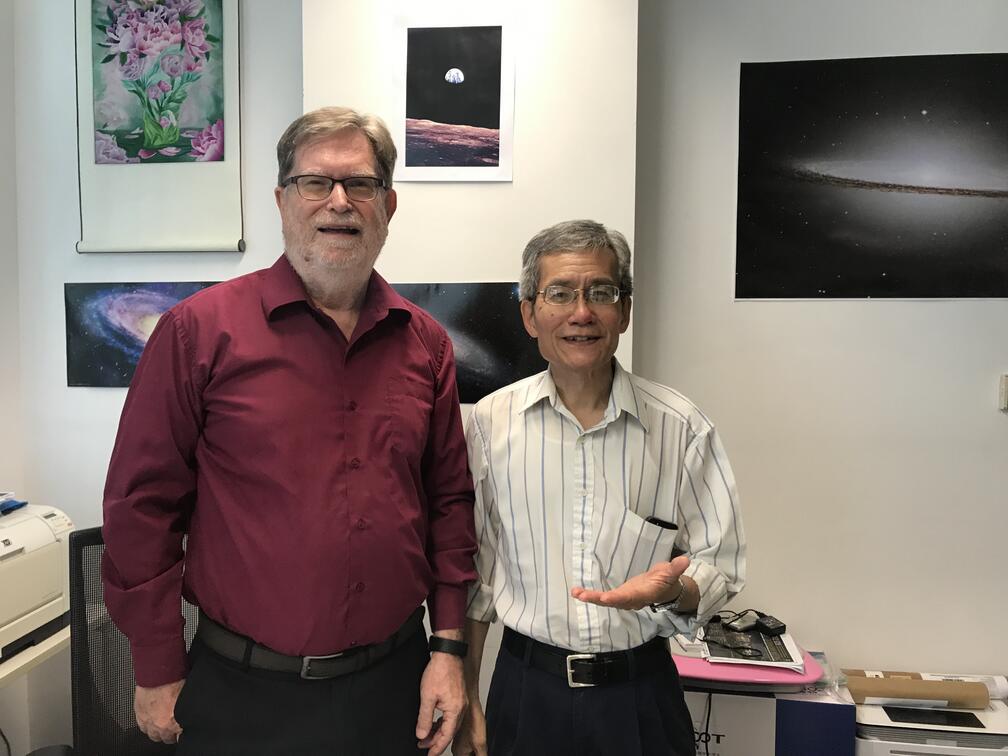}
\end{center}
\caption{%
\textbf{%
With Kam-Biu Luk at HKUST.}
[Photo credit: N\'ora Csisz\'ar]
}
\end{figure}

George was a wonderful colleague and friend. In the early days, his office at 
Lawrence Berkeley Laboratory (LBL) was only two offices 
from mine. George came over 
from time to time to educate me about cosmology, about the COBE analysis, and about the importance of having a Plan B in 
life. After a reshuffling of offices, by chance George’s new office was right across the hallway from mine. We continued 
to have fun talking about science, and yes, more about Plan B in life. I still remember the morning when the Nobel Prize 
was announced in 2006. It was eight in the morning. George told me that he did not get any sleep as his phone kept on ringing. 
I told him to swallow his breakfast fast before the film crew showed up. Well, it was too late. George’s breakfast became 
his lunch. What a memorable day!

George and I co-taught the lower-division electricity and magnetism on 
the UC Berkeley campus for a few years. We asked for this 
arrangement so that we could ping-pong our hectic traveling schedules. George certainly was a fantastic teammate and the 
students loved him.

It is a fortuity that we both got connected again as colleagues at the 
Hong Kong University of Science and Technology 
(HKUST). Although we did not overlap as much as we did at 
Berkeley, my wife Saidee and I greatly appreciated the opportunities of having lunch and dinner gatherings with N\'ora 
and George that we did not have in the States.

A few months after the passing of George, Tony Spadafora asked me to help with going through some of the scientific documents stored in 
George’s LBL office to decide on what should be sent to the LBL archive. In the process, I learned that George had spent several years 
in the nineties designing the next-generation, large-scale facility to succeed AMANDA for detecting ultra-energetic cosmic neutrinos. 
From this work emerged the IceCube Neutrino Observatory at the South Pole. He played an active role in this effort. For example, 
he wrote codes to estimate the rates of cosmic ray muons and neutrinos. Clearly George loved science and was driven by science.

I was extremely fortunate to have overlapped with George in space and time. His upbeat attitude was certainly contagious and greatly 
appreciated. He is dearly missed.

\begin{figure*}[h]
\begin{center}
\includegraphics[width=0.995\textwidth ]{%
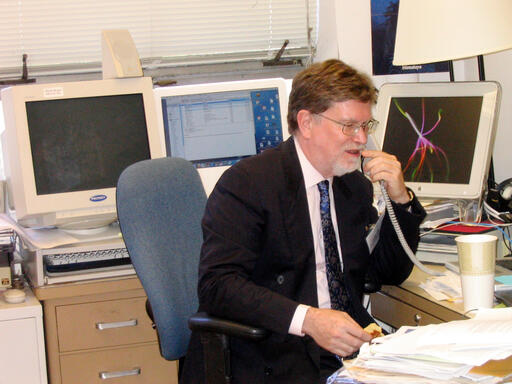}
\end{center}
\caption{%
\textbf{%
George Smoot in his LBL office speaking
with a reporter.} (3 October 2006)
[Photo credit: Kam-Biu Luk]}
\end{figure*}

%% file: mather.tex
\phantom{singini}

\begin{figure}[h] 
\begin{center}
\includegraphics[width=0.7\textwidth]{%
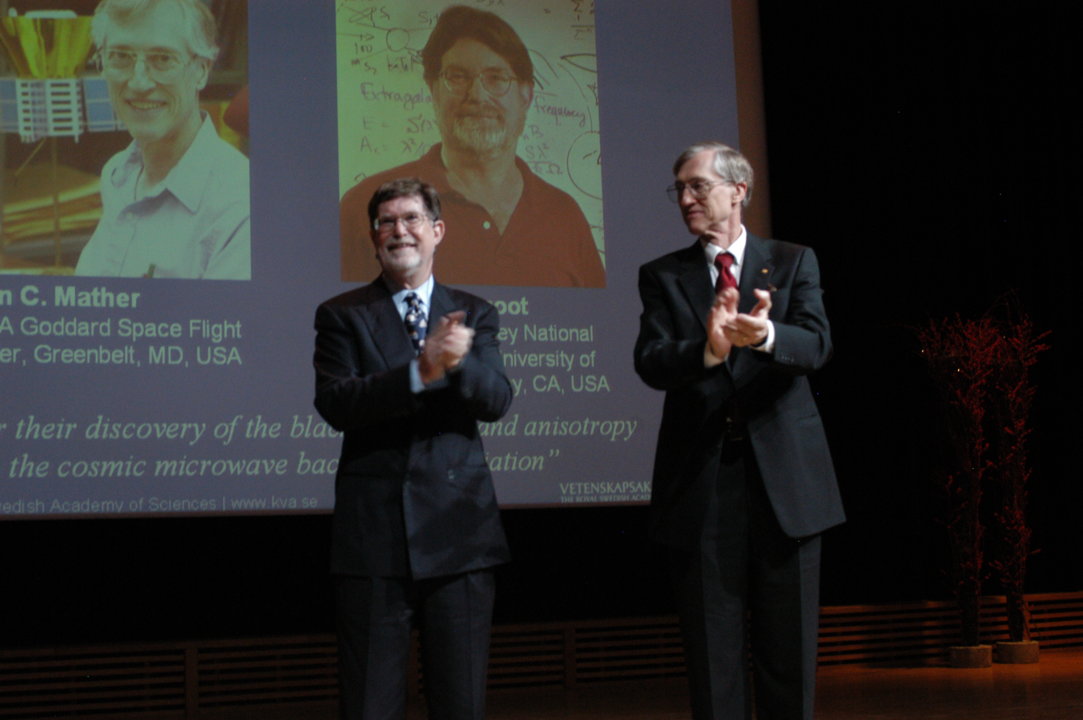}
\end{center}
\caption{%
\textbf{%
George Smoot and John Mather after delivering their Nobel Lectures at the Aula Magna.}
(Stockholm University, December 8, 2006).
[Photo credit: Mirek Labedzki. Nobel Media AB]}
\end{figure}

I think I met George in 1976, the year that I moved to NASA Goddard in Greenbelt, Maryland,  
to start the Cosmic Background Explorer (COBE) mission. In 1974, 
NASA had received three proposals to measure the cosmic microwave background radiation, my team's proposal, one from 
Berkeley, and one from Jet Propulsion Laboratory. The Berkeley team came from the lab of Luis Alvarez, where Rich Muller 
George Smoot, and Marc Gorenstein had been working on the CMB. NASA asked us to form a new team from members of the 
three competing teams, so we did. With NASA Headquarters backing from Nancy Boggess, we started talking. After we came to a 
general agreement about what the COBE mission should do, we had to decide who would do what. We elected George to lead the 
DMR, the Differential Microwave Radiometers, that would measure the hot and cold spots in the map of the cosmic background 
radiation. George had ideas, he had enthusiasm, he was young, which was very important as it turned out, and he had the 
backing of a major group at Berkeley. We all agreed with the general concept and agreed to work as a team and share in the 
publications and glory.

Well, nothing’s ever simple. We didn’t quite agree about things. We asked George to work with an engineering team at 
NASA 
Goddard, to make sure we were all working under one roof with experienced space people, and we hired a Deputy PI, Chuck 
Bennett. This was a bit tricky for everyone; who’s in charge here? Sometimes, George was pretty annoyed with our 
engineering team, and vice versa. We had funding shortages and more time, so we took advantage to improve the DMR design. Then 
we had to redesign the mission after the Space Shuttle Challenger disaster
in 1986, with more time and more work, but we would be 
the first scientific mission after the Challenger. Finally we launched it, November 18, 1989, 36 years ago now, and it 
worked beautifully. Just eight weeks after launch, we showed our first data (the CMB spectrum) to the 
American Astronomical Society and 
got a standing ovation. Altogether 1500 people worked on 
COBE, and they all knew they’d done something special.

But interpreting the DMR data was really hard. Ned Wright made us the first map of the hot and cold spots and we knew we were 
onto something, but we knew how terrible it would be if we were wrong. (Cold fusion and polywater were not so distant 
memories.) We knew we had problems with magnetic field sensitivity, and with diffracted light from the Earth getting over 
the edge of the sunshade into the antennas. We checked and rechecked, under the leadership of Chuck Bennett, while George 
went to Antarctica to work on a different experiment. Fortunately, we had internal confirmation from another instrument, a 
balloon payload launched by other team members, and they saw the same spots in the area they could study. Finally, we, the 
whole team, were sure enough, and George got to present the results. It was a worldwide sensation, with front page pictures 
in newspapers around the world. We were all so proud, and Stephen Hawking said it was the ``most important scientific 
discovery of the century, if not of all time.'' Higher praise we could not imagine.

George also got into hot water with the science team for not following the team agreements about publication and 
journalists, and for the book he wrote with Keay Davidson, \textit{Wrinkles in Time,} 
in 1994. As a result the science team said I had to 
write a book, which I did with John Boslough called \textit{The Very First Light} in 1996.
In the end the books were part of the tale 
that the Nobel committee had to consider.

After that, Chuck Bennett and Dave Wilkinson led the team that built the WMAP satellite, the Wilkinson Microwave Anisotropy 
Probe, and they confirmed our maps from the DMR, with better equipment and a sharper map. The WMAP was launched in 2001. It 
was followed by another even more sensitive payload, the Planck mission from ESA, launched in 2009. Again they confirmed 
the results from COBE and WMAP.

So we shouldn’t have been too surprised when we got the calls from Stockholm
in 2006. We all knew we’d done something worthy of a 
Nobel Prize, 
if we were right. The Nobel committee said it was the beginning of cosmology as a precision science, and now it is, 
with a standard model, and parameters with error bars of a few percent. There are still some issues with the standard 
model, because there are at least two ways to estimate the Hubble constant, and they don’t agree within the error bars. So 
maybe we’re missing a piece of the model. And of course the model requires two surprising items that can’t be measured in 
the laboratory yet: cosmic dark matter and cosmic dark energy. Onwards!

So thank you George for your leadership of the DMR team. It was an exciting adventure to work with you. And it turned out 
beautifully.


%% file: perlmutter.tex
\phantom{singini}

\begin{figure}[h] 
\begin{center}
\includegraphics[width=0.7\textwidth]{%
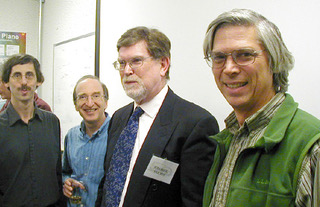}
\end{center}
\caption{%
\textbf{%
Eric Linder, Saul Perlmutter, George, and Kevin Lesko at the 2006 INPA celebration of George’s Nobel Prize.}
[Photo credit: UC Berkeley]
}
\end{figure}

Since I was a graduate student to today, George was always a kindly presence in my life. My PhD research group was 
under Rich Muller, and when I joined that group based at Lawrence Berkeley Laboratory (LBL), it was still closely affiliated with George’s group, 
coming out of their previous work together. So I got to know George first in that role, and he was already an 
encouraging force. In fact, I remember that he helped me with a practice exam for the orals and qualifying exams. 
George’s friendly, supportive role continued all the way through the years of his most visible cosmology 
discoveries---which were so important in shaping the field that I work in---and through the following decades 
even as he spent less and less time in Berkeley.

Through most of these decades my image of George is almost always of him dropping by to chat. He always had interesting 
ideas he was currently thinking about (be it surprising ideas of fundamental cosmological physics or expectations of 
the loss of jobs when AI and robots took them---and this was some 20 years ago). In this broad range of interests and 
the pleasure in applying scientific analytic tools to any topic of interest (scientific or not), he also represented 
for me a model of that generation of physicists who I saw around me at LBL and on campus at Berkeley.

And George always had advice for me when he came by. This was partly because he still had an advisor role and partly 
because it was one of the ways that he enjoyed thinking about the world. Even after he had left Berkeley, every now and 
then I would know he was in town because he would drop by, say hi, and give me his latest advice. Actually, I think it 
was almost always great advice---but I think I almost never figured out how to follow the advice! I don’t know if 
others have had that experience.

George pursued many directions, both intellectual and international, and this meant he always had stories to tell from 
his visits to Korea, the former Soviet Union, China, or of course France, his primary location. Sometimes the stories 
were about a new idea for a science experiment or project, or the development of some new institute, but sometimes it could 
be an investment in an air purifier company.

Along the way, George got me involved in one of the iMax cosmology films---and I had a chance to see how the outreach 
work was clearly something he really enjoyed.

My most recent visits with George were in Paris---and I should finish by saying that not only was Paris clearly a great 
intellectual home, but also that N\'ora made it a personal home. I think we all feel for her loss.

I’ll miss George and always half expect to see him dropping by with stories, thoughts, and ideas.

%% file: federico_piazza_smoot_memory.tex
%
%
%
%
%

In the list of people to whom I owe the opportunity to move forward in my career in physics, George Smoot 
surely stands very close to the top.

Already a relatively senior postdoc with an uncertain future, in 2009 I was offered the chance to become a 
member of the newly formed Paris Center for Cosmological Physics (PCCP) by Pierre Bin\'etruy and George Smoot. 
My Paris years were full of events, happy memories, and important collaborations that allowed me to land, soon 
after, a permanent position. Whether Pierre and George's choice was good for physics is still unclear, but it 
was certainly decisive for me.

Going to lunch with George in the neighbourhood of the Biblioth\`eque Fran\c{c}ois Mitterrand was 
unfailingly an adventure. We would try to order some food. Despite his American accent and my Italian accent,
we could always make ourselves understood by the waiter. French words coming out of George's mouth were always 
accompanied by a funny expression somewhere between disbelief and pride. The rapidity of his flow of thought, 
the wideness of his mental connections, made the discussions precious, and at times hard to follow. He loved 
paradoxes, simple anecdotes, and dimensional analysis.

I remember once he was particularly amused by one remark that I made: Immortality was not completely excluded, 
since roughly half of the human beings who ever existed are still alive. ``I think it is closer to one 
tenth,'' George replied laughing, ``but you are basically right: it must still be a large fraction of the total.''

George was clearly an exceptional human being. I wish I could have shown him more gratitude for our time 
together (and for having hired me!).


%% file: rittner.tex

Those who spent time with George often knew him in a particular context---in part because George had both a 
need and an ability to compartmentalize his interactions and relationships. It is a common trait among people 
with complex lives, tremendous responsibilities, and commitments. I’m not exactly sure where I and my late 
husband, Peter, fell into those compartments but I like to think of us as friends. I met George at Logan 
airport in 1970. He was heading to Berkeley for a fellowship with Luis Alvarez and was vibrating with 
excitement. Peter and I were heading to Miami where Peter’s family lived. They hadn’t seen each other in the 
two years since Peter left MIT. They had a lot of catching up to do. As we were about to leave, I said, ``Nice 
beard.'' He laughed and said, ``I grew it so I look older than the graduate students at Berkeley.''

That initial encounter at Logan airport proved typical of my over 50 year friendship with George. He seemed as 
interested in what I thought as what Peter thought as we talked about what we were doing, reading, 
international politics, and movies---all in a rapid pace since there was so little time before our departures. 
As he walked away to catch his plane, Peter said to me, ``I’d put money that he wins a Nobel.'' 
``I heard that,” George replied.

On the plane to Miami, Peter told me how he and George connected at MIT. They had rooms on the same hall on 
East Campus and theirs was a friendship that formed quickly and lasted their lifetimes. The superficial 
intersect was literature and mathematics, and they shared a talent for taking seriously 
what they were thinking and working on 
but never themselves. At the heart of their connections was that both were born 500 
years too late. They were Renaissance men with exceptionally wide and intense areas of interests and the 
pleasures of conversations, as they explored those wide ranges of topics and thoughts. It was clear that both 
were voracious readers, and when they were at MIT, Peter spent all of his discretionary cash at the Harvard 
Coop. George would stop by Peter’s room in the evenings to talk and before he left would browse Peter’s 
books. ``Mind if I borrow this?'' was a common exit line. They made their way through Peter’s library 
encompassing economics, philosophy, politics, essays, some poetry, and contemporary novels. Peter said that 
within a day or two he’d bring back whatever he borrowed having read it, consumed it, critiqued it, and 
compared its merits and deficits to other writers of the same genre.

George called Peter to let him know that he was prescient as he’d just been notified that he was now a Nobel 
Laureate in 2006. Peter wept as he congratulated him. Their friendship endured and thrived until 2023 when Peter 
died. Their chances to talk were made easier once long distance phone calls were replaced by mobile phones, 
emails, Zoom, and Facetime---and they did so as often as they could.

Most calls and all visits with George had to be arranged in advance because George’s schedule was so complex, often 
needing to account for time zone differences and George’s pattern of altering his sleep-wake cycle to 
minimize jetlag. First we’d or he’d send an email asking, ``Where are you?'' ``Am in Pamplona today and tomorrow,
and then go to Zurich for a couple of days, and then to Lindau,'' he typically replied to one inquiry before we 
settled on a time to talk. Once the pandemic started, it was easier to plan calls but then again, his sense 
of being trapped became obvious. George loved being fully emerged in the whole expansive world (commenting 
once he’d been on every continent but not to the North Pole), but as the pandemic unfolded, he sheltered in Paris. ``Have 
moved early to Paris,'' he reported. ``Problem is that it is relatively cooler (read need heat and warm clothes 
and drinks of soup) in Paris\ldots So now I am an American in Paris except in the bathroom where I am European!''

Nóra’s entrance into George’s life ten years ago changed George in many wonderful ways. Paris was more than 
just another address between wherever he was and Berkeley. It was where ``they'' made their home and for the 
first time his emails were filled with ``home,'' ``we,'' and ``us'' as his personal life became more 
central to his sense 
of self. 

George celebrated the waning days of the COVID shutdown. ``They'' were finally eating in restaurants. ``\ldots you 
could get a table at a restaurant and eat well if you have the money.'' But as was typical, he concluded with, 
``Now I am thinking that the poor will certainly suffer, and the poor in poor countries especially. However 
even some in moderately well off countries can suffer too due to poor leaders and war makers.'' Likewise, many 
of his letters were on poverty, oppression, the world, and USA politics. Classic were his worries during the 
2019--2020 student protests in Hong Kong that focused less on his own safety than on the risks to his students 
and on how worried he was for them.

George loved teaching. We shared many similar concerns as university faculty members. Because I was in the 
social sciences, what we shared was obviously not discipline focused but the desire to ensure the best for 
students including securing resources to support them. Typical of these exchanges was an email in 2025: ``I 
have just, I hope, completed one set of short lectures. These were on questions opened by an interesting 
science fiction book that was a best seller in China, \textit{The Three Body Problem} by Liu Cixin, which is very good 
and inventive and is being revived as an animated series in China. A lot of it is about the ecosystem and 
threat of intelligent life in the Universe. Sent the videos off and hopefully they will like them I will not 
have to redo. Now to catch up a bit on the other things I put off while doing that.''

Before his death, his plans to retire were marked by an urgent need to complete the fellowships of two 
students in Hong Kong, because they were doing exciting work and equally importantly because of a keen sense of 
the political realities for those students. We talked often about the importance of the ``next generation'' of 
researchers, the next questions they would ask, and the next roles they would have in society and in 
universities. In summer 2025 he wrote to me: 
\begin{quote} Einstein published his four papers for Germany in November 
1915 and the field equations paper was the last on Nov 25th, 1915. We tried to coax them to come out with a 
result on that date for the 100th anniversary but they wanted to finish the science run (Jan 12) and then 
make all the arrangements. Still it is a big deal but already 100 years since general relativity and prediction of 
gravitational waves, 50 years since the first idea of detectors that might work and 40 years since Rai Weiss 
wrote his paper on how to actually design and size LIGO and a lot of work.  Sometimes progress in science is 
not so fast.
\end{quote} 

The last time I saw George in Paris, he had shaved. I asked him why and he said: ``I grew it so I would look 
older. I looked in the mirror one day and saw that I am. So I shaved it.'' I so wish he had lived to be old 
enough to outlive me. I will miss him.

%% file: seljak.tex
\phantom{singini} 

\begin{figure*}[h]
\begin{center}
\includegraphics[width=0.5\textwidth ]{%
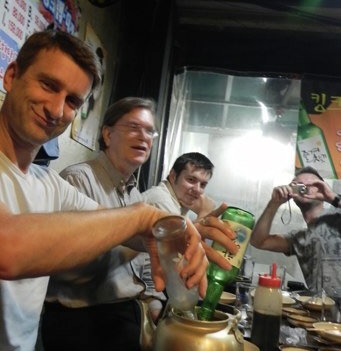}
\end{center}
\caption{%
\textbf{%
Uro\v s Seljak, 
George, Zvonimir Vlah, and Nico Hamaus
dining in Seoul, South Korea.}
[Credit: Uro\v s Seljak]}
\end{figure*}

Looking back at my career, I realize how many of my professional events were connected to George in one way or 
another. It was in Spring 1992, just after 
Cosmic Background Explorer (COBE) results came out, that I declared to my PhD advisor: I want 
to work on the theory of the Cosmic Microwave Background (CMB). It is the future. My first paper was a re-analysis of COBE 
DMR correlation function, for which he received the Nobel Prize in 2006. Many years later, George recruited me to come 
to Berkeley in 2006, and it was during this period that I got to know him well through our conversations of 
the pros and cons of UC Berkeley and Lawrence Berkeley National Laboratory (LBNL). He was persuasive and he was a fresh Nobel prize winner. And indeed 
I joined the Berkeley faculty and LBNL in 2008.

At the time George had just established Berkeley Center for Cosmological Physics (BCCP), using his Nobel Prize 
as seed money. BCCP started with a bang, recruiting some of the best postdocs in the world and rapidly 
establishing itself as the premier center for cosmology worldwide. George also participated in the fund 
raising for BCCP in the following years. It was an enormous privilege for me when George passed the baton 
of the BCCP directorship to Saul Perlmutter and myself. BCCP continues to be a vibrant center for cosmology, 
recruiting postdocs, and supporting workshops---and we all owe a big thank you to George for establishing 
it.

In 2008, George also recruited me to join him in participating in the World Class University initiative of 
the South Korean government. The idea was to spend two months in South Korea each year, establishing a research group at 
one of the South Korean universities. George had some connections with Ewha Women's University, so we jointly applied 
and received the grant. This is how that adventure began and continued for five years. And what an 
adventure it was, establishing a research group in a foreign country! I suspect this is when George got the 
travel bug, which then took him 
all over the world  
to establish many other research centers.
We barely saw 
him in Berkeley ever since.

But the things I remember most about George are the little anecdotes that I still enjoy sharing among 
friends. For example, once I asked him what his plans were with the \$1 million USD he 
won in 2008 at the 
TV show ``Are you smarter than a fifth grader?'' and whether he was serious about opening a Lube shop in Vegas. 
His answer to all of the above was: it’s complicated and I cannot speak about it. I suspect he never saw 
the money, and I am not sure the Lube shop in Vegas was his idea either. George was also immortalized around 
the same time by appearing in the cult TV show ``The Big Bang Theory.'' Both of these TV appearances were 
arranged by the BCCP executive director at the time. Alas, she left BCCP shortly thereafter and the tradition of 
BCCP directors appearing on TV shows died with it.

When I showed up in Seoul, South  Korea for the first time, George was present everywhere in the subway, in a video 
ad for an air humidifier (I think). So for a brief period he was the most famous non-Korean in South Korea, seen by 
pretty much everybody there. 

%

%% file: mehdi.tex

\phantom{singini} 

\vskip -1cm 

\begin{figure}[h]
\begin{center}
\includegraphics[width=0.65\textwidth]{%
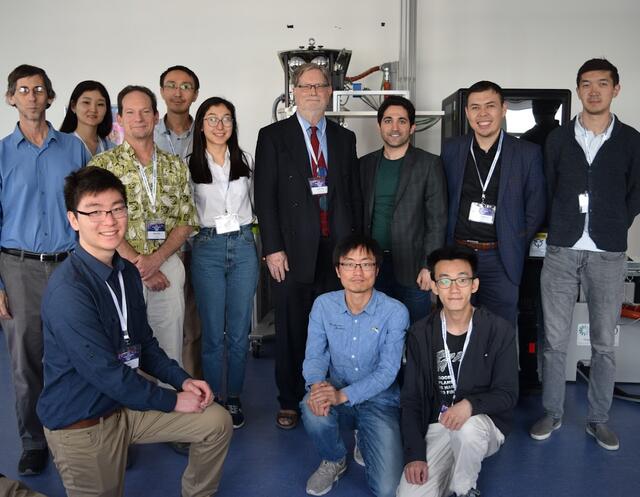}
\end{center}
\caption{%
\textbf{%
ECL (Energetic Cosmos Laboratory) Group in Kazakhstan (2019).}
(Standing: left to right) 
Eric Linder, 
Zhanel Tagayeva, 
Bruce Grossan, 
Zhanat Maksut, 
Marzhan Bekbalanova, 
George Smoot, 
Mehdi Shafiee,
Bekdaulet Shukirgaliyev, 
Zhaksylyk Kazykenov. 
(Sitting, left to right): 
Sean (Siyang) Li, 
Jie Hu, 
Albert Wai Kit Lau. 
[Photo credit: ECL (Energetic Cosmos Lab)]
}
\end{figure}

\begin{figure}
\begin{center}
\includegraphics[width=0.60\textwidth]{%
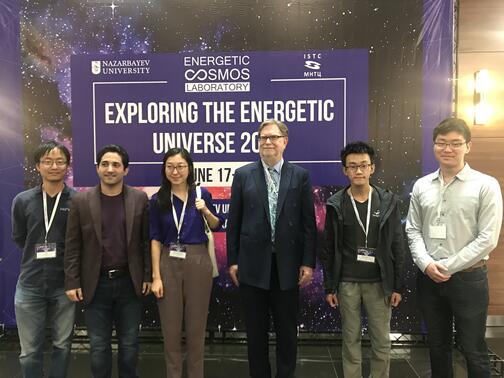}
\end{center}
\vskip -5mm
\caption{%
\textbf{%
ECL event. (June 17-19, 2019).}
Left to right:
Jie Hu, Mehdi Shafiee, Marzhan Bekbalanova, George Smoot, Albert Lau, and Sean Li.
[Photo credit: N\'ora Csisz\'ar]
}
\end{figure}

I first met Prof.~George Smoot
in August 2017, when I joined the Energetic Cosmos Laboratory (ECL) in Kazakhstan as his postdoc. On my very 
first day after arriving in Astana, Kazakhstan, 
an ECL conference was being held, followed by a trip to a nearby city about 250 km 
away. I had the remarkable opportunity to sit next to George throughout the journey. This became an exceptional chance to 
meet him for the first time, to talk extensively, and to begin learning from him immediately. During that conversation, 
George described a clear roadmap for my scientific development and outlined what he expected me to accomplish. The tasks he 
assigned gave me long-term direction and commitment, and I remain proud of having pursued them to completion. Under his 
supervision, I worked on Microwave Kinetic Inductance Detectors (MKIDs), a field for which he had deep vision and 
enthusiasm. George believed strongly that MKIDs represent the future of imaging technology and that they would eventually 
replace many existing detector technologies, especially for understanding the universe through fast transient signals. He 
also believed that MKIDs were not limited to astronomy but had significant potential in other fields such as 
superconducting quantum computing and even biology.

In recent discussions, we explored how MKIDs could be applied to measure phenomena related to radiation effects in quantum 
systems and biological samples. As George expressed:
\begin{quote}
\onehalfspacing%
I think that cosmic rays and natural radioactivity from the terrestrial environment are at a level to interrupt qubits in 
quantum computing unless we either shield or make the qubits much smaller in volume. Stem and T-cells are roughly 30 
microns in size and get hit by radiation about three times a year at sea level according to my calculations. When alive 
they can make some repairs to the damage caused by ionizing radiation. However when frozen, I think that the damage track 
will simply be frozen in and added to by subsequent years. Thus, if we want to take the best possible stem cell from 
umbilical cord blood, then freeze it and store it for 40 or 50 years, we need to protect them. If one takes from 20+ year 
olds (peak immunity) blood and separates and stores T-cells until they start developing cancer, again we want to store for 
decades. So two key technologies will be limited by natural radiation backgrounds, which need careful observational 
measurements and take place in shielded locations.
\end{quote}

He further emphasized the broader relevance of this research:
\begin{quote}
\onehalfspacing%
I think that the mKIDs can only monitor and determine the event and energy deposition rate. However for quantum 
computing they can also register large events or be coupled with scintillators and SIPMs to provide notice of large 
disruptive events. However I think it very worthwhile to provide mKIDs that replicate the area and volume of actual 
physical qubits and measure the spectra of events from cosmic and local radioactivity. If these are serious qubit disruptions, 
there will be an issue in reducing them by undergrounding with an overburden to reduce the cosmic rays by $10^4$ and shielding 
against gamma rays primarily from the thorium in the rocks. I am looking into this a bit now theoretically.
\end{quote}

George was not only my mentor. He also taught me how to navigate the scientific community, how to stay motivated, how to 
find one’s own path, and how to think critically and independently. His guidance shaped both my scientific direction and my 
professional mindset.

We will all miss him deeply.

%% file: spadafora.tex
As someone who has spent a long time in Berkeley, I've had the privilege of seeing many worldclass physicists in action, 
but without a doubt George Smoot stands out as one of the most brilliant and colorful people around.

I first got to know George personally when I switched from particle physics research to helping Bernard Sadoulet run the 
Center for Particle Astrophysics (CfPA), based at UC Berkeley. George was a senior scientist at the Center, and he 
enlivened (to put it mildly) every meeting with insights and opinions, which he didn't hesitate to share.  With the recent 
discovery of dark energy, with dark matter detection experiments getting going, and with the 
Cosmic Microwave Background (CMB) 
revealing the properties 
of the universe, these were exciting times, and George was there, at the center of things.

The CfPA ended in 2001, and we had proposed new centers, but without success.  Then, the great excitement was the 
announcement that George was sharing the Nobel Prize with John C. Mather in 2006. 
I vividly remember the unforgettable day of the 
announcement. George was gracious in receiving the prize, sharing the excitement with all of us, and inspiring us with his 
enthusiasm for understanding the universe. Then, on winning the Nobel prize, on the spot he single-handedly launched a new 
center, the Berkeley Center for Cosmological Physics (BCCP), 
filling the gap left by the ending of the CfPA.  This ``cutting the 
Gordian knot'' was characteristic of George---unlike anything I've seen before or since in science administration.

George had always been passionate about support for young scientists.  He welcomed and inspired students at all levels, 
and touched the lives of many people, who have gone on to various careers.  Nothing exemplified this more than his 
donating, without hesitation, 
his Nobel prize money to the new center as an endowment to support postdoctoral 
fellows. At the same time, he engaged in education and outreach with much enthusiasm and brought in staff to help with 
this.  He joined with another outreach effort to start a summer program for high school teachers and students focused on 
particle physics and cosmology. This program is still going strong, now in its twentieth year. Through it, George touched 
the lives of many students and teachers, inspiring them with the joy of learning. In the early years, George gave 
introductory talks in which he'd start off with simple topics but soon would go deeper into the subject and you could see 
his brilliant mind at work.  He couldn't resist sharing his deep understanding of physics and making connections between 
ideas, leaving all of us to realize there was a lot more for us to learn. But this was done in a good way, inspiring to 
all.

After launching the Berkeley center, George went on to create similar centers all over the world. Although he became an 
internationally known scientific personality, he stayed in touch with us in Berkeley. We kept his office for him, holding 
out hope he'd return and spend time with us again.  Despite his global travels, he would always answer emails promptly.  
These later exchanges were a delight to read, letting us know what he was up to and telling us of his latest projects, 
which could be anywhere on the globe.  George would offer insights into both scientific and general trends, always showing 
deep understanding. He was quick to offer all sorts of advice---his last email to me had advice on investing for 
retirement. These were a delight to read (even if I can't say I took advantage of his advice).

Looking back, George was an inspiring mentor and friend. His love of knowledge, which was not limited to physics, and his 
love of life greatly enriched all our lives. It is consolation to know that many people all over the world have been 
inspired by George. He will be greatly missed.

\begin{figure*}[h]
\begin{center}
\includegraphics[width=0.5\textwidth]{%
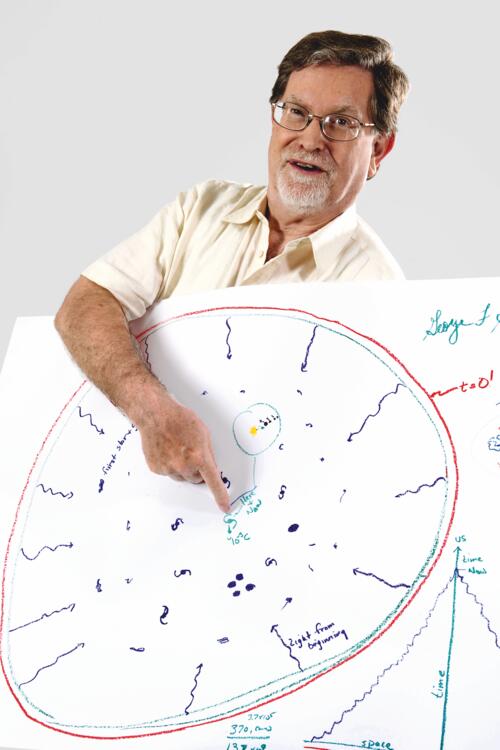}
\end{center}
\caption{%
\textbf{George Explaining the Essence of the Big Bang.}
[Photo credit: Volker Steger. Courtesy of Lindau Nobel Laureate Meetings]
}
\end{figure*}

{
\begin{quote} \small\setlength{\baselineskip}{12pt}
This is a drawing of the entire universe! Wow!\\
Smoot wants to carry the drawing on his shoulder, just like Atlas carried the world\\
on his shoulder. But paper just does not have the necessary rigidity,\\
and we decide to do something else for the shot.\\
I ask Smoot where we are in the drawing. He points to the centre,\\
some place close to the sun. “Is that you?”\\
“No“, he says, “it's the observer, he is always in the centre!”\par
\end{quote}
}

%% file: pintos.tex

\phantom{singini}

\begin{figure}[h] 
\begin{center}
\includegraphics[width=0.7\textwidth]{%
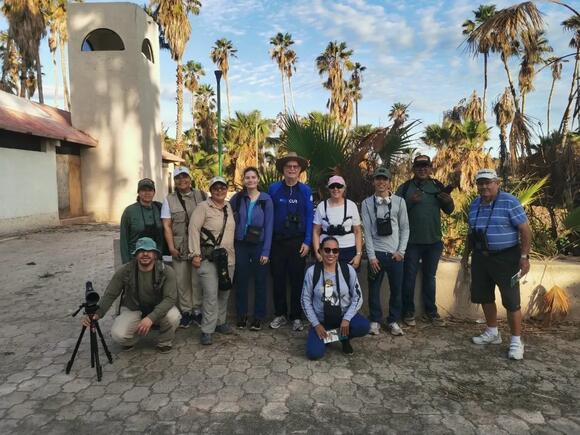}
\end{center}
\caption{%
\textbf{%
George Smoot, N\'ora Csisz\'ar, Graciela Tiburcio Pintos with others at the Christmas Bird Count (December 23, 2023).}
[Photo credit: Graciela Tiburcio Pintos]
}
\end{figure}

On December 4, 2018, I had the opportunity to meet Dr. Smoot in La Paz, Baja California Sur, Mexico, 
during his keynote address 
``Wrinkles in Time.'' It was amazing to hear him speak. At the end of the event, a photo with my daughter and his signed 
book was more than I could have asked for.

But fate had a surprise in store for me. Five years later (on December 13, 2023), Dr. Omar López-Cruz arrived in San José 
del Cabo with a couple of friends who needed 
a ride to the nearby lower
Cabo San Lucas. I gladly offered to pick them up at the airport and 
drive them. Imagine my surprise when I learned that Dr.~Smoot and his partner, N\'ora, were with them. From that moment on, we 
forged a friendship, one marked by adventures that immersed us in the culture and wildlife of Baja California Sur, far 
removed from the scientific realm.

George, as he asked me to call him, despite being a great speaker, was a quiet person who spoke only when necessary, but 
when he did, his opinions were forceful. Passionate about Mexican food and baseball, he enjoyed Mexico so much that he had 
a residence in Los Cabos, where he spent the winter. N\'ora is a kind, polite woman, always attentive to George's needs. 
They were a great example of a couple where love and respect were evident.

The day of his arrival was marked by a seafood lunch and his participation in a sea turtle release. After that, there were 
days filled with numerous activities. To be honest, I never thought he would accept the invitations I sent him. But 
without hesitation, he was present for the Christmas bird count. We went whale watching, swam with whale sharks, and ate at 
every regional restaurant we could. They even came to my house to enjoy a meal with my parents, who prepared the typical 
Veracruz-style Arroz a la Tumbada (rice cooked in broth with shrimp), which George loved so much that he readily asked for 
takeout so he could keep eating it later. My mother, of course, packed him his ``itacate'' (as we say in Mexico to refer to 
the containers of food we use to take our lunch to the office). Without a doubt, he was never tired of traveling, eating 
delicious food, and having so many adventures.

During a radio interview, George admitted that in all his years living in Los Cabos, he had never seen as much of the 
place as he had during that season. And impressed, he recounted all the adventures and places we had visited, publicly 
thanking me (the latter was unnecessary). But I recognize a simple and grateful man (beyond being a Nobel laureate), a 
lover of Mexico, still capable of being awestruck by nature.

``We are stardust,'' Dr. Carl Sagan once said… George didn't die, he simply returned to the place from which we came. 
Today he is in the cosmos---the place he dedicated his entire life to studying---and appears every night as a star.

\begin{figure}[h] 
\begin{center}
\includegraphics[width=0.9\textwidth]{%
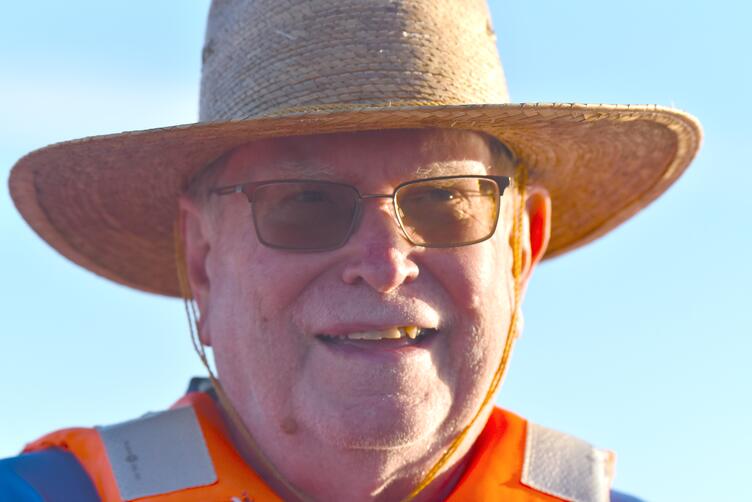}
\end{center}
\caption{%
\textbf{%
Jorge el mexicano en Los Cabos.}
[Photo credit: Graciela Tiburcio Pintos]
}
\end{figure}

%% file: tye.tex

\phantom{tom} 

\vskip -5mm

\begin{figure}[h]
\begin{center}
\includegraphics[width=0.7\textwidth]{%
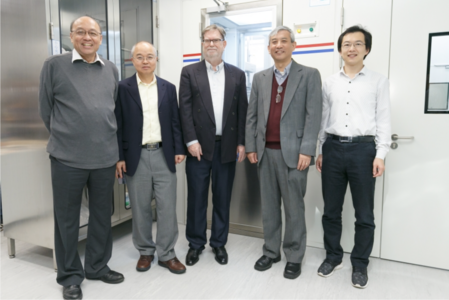}
\end{center}
\caption{%
\textbf{George and Henry Tye at IAS HKUST.}
(left to right): 
Henry Tye, Yang Wang,  George Smoot, Penger Tong, 
and Yi Wang. 
[Photo credit: HKUST] 
}
\end{figure}

\vskip -5mm

I heard of George relatively late. About 1980, after Alan Guth proposed how the 
universe began, namely the inflationary universe (with some help from me), we chatted 
about the future development of cosmology. At that time, we believed that the 
inflationary universe idea will never be tested in our lifetime. So it was a shock to 
me when the temperature variation measurement from COBE, led by George, came out in 
1992. This amazing result and the advanced technology it relied on opened the door for 
testing the inflationary universe paradigm, which has been a very active area of 
cosmological research ever since. So by 1992, George became a hero of mine.

The time I actually interacted with George began in 2014. When I became the director 
of the Institute for Advanced Study (IAS) of the Hong Kong University of Science and 
Technology (HKUST), I actively recruited top scientists and scholars to the IAS. I was most 
fortunate that George was willing to spend 
some time in Hong Kong. During his time at IAS, HKUST, 
we discussed science, and even collaborated on projects on wave dark matter. It was 
always fun to pick his brain for new ideas. George was very broadly knowledgeable and 
interacted well with everyone, providing a lively atmosphere for the group.

Outside of science, we also socialized a lot. We often grabbed some take-outs and ate 
lunch together. Together with N\'ora, my wife Bik, and others, we explored a variety of 
food in Hong Kong. George never turned down a chance to enjoy the seafood in the nearby 
Sai-Kung.

We planned to meet again in the fall 2025 in Hong Kong, but then we were shocked to learn of 
George’s sudden passing. He’ll be sorely missed.

%% file: annWhetstone.tex
\begin{quote}

\Large 
\setlength{\baselineskip }{16pt}
\setlength{\parskip }{0pt}
{\it 
While George Smoot will be forever celebrated for his contributions to 
science, he’s remembered differently by some who knew him from an earlier time. 
}

\end{quote}

\begin{figure}[h]
\begin{center}
\includegraphics[width=0.65\textwidth]{%
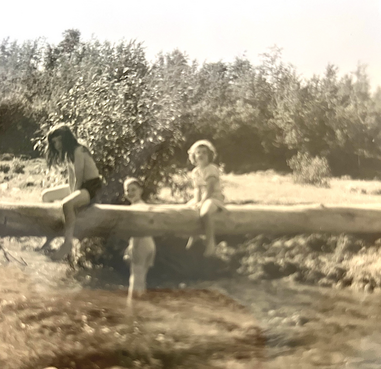}
\end{center}
\caption{%
\textbf{%
George in Alaska. 
}
(left to right): Ann, George, and Barb Whetstone.
[Photo credit: Ann Whetstone]
}
\end{figure}

I’ll always remember George as my oldest and dearest friend---a friendship that spanned 75 years.

It was in early 1950s Alaska, where our fathers worked for the United States Geological 
Survey, that we first met. George was six years old; I was seven.
Our friendship was cemented there. At the time, Alaska was America’s frontier. Our poor 
mothers---both Southerners---must have wondered about raising their children in such a cold, 
inhospitable place. Perhaps that’s why our families became inseparable. And no one more 
than George and me.

There were days of building forts and blazing trails—or so we thought. We scoured the 
ground of an Army artillery range looking for, and often finding, shell casings and 
occasionally, much to our delight, even unexploded ammunition, which we attempted to 
explode with rocks (not perhaps an early indication of intelligence). In our eyes, it 
didn’t get much better than that.
We chopped down trees---probably again to our mothers' consternation---with the intention of 
building a raft that would take us to heaven knows where. (It sank quite quickly.)
We dared each other to jump off roofs into snowbanks and rode our bikes (``look no hands'') 
down steep terrain, not without some risk as this was the Alaska of unpaved roads.
In the winters, which were long, dark, and very, very snowy, we raced our sleds down the 
road on Lazy Mountain.
Using our pocketknives---everyone had them back then—we became blood ``brothers.'' 

We were best friends---but not to say we didn’t have an occasional falling out.
One resulted in my shooting George in the thigh with my BB gun—after words to the effect 
that ``girls can’t do that.''
We both learned something from that experience.  I learned to never aim a gun at your 
friend; my parents immediately relieved me of that weapon. George learned that, indeed, 
girls could do anything boys could---knowledge that I think he always remembered.
The gun incident did nothing to tarnish our friendship. Indeed, over the years, George 
seemed to take delight in telling the story and showing the scar.

When both our families left Alaska in the mid-50s for Ohio, our lives changed---mostly 
intersecting only at holidays, which our families continued to celebrate together.
While we no longer attended the same schools, George and I both embraced the academic path. 
Two kids from a small Alaska school, where sometimes wandering moose ruined recess, went on 
to receive their doctorates—mine in history and George’s, of course, in the sciences.
Although we lived on different coasts, we never lost touch: phone calls, emails, Skype, and flying visits. And 
then when George and N\'ora settled in Paris, longer visits together with Paula. We were luckier than we 
knew.

\begin{figure*}[h]
\begin{center}
\includegraphics[width=0.7\textwidth]{%
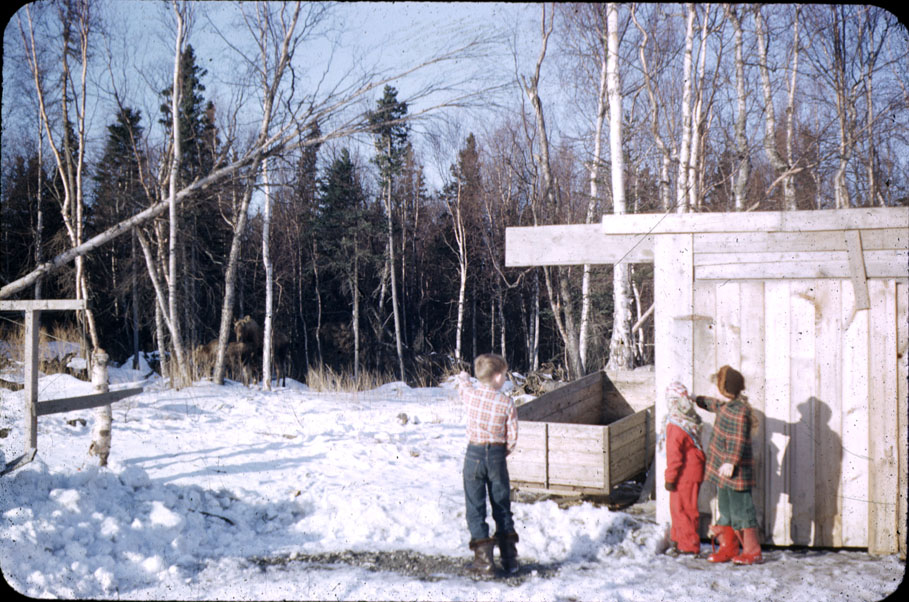}
\end{center}
\caption{%
\textbf{%
George pointing at mama and baby moose in Alaska.}
[Credit: Ann Whetstone]
}
\end{figure*}

\begin{figure}[h]
\begin{center}
\includegraphics[width=0.65\textwidth]{%
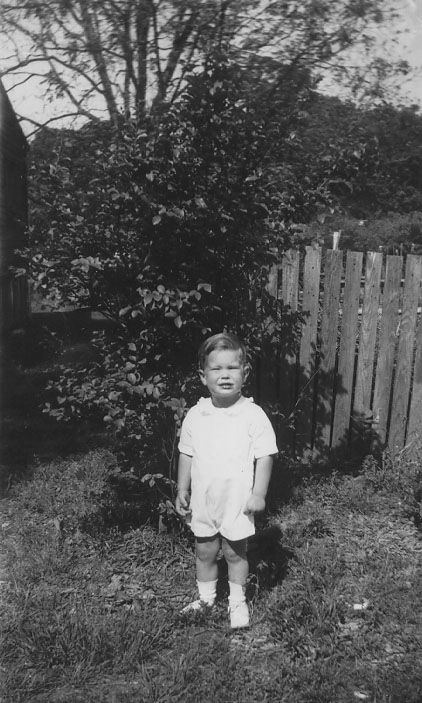}
\end{center}
\caption{%
\textbf{%
George at 2 years.}
[Photo credit: Ann Whetstone]
}
\end{figure}

\begin{figure}[h]
\begin{center}
\includegraphics[width=0.65\textwidth]{%
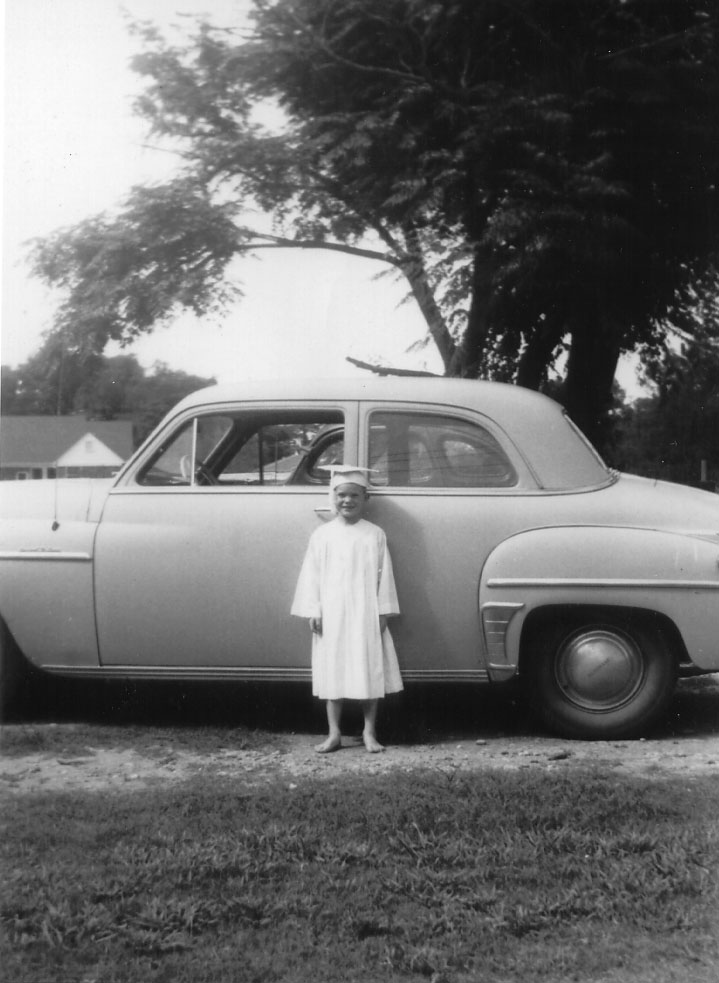}
\end{center}
\caption{%
\textbf{%
George graduating from kindergarten.}
[Photo credit: Ann Whetstone]
}
\end{figure}
\clearpage

%% file: alex.tex
\phantom{Alex}

\vskip -5mm 

\begin{figure}[h]
\begin{center}
\includegraphics[width=0.7\textwidth]{%
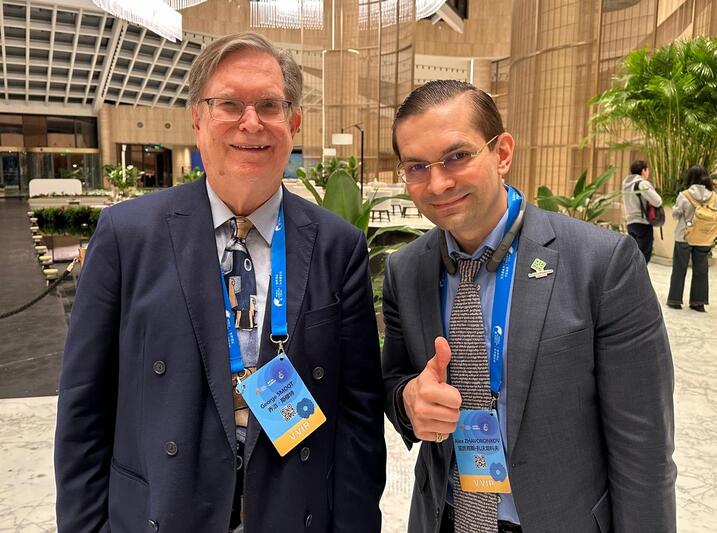}
\end{center}
\caption{%
\textbf{%
With Alex Zhavoronkov at WLA China 2024.}
[Photo credit: N\'ora Csisz\'ar]
}
\end{figure}

In November 2023, the World Laureates Association (WLA) event became the setting for an unexpected crossover between two 
very different fields of science: the origins of the universe, and the future of human longevity. It was there that I 
watched Dr. George Smoot, the Nobel laureate famous for capturing the ``first light'' of the infant universe, turn his 
attention to a completely new frontier. He didn’t approach the problem of aging like a retired dignitary looking for a 
hobby. Instead he brought the exact same data-driven intensity to biology that he had once used to map the cosmos. The 
world of longevity biotechnology celebrated when Dr. Michael Levitt, the ``father of computational biology'' and 2013 Nobel 
laureate in Chemistry, developed an interest in longevity research. And for those of us in the longevity field, having a 
master of experimental physics turn his gaze toward the mechanics of life extension was going to be another major 
milestone.

This shift quickly led to a close relationship between us. George and I met at
Hong Kong University of Science and Technology (HKUST), 
where our conversation skipped 
the usual pleasantries and dove straight into the practical mechanics of extending human life.

I’ll never forget how fast George moved that day. We were discussing rapamycin—a key drug in our longevity research—and he 
wasn’t just politely nodding at the data. Before our tea had even cooled, he was already on his phone, looking up the 
rapamycin protocols. His energy and curiosity were contagious. That lunch changed our relationship instantly. We went from 
being formal academic acquaintances to colleagues racing against the clock. His wonderful partner in life and 
science, N\'ora Csiszár, was always there to help and facilitate the work.

When George returned home to Paris, this initial spark turned into a frequent, highly technical correspondence. I 
sometimes needed a few days to process the lengthy highly-technical emails and respond. Together we began treating the 
microscopic vulnerabilities of the human cell with the exact same seriousness he usually reserved for the physics of the 
universe.

\noindent
\textbf{The Man Behind the Science: Jokes and ``Seeing Gods''}

To the public, George was a rarity: a Nobel laureate who was completely approachable. He happily went on 
\textit{Are You Smarter Than a Fifth Grader?} (and won the top prize) and made cameo appearances on 
\textit{The Big Bang Theory.} He didn’t do these things 
for his ego. He did them because he believed scientists needed to be visible, fun, and human to inspire the next 
generation. He proved that having a profound intellect didn't mean you had to lose your sense of play.

%

Yet underneath his quick wit was a deep sense of awe. When his team discovered the ``lumps'' in the cosmic microwave 
background in 1992—the very irregularities that allowed galaxies and life to form—he told the press it was ``like seeing 
God.'' It wasn't ego talking. It was his deep reverence for the underlying order of the universe. In his final years, he 
brought that same reverence to our work in biology, looking at the human brain and body as beautiful, solvable systems 
waiting to be decoded.

\noindent
\textbf{%
Engineering the Future: Digital Clones and Cryobiology}

Throughout 2024 and 2025, George began applying the practical, ``can-do'' engineering style of the Lawrence
Berkeley Laboratory (LBL) to the 
concept of human preservation. He was no longer just asking how the universe began. He was asking how human intelligence 
could survive its biological limits.

In our discussions, we frequently debated the best path forward—specifically, ``Digital Clonings'' versus ``Hardcore 
Cryobiology.'' For George, who had spent decades correcting errors in satellite data, this was a strict engineering 
problem. He argued that while freezing a body might work for traveling around our solar system, traveling to other stars 
would require ``mind downloading.'' He viewed the human mind as a complex neural network that could eventually be mapped 
digitally and later paired with cloned biology.

He pushed our team to think bigger about physical preservation. He thought modern cryogenic systems were ``antique'' and 
sketched out ideas for highly ambitious units that could manufacture their own liquid nitrogen, backed up by failsafe 
argon systems to ensure they never lost power or temperature.

\noindent
\textbf{%
The Cryobiology Emails: From Science Fiction to Serious Physics}

George’s knack for practical engineering really shined in a fascinating email thread we had between December 2024 and 
April 2025. Realizing the massive physics challenges involved in successfully freezing and thawing human tissue, I brought 
two more experts into our conversation: pioneering computational and quantum chemist Dr. Alan Aspuru-Guzik and longevity 
scientist Dominika Wilczok.

In December 2024, I formally introduced Alan to George, kicking off a discussion about how to use high pressure and 
microwaves to quickly warm frozen tissue without forming deadly ice crystals.

George replied later that day, injecting his trademark humor into our high-level physics debate. To illustrate how hard it 
is to instantly freeze a thick biological object, he proposed something he called the ``DASAR''—Darkness Amplification 
through Stimulated Absorption of Radiation. He described it as a reverse-laser that would shoot a beam of pure cold. He 
wrote about it with such deadpan authority that the next day, Alan genuinely replied: ``Nice to meet you! The thread is 
very interesting. I will read up on DASAR. Do you have good references to it?'' All of us started doing research on 
DASARs. This idea seemed to be very fresh---we never heard of anything like this before.

George quickly apologized, admitting to us, ``I made up the DASAR as a joke \ldots It is what we would like to have to 
`instantly' freeze things like a laser or microwave can heat.''

Jokes aside, we quickly got down to business. By late December 2024, George realized that microwaves wouldn't warm tissue 
evenly enough. Instead, he proposed injecting magnetic nanoparticles (MNPs) into the tissue and using an alternating 
magnetic field to heat them via induction. As he explained to us, magnetic fields could penetrate biological mass much 
deeper and more evenly than specialized microwaves, offering a real engineering solution to a massive biological hurdle.

\noindent
\textbf{%
A Global Mentor}

George was deeply invested in protecting scientific talent, no matter where it came from. When geopolitical conflicts 
displaced around 60 top Russian biotechnology and AI experts, he didn't just offer his sympathies. He drew up a practical 
blueprint to create a ``Free Trade Zone for biomedical research'' in Tijuana, Mexico. He envisioned a safe haven where these 
``AI refugees'' could work as part-time professors and full-time innovators, free from the bureaucratic red tape that 
usually slows down science.

To Alan, Dominika, myself, and our broader team, George was a warm, hands-on mentor. He loved to test our problem-solving 
skills with ``Fermi questions''—like asking us to calculate exactly how much dinosaur breath is in a single human 
inhalation. It was his way of teaching us how to cut through distractions and find the real physics in any problem.

He lived his principles. He stayed active, famously walking two miles ``up the hill'' to his office at Berkeley Lab well 
into his later years. He was also incredibly generous, donating \$500,000 of his Nobel Prize money to endow the Berkeley 
Center for Cosmological Physics, ensuring the next generation of scientists had the resources they needed.

\noindent
\textbf{%
The Final Calculations and a Living Legacy}

By early 2025, George was focused on what he called ``lifespan terminal escape velocity.'' In an April email to our team, he 
confidently predicted that major leaps in AI modeling and stem cell technology would make the need to freeze entire humans 
obsolete within just five years.

His mind remained incredibly sharp. Right up until his final months, he was doing complex 
math on 30-micron stem cells, calculating that they are hit by background radiation ``three times a year at sea level.'' He 
warned us that long-term cryopreservation would require deep underground storage, carefully shielded from environmental 
radiation like the thorium found in everyday rocks, to prevent microscopic damage from accumulating in frozen cells. He 
was quite literally calculating the math needed to save others' lives as he neared the end of his own without knowing it.

George passed away in September 2025, right as this groundbreaking work was gaining momentum. The silence in my inbox was 
heavy. When our team gathered online for his memorial, and in my subsequent conversations with his partner, Nóra Csiszár, 
the sheer human weight of his loss really hit me.

But we refused to let the project stop.
Just a few weeks later, on October 16, Dominika emailed Alan and me to propose a formal research paper on using drug 
discovery models to design better cryoprotectants—a direct continuation of our emails with George. She noted sadly, ``Our 
conversation with George Smoot will sadly not follow. He passed away last month.''

Alan wrote back immediately with enthusiastic support: ``Let’s make this research project move the needle in cryobiology in 
George’s memory.'' Over the next several months, we worked tirelessly to turn those email debates into a published 
perspective paper. What had started as a brainstorming session with George became a living tribute to his genius.

Seeing Dominika, Alan, and the rest of our team seamlessly pick up the torch has been incredibly moving. Our paper on 
cryoprotectant design isn't just a research goal anymore; it’s a living continuation of George’s vision. And I am deeply 
saddened that he did not live long enough to enjoy the benefits of advances in cryobiology he contributed to.

George Smoot was a true Renaissance man. He was someone who could break down the radiation spectrum of the Big Bang and 
then casually discuss the cultural impact of a Taylor Swift concert in the very next breath. He was brilliant, totally 
uncensored, and fundamentally kind. As we move into the ``Golden Age of Longevity'' that he so clearly saw coming, we carry 
his light with us. I owe a massive debt of gratitude to a mentor who taught me to look at the stars in the sky and the 
cells in our bodies with the exact same unwavering curiosity.

George, we are going to continue the work. The escape velocity you predicted is now ours to achieve.

%% file: epilogue.tex
\phantom{Nora} 


\begin{figure*}[h]
\begin{center}
\includegraphics[width=0.7\textwidth]{%
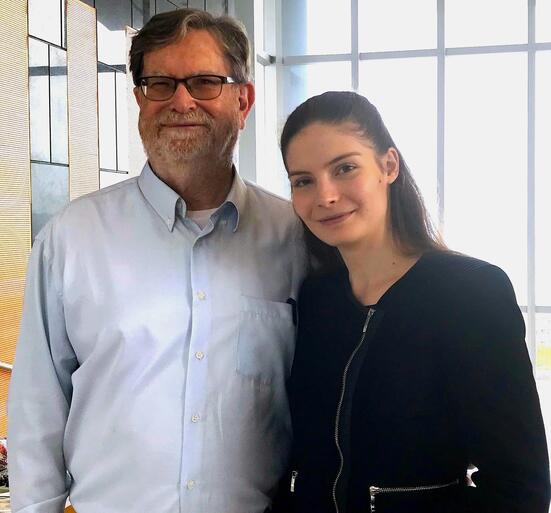}
\end{center}
\end{figure*}

\begin{center}
\huge In Loving Memory of George
\end{center}

Thank you so much to everyone for writing such beautiful essays about George. I can feel that
you are all missing him dearly\ldots 
Those of us who had the privilege to know George will miss his laugh, curiosity, playfulness \ldots 
his passion for knowledge and life, even when he felt disappointed by certain people and
world politics \ldots  his warm, caring, personal side, beside the rigorous, truth-seeking scientific
side.

He was a hardworking and 
generous man. Not only was he committed to science—working long
hours, coming up with new ideas, brainstorming with his colleagues, guiding his post docs. He
was also committed to giving back to the community when he was able.
As a teenager, he had three jobs (paperboy, mowing lawns, and golf caddy) in order to earn
money to get into MIT. With all the money that he saved up, it was only enough for one semester.
Once he was at MIT, he could get the necessary help to continue his education, but he always
remembered the challenges he faced as a poor student. As a result, he wanted to help young
people to not struggle financially and be able to focus on studying instead. He was involved in
many fundraising activities, and he donated his share of the Nobel Prize money to support the
next generation of scientists. He always guided his students with grace, encouraging them to
pursue their own research and to work on the hard questions of life, science, nature, and the
universe.

George wanted to leave the world a better place than he found it. I believe that he did. He
also wanted to encourage everyone to keep doing the best they can and strive to make things
better for humanity. He believed in science, international cooperation, and global problem
solving.

What I personally will miss is his kindness, care, and generosity---how he always protected me
and loved me throughout his last decade of his life 
spent together.
I remember the first time we met in 2015 at the Milan World Expo at the Hungarian Pavilion. It
was not a physics or cosmology related event. The theme of the conference was `How to Care for People and the Planet.'
This was something that he was striving for by using his scientific
expertise and know-how. He wanted to educate both the public and the people in charge
about possible future outcomes and how to get there.

He talked about how science can help humanity to overcome future challenges and showed
different ways we can actually handle the various crises currently affecting our planet. He
had tons of ideas on how to solve the big issues, such as climate change, pollution, war, and famine.
But his main point was always the same: we need to stop fighting and pull together as a
global society, using science as a way to resolve the ongoing problems of the world.
I remember watching him present with such passion and clarity. After the talk, I went to get
some food from the buffet and he just came up to me, introduced himself and asked me
about the various dishes at the table, and which I recommend. I was too shy initially, but he
made me feel comfortable and got me to laugh at his jokes and puns.

He had to give some interviews, but at dinner he wanted to sit with me and talk. He wanted to
know about my research in psychology and how important it is to understand the human mind and
consciousness.
He wanted to know how my thesis defense was going, so he gave me his
business card and asked me to reach out to him when I was back at home. I took his 
advice and suggestions when I successfully defended my diploma, and I was happy to tell him
that I received my MA in Psychology. To celebrate, he invited me to come to Paris, which I
accepted and the rest is history.

Throughout the years, we not only lived but we worked together on many projects, including
health devices such as portable air quality monitors, air purifiers, oscillating beds, and
couches. We were a great team as we could bounce ideas off of each other and find new ways
to make things work. Our scientific backgrounds were different (him being a physicist, me
being a psychologist), but that is exactly why we could work together effortlessly as we
complemented each other. He always encouraged me to also work on my own projects and
make my dreams come true with hard work and dedication---just as he did.

George and I shared so many small routines that added up to a beautiful life. The way he
smiled at me across the table and said something witty or funny to make my day brighter. He
made every ordinary day extraordinary because he was there. I still look for him in familiar
rooms, at the open air market, or on the streets we walked down together countless times.
Sometimes I can still hear his voice, his 
laughter, his encouragement.

His death was sudden and unexpected. There is no way to prepare for this kind of absence.
Losing a partner not only means saying goodbye to that person with whom you built your life
together. It also means saying goodbye to the person who you have been as a partner of
that person. Any love connection leaves a deep imprint. The deeper the love, the deeper the
grief.

I know that he is with me. He is with all of us who loved him, who learned from him, who
laughed with him, who cared about him.

Thank you everyone for honoring George’s legacy and memory.

Memory. Such a strange word \ldots I think that when someone becomes a memory, the memory
becomes a treasure. And I will keep this treasure in a special place in my heart \ldots  as I hope
you all will as well.


\begin{figure*}[h]
\begin{center}
\includegraphics[width=0.9\textwidth]{%
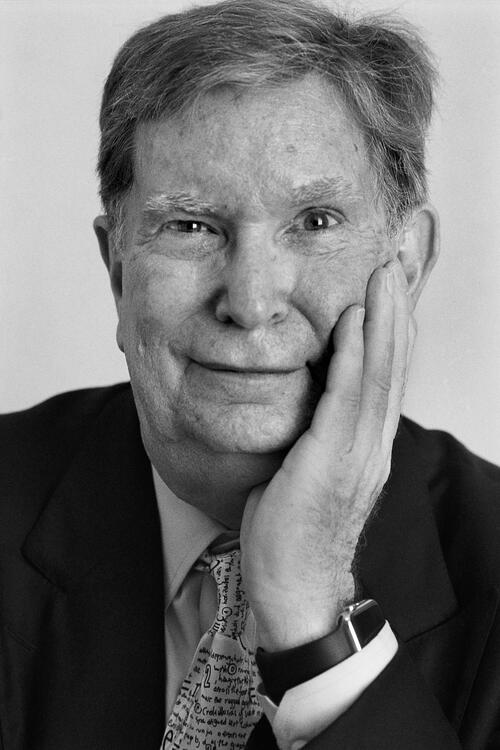}
\end{center}
\caption{%
[Photo credit:Peter Badge. Courtesy of  Lindau Nobel Laureate Meetings Foundation]
}
\end{figure*}

\clearpage